\documentclass[twocolumn]{aastex63}
\usepackage{amsmath,amstext}
\usepackage{mathtools}
\usepackage{afterpage}
\usepackage{lipsum}
\usepackage{capt-of}
\usepackage{relsize}
\usepackage{newtxtext,newtxmath}
\usepackage[all]{hypcap} 
\usepackage{breqn}
\usepackage{verbatim}
\usepackage{graphicx}
\usepackage{footnote}
\usepackage{float}
\usepackage{amsmath}
\usepackage{enumitem}
\usepackage{amssymb}
\usepackage{array}
\usepackage{gensymb}
\usepackage{mathtools}
\usepackage{relsize}
\usepackage{lipsum}  
\usepackage[graphicx]{realboxes}
\usepackage{natbib}

\usepackage{lipsum}
\usepackage{rotating}
\usepackage{nicefrac}
\newcommand\blfootnote[1]{%
  \begingroup
  \renewcommand\thefootnote{}\footnote{#1}%
  \addtocounter{footnote}{-1}%
  \endgroup
}
\usepackage{color, colortbl}

\newcommand{\astrothreed}{ARC Centre of Excellence for All Sky Astrophysics in 3 Dimensions (ASTRO 3D)}
\newcommand{\Cambridge}{Institute of Astronomy and Kavli Institute for Cosmology, Madingley Road, Cambridge, CB3 0HA, UK}
\newcommand{\CAPS}{Center for Astrophysical Surveys, Urbana, IL, 61801, USA}
\newcommand{\Carnegie}{The Observatories of the Carnegie Institution for Science, 813 Santa Barbara St., Pasadena, CA 91101, USA}
\newcommand{\CIERA}{Center for Interdisciplinary Exploration and Research in Astrophysics (CIERA), 1800 Sherman Ave, Evanston, IL 60201, USA}
\newcommand{\DARK}{DARK, Niels Bohr Institute, University of Copenhagen, Jagtvej 128, 2200 Copenhagen, Denmark}

\newcommand{\Einstein}{NASA Einstein Fellow}

\newcommand{\IfA}{Institute for Astronomy, University of Hawaii, 2680 Woodlawn Drive, Honolulu, HI 96822, USA}
\newcommand{\Illinois}{Department of Astronomy, University of Illinois at Urbana-Champaign, 1002 W. Green St., IL 61801, USA}
\newcommand{\JHU}{Department of Physics and Astronomy, The Johns Hopkins University, Baltimore, MD 21218}

\newcommand{\Melbourne}{School of Physics, The University of Melbourne, VIC 3010, Australia}

\newcommand{\NCSA}{National Center for Supercomputing Applications, Urbana, IL, 61801, USA}

\newcommand{\Northwestern}{Department of Physics and Astronomy, Northwestern University, Evanston, IL 60208, USA}
\newcommand{\NSF}{National Science Foundation Graduate Research Fellow}

\newcommand{\STScI}{Space Telescope Science Institute, Baltimore, MD 21218}
\newcommand{\Thacher}{The Thacher School, 5025 Thacher Rd, Ojai, CA 93023}
\newcommand{\Toronto}{David A. Dunlap Department of Astronomy and Astrophysics, University of Toronto, 50 St. George Street, Toronto, Ontario, M5S 3H4 Canada}
\newcommand{\UCSC}{Department of Astronomy and Astrophysics, University of California, Santa Cruz, CA 95064, USA}
\newcommand{\UCSD}{Center for Astrophysics and Space Science, University of California San Diego, La Jolla, CA 92093, USA}

\newcommand{\TCD}{School of Physics, Trinity College Dublin, The University of Dublin, Dublin 2, Ireland}

\usepackage{xtab}

\begin{document}

\title{\vspace{-3em} An Early-Time Optical and Ultraviolet Excess in the type-Ic SN~2020oi}
\author[0000-0003-4906-8447]{Alexander~Gagliano}\blfootnote{Corresponding author: Alex~Gagliano\\ \href{mailto:gaglian2@illinois.edu}{gaglian2@illinois.edu}}
\affiliation{\Illinois}
\affiliation{\NCSA}
\affiliation{\CAPS}
\affiliation{\NSF}

\author[0000-0001-9695-8472]{Luca~Izzo}
\affiliation{\DARK}

\author[0000-0002-5740-7747]{Charles~D.~Kilpatrick}
\affiliation{\Northwestern}
\affiliation{\CIERA}

\author[0000-0001-6350-8168]{Brenna~Mockler}
\affiliation{\UCSC}

\author[0000-0002-3934-2644]{Wynn~Vicente~Jacobson-Gal\'an}
\affiliation{\Northwestern}
\affiliation{\CIERA}
\affiliation{\NSF}

\author[0000-0003-0794-5982]{Giacomo~Terreran}
\affiliation{\Northwestern}
\affiliation{\CIERA}

\author[0000-0001-9494-179X]{Georgios~Dimitriadis}
\affiliation{\UCSC}
\affiliation{\TCD}

\author{Yossef~Zenati}
\affiliation{\JHU}

\author[0000-0002-4449-9152]{Katie~Auchettl}
\affiliation{\Melbourne}
\affiliation{\astrothreed}
\affiliation{\UCSC}
\affiliation{\DARK}

\author[0000-0001-7081-0082]{Maria~R.~Drout}
\affiliation{\Toronto}
\affiliation{\Carnegie}

\author[0000-0001-6022-0484]{Gautham~Narayan}
\affiliation{\Illinois}
\affiliation{\NCSA}
\affiliation{\CAPS}

\author[0000-0002-2445-5275]{Ryan~J.~Foley}
\affiliation{\UCSC}

\author[0000-0003-4768-7586]{R.~Margutti}
\affiliation{\Northwestern}
\affiliation{\CIERA}

\author[0000-0002-4410-5387]{Armin~Rest}
\affiliation{\STScI}
\affiliation{\JHU}

\author[0000-0002-6230-0151]{D.~O.~Jones}
\affiliation{\UCSC}
\affiliation{\Einstein}

\author[0000-0003-2094-9128]{Christian~Aganze}
\affiliation{\UCSD}

\author[0000-0002-6298-1663]{Patrick~D.~Aleo}
\affiliation{\Illinois}
\affiliation{\NCSA}
\affiliation{\CAPS}

\author[0000-0002-6523-9536]{Adam~J.~Burgasser}
\affiliation{\UCSD}

\author[0000-0003-4263-2228]{D.~A.~Coulter}
\affiliation{\UCSC}
\affiliation{\NSF}

\author[0000-0003-0398-639X]{Roman~Gerasimov}
\affiliation{\UCSD}

\author[0000-0002-8526-3963]{Christa~Gall}
\affiliation{\DARK}

\author[0000-0002-4571-2306]{Jens~Hjorth}
\affiliation{\DARK}

\author[0000-0002-5370-7494]{Chih-Chun~Hsu}
\affiliation{\UCSD}

\author[0000-0002-7965-2815]{Eugene~A.~Magnier}
\affiliation{\IfA}

\author[0000-0001-9846-4417]{Kaisey~S.~Mandel}
\affiliation{\Cambridge}

\author[0000-0001-6806-0673]{Anthony~L.~Piro}
\affiliation{\Carnegie}

\author[0000-0002-7559-315X]{C\'{e}sar Rojas-Bravo}
\affiliation{\UCSC}

\author{Matthew~R.~Siebert}
\affiliation{\UCSC}
\affiliation{\NSF}

\author{Holland~Stacey}
\affiliation{\Thacher}

\author[0000-0002-3019-4577]{Michael~Cullen~Stroh}
\affiliation{\Northwestern}
\affiliation{\CIERA}

\author[0000-0002-9486-818X]{Jonathan~J.~Swift}
\affiliation{\Thacher}

\author[0000-0002-5748-4558]{Kirsty~Taggart}
\affiliation{\UCSC}

\author[0000-0002-1481-4676]{Samaporn~Tinyanont}
\affiliation{\UCSC}

\collaboration{1000}{(Young Supernova Experiment)}

\begin{abstract}
We present photometric and spectroscopic observations of Supernova 2020oi (SN~2020oi), a nearby ($\sim$17~Mpc) type-Ic supernova (SN~Ic) within the grand-design spiral M100. We undertake a comprehensive analysis to characterize the evolution of SN~2020oi and constrain its progenitor system. We detect flux in excess of the fireball rise model $\delta t \approx 2.5$~days from the date of explosion in multi-band optical and UV photometry from the Las Cumbres Observatory and the Neil Gehrels \textit{Swift} Observatory, respectively. The derived SN bolometric luminosity is consistent with an explosion with $M_{\rm ej} = 0.81\,\pm\,0.03\;M_{\odot}$,  \textcolor{black}{$E_{k}= 0.79\,\pm\,0.09\;\times\;10^{51}\;\rm{erg}\,\rm{s}^{-1}$}, and $M_{\rm Ni56} = 0.08\,\pm\,0.02\;M_{\odot}$. Inspection of the event's decline reveals the highest $\Delta m_{15,\rm{bol}}$ reported for a stripped-envelope event to date. Modeling of optical spectra near event peak indicates a partially mixed ejecta comparable in composition to the ejecta observed in SN~1994I, while the earliest spectrum shows signatures of a possible interaction with material of a distinct composition surrounding the SN progenitor. Further, \textit{Hubble Space Telescope} (\textit{HST}) pre-explosion imaging reveals a stellar cluster coincident with the event. From the cluster photometry, we derive the mass and age of the SN progenitor using stellar evolution models implemented in the \texttt{BPASS} library. Our results indicate that SN~2020oi occurred in a binary system from a progenitor of mass $M_{\rm ZAMS} \approx 9.5\,\pm\,1.0\;M_{\odot}$, corresponding to an age of $27\,\pm\,7$~Myr. SN~2020oi is the dimmest SN~Ic event to date for which an early-time flux excess has been observed, and the first in which an early excess is unlikely to be associated with shock-cooling.

\end{abstract}

\section{Introduction}\label{Intro}
Core-collapse supernovae (CCSNe) are both common \citep{2019Modjaz} and vital in shaping the chemical evolution of the universe \citep{2020vandeVoort_enrichment}; however, many questions remain concerning the nature of their progenitor systems and their behavior immediately before explosion. 
The final state of a progenitor star likely plays a decisive role in the large observed diversity of CCSNe, influencing their total luminosities \citep[e.g., for SN IIP;][]{2021Barker_IIP_Progenitors}, the composition of their ejecta \citep{1996Thielemann_CCEjecta}, and the compact remnant that remains when the ejecta clear \citep{2012Ugliano_RemnantMasses}. These questions have motivated decades of targeted searches for the progenitors of CCSNe \citep{1994Aldering_1993JProgenitor, 2003Smartt_2001du, 2009Smartt_CCProgenitors, 2014VanDyk_IIbProgenitor, 2015Smartt_CCProgenitors, 2017Kilpatrick_2016gkg, 2017Kochanek,2018VanDyk_Progenitor, 2018Kilpatrick_potentialProgenitor, 2019ONeill_ProgenitorIIP, Kilpatrick21:19yvr}, beginning with the type-II SN~1987A \citep{1987AWest_Progenitor}. Nevertheless, despite a wealth of high-resolution pre-explosion imaging within nearby galaxies, only a few progenitors have ever been directly observed.

In the absence of direct detections of CCSN progenitors, \textcolor{black}{multiple} lines of indirect evidence have proven fruitful. The first of these is the host galaxy and local environment of the supernova (SN). Owing to the short-lived nature of core-collapse progenitors ($\lesssim50$~Myr for single stars), stellar populations spatially coincident with the SN are likely to share a formation history. As a result, tight constraints can be placed on the age and mass of a progenitor system by comparing stellar evolution models to resolved photometry from stars near the SN site \citep{2017Maund, 2018Williams_NearbyStars}. This method has also been successfully applied to other SN classes with similarly short-lived progenitor systems \citep[e.g., for SNe~Iax;][]{takaro2019constraining}. Host-galaxy spectroscopy can also be used to derive local properties of underlying stellar populations \citep{2015Kuncarayakti_HostIFU,2016Galbany,2018Kuncarayakti_IFU,2019Meza_ASASSN14jb}. 

Complementing local environment studies, early-time observations are a critical tool in our investigation into the progenitors of \textcolor{black}{CCSNe}. In a handful of events, high-cadence observations have facilitated the detection of the X-ray or UV emission associated with shock breakout \citep{Campana2006,Soderberg08,Modjaz2008,2016Garnavich_KeplerBreakout,2018Bersten}, during which the explosion shock traveling at velocity \textcolor{black}{$v_{S}$} escapes the edge of the progenitor star (or the circumstellar medium, if the environment is particularly dense) where the optical depth is \textcolor{black}{$\tau \approx v_{S}/c$} \citep{2017Barbarino_SBOIc,2018Bersten,2019Xiang_2017ein}. As the shock front cools, its associated emission may further extend into optical wavelengths. Because shock breakout occurs at the edge of the \textcolor{black}{progenitor}, the signal uniquely encodes \textcolor{black}{its} pre-explosion radius and surface composition \citep{2017Waxman_2020oi}. \textcolor{black}{Panchromatic photometry and spectroscopy obtained in the first few days of an explosion can also reveal the presence of circumstellar material by its interaction emission or distinct composition, respectively, encoding the pre-explosion mass-loss history of the progenitor star.} \textcolor{black}{In the absence of this early emission, photometric and spectroscopic modeling of later explosion phases still provides valuable insights \citep[e.g.,][]{2011Drout, 2015Morozova_SNECModels, 2016Lyman_bolo, 2017Jerkstrand_NebularPhaseSNe, 2018Taddia}.}

Type-Ic supernovae (SNe~Ic) are a class of core-collapse phenomena for which progenitor searches in recent years have motivated new questions. These explosions are characterized by an absence of hydrogen and helium lines in their spectra, indicating pre-explosion stripping of the stellar envelope. The loss of hydrogen from the outermost layers of the progenitor star is believed to occur either through Roche-lobe overflow onto a stellar companion (in the case of a binary system) or through stellar winds originating from a single progenitor \textcolor{black}{\citep{Yoon+10_BinaryProg, 2011Smith_ObservedRates, Yoon15}}. Both channels result in a Helium star that loses its remaining envelope through line-driven winds \citep{2014Smith_MassLoss, 2017Yoon_WolfRayetEvolution}, but their relative roles in driving type-Ic and type-Ib (in which only hydrogen has been stripped) explosions remain unknown.

\textcolor{black}{The progenitor mass required for explosion as an SN~Ic is lower for binary than for single systems, and} constraints have \textcolor{black}{often} favored \textcolor{black}{the low-mass solution} \citep{2011Drout, 2013Cano, GalYam2017}\textcolor{black}{; further,} the dearth of progenitor detections disfavors single massive stars \textcolor{black}{whose comparatively bright flux should be detectable above the magnitude limit of the observations \citep{2008Eldridge_BinaryRates, 2013Groh_iPTF13bvn, 2014Kelly_2014J}}. Nevertheless, detailed investigations into individual objects have revealed unique exceptions: pre-explosion photometry obtained by \citet{2013Cao_iPTF13bvn} for the type-Ib SN~iPTF13bvn was found to be consistent with models for a single massive Wolf-Rayet \citep[although this interpretation has been challenged; see][]{2016Folatelli_iPTF13bvn}. 
 Further \textcolor{black}{complicating} these efforts, the nature of the SN~Ic progenitor system is often ambiguous from pre-explosion photometry, as exemplified by the type-Ic SN~2017ein \citep{2018Kilpatrick_potentialProgenitor, 2018VanDyk_Progenitor}.

 Uncovering the true nature of the type-Ic progenitor system is critical to understanding what conditions give rise to normal SNe~Ic and the more energetic broad-lined type-Ic (Ic-BL) \textcolor{black}{events}. Type-Ic-BL are the only SNe that have been unambiguously associated with long-duration Gamma-Ray Bursts (LGRBs) \citep{MacFadyenWoosley99,2003Hjorth_Followup,Nagataki+18,Zenati+20}, but we do not know if these phenomena arise from distinct explosion mechanisms or if there is a continuum of stripped-envelope scenarios varying in progenitor mass, explosion velocity, and explosion geometry \citep{2011_Pignata2009bb,2011Taubenberger_Luminosity}. \textcolor{black}{Because LGRB emission occurs within a narrow opening angle while SN radiation is isotropic, this picture is further complicated by the possibility of undetected "choked" or off-axis jets arising from SNe Ic-BL \citep{2015Urata_GRBAfterglow, 2020Izzo_GRBafterglow}.} Can single Wolf-Rayet stars yield ``normal'' type-Ic explosions, or are these events the endpoint of binary interaction, with Wolf-Rayet stars only responsible for GRB-SNe and SNe~Ic-BL? Accurate progenitor mass and age estimates will be key for distinguishing these two formation channels and validating models for the physical environments that give rise to SNe~Ic, SNe~Ic-BL, and \textcolor{black}{L}GRBs  \citep{2003Mazzali_2003dh, 2006Woosley_GRBs}.

\textcolor{black}{In this work, we undertake an analysis of SN~2020oi to shed light on the nature of its progenitor system.} SN~2020oi was discovered by the Automatic Learning for the Rapid Classification of Events (ALeRCE) transient broker on January 7th, 2020 at 13:00:54.000 UTC \citep{2020Forster} from the alert stream of the Zwicky Transient Facility \citep[ZTF;][]{2019Bellm_ZTFAnalysis}. It was classified as a type-Ic SN by the authors two days later using the Goodman Spectrograph at the Southern Astrophysical Research Telescope \citep{2020Siebert_oiClass}. The event occurred at $\alpha, \delta$ = 185.7289$\degree$, 15.8236$\degree$ (J2000), \textcolor{black}{$\sim4.67\arcsec$} North from the nucleus of the SAB(s)bc spiral galaxy Messier 100 (M100/NGC 4321) presiding at a distance of $17.1\,\pm\,1.8$ Mpc \citep{1994Freedman_Cepheids}. SN~2020oi is the seventh SN discovered in M100, preceded by the unclassified SNe 1901B, 1914A, and 1959E; and the type-IIL SN 1979C \citep{1980Carney_1979C}, type-Ia SN~2006X \citep{2006Quimby_2006X}, and calcium-rich transient SN~2019ehk \citep{2020Jacobson-Galan}. As the most recent in this series of observed M100 explosions spanning over a century, SN~2020oi has been continuously monitored since its discovery, and a wealth of pre-explosion data have been collected on its local environment. For these reasons, SN~2020oi represents an ideal event for constraining SN~Ic progenitor properties and explosion physics.

Because of the close proximity of M100, redshift-based distance estimates are likely to be biased by the peculiar velocity of the galaxy. Archival estimates for the distance to M100 range from 13~Mpc to 20~Mpc \citep[e.g.,][]{2007Smith_M100Dist,2008Tully,2016Tully}. In this paper, we assume a redshift-independent distance of 17.1 Mpc corresponding to the distance derived from Cepheids \citep{1994Freedman_Cepheids}. We note that the distance adopted in the analysis for the Ca-rich transient SN~2019ehk in the same galaxy was $d\approx16.2$~Mpc, while the distances used in the previous analyses of SN~2020oi were 14~Mpc, 16.22~Mpc, and 16~Mpc, respectively \citep{2020Horesh_oi,2020Rho,2021Tinyanont}. Although these values are roughly consistent, they will be the source of some discrepancy between the SN parameters derived in this work and those from the previous studies.
  
We have observed a bump lasting $\sim1$~day and beginning $\sim2$~days from the time of explosion in nearly all bands of our optical and UV photometry. In $gri$ bands, we observe a brief increase and decrease in flux; in $u$ band, we observe only \textcolor{black}{a} flux decrease (see Fig.~$\ref{fig:FullPhotometry}$). The coincidence of this phenomenon across bands suggests a high-temperature component to the early-time photometry of SN~2020oi above the standard SN rise.

Early-time bumps such as the one observed in the SN~2020oi photometry are extremely rare among spectroscopically-standard SNe~Ic, particularly when observed in multiple bands and across multiple epochs. Early-time ATLAS data revealed emission in excess of a power-law rise for SN~2017ein, which was interpreted as the cooling of a small stellar envelope that was shock-heated \citep{2019Xiang_2017ein}. A decrease in $V$-band flux in the first photometric observations of SN~LSQ14efd \citep{2017Barbarino_SBOIc} was similarly attributed to shock-cooling. An extended ($>500\ R_{\odot}$), low-mass ($0.045\; M_{\odot}$) envelope, potentially ejected by a massive Wolf-Rayet progenitor pre-explosion, was proposed to explain the luminous first peak in SN~iPTF15dtg \textcolor{black}{\citep{2016Taddia}}. The multi-wavelength coverage of the SN~2020oi bump, coupled with the classification spectrum obtained immediately following its decline, together comprise a rich dataset for investigating the early-time behavior of SNe~Ic. 

In this paper, we describe the photometric and spectroscopic coverage of SN~2020oi spanning $\sim1$~year of observations and the corresponding constraints that these data provide for the progenitor of this SN~Ic. Further, we provide a detailed spectroscopic analysis of M100 and the region immediately surrounding SN~2020oi using pre-explosion Integral Field Unit (IFU) spectroscopy \textcolor{black}{with the Multi Unit Spectroscopic Explorer (MUSE) mounted on the European Southern Observatory Very Large Telescope}. By presenting a comprehensive picture of the most rapidly fading SN~Ic observed to date, this work will shed additional light on the full diversity of stripped-envelope explosions and their origins. 

Three previously published works have investigated this SN: \citet{2020Horesh_oi}, who reported evidence of dense circumstellar material from radio observations; \citet{2020Rho}, who modeled near-IR spectroscopy to derive the presence of carbon monoxide and dust; and \citet{2021Tinyanont}, who presented spectropolarimetric observations suggesting SN~2020oi is unlikely to be an asymmetric explosion. None of these studies investigated the early-time excess reported here, nor did they attempt an analysis of the explosion environment from host-galaxy spectroscopy.

Our paper is laid out as follows. In \textsection\ref{Data}, we outline the photometric and spectroscopic observations collected for SN~2020oi, which span optical, UV, and X-ray wavelengths. We use the notation $\delta t$ to refer to the number of days from the explosion time of MJD \textcolor{black}{58854.0}, which is determined \textcolor{black}{using a fireball rise model outlined} in \textsection\ref{sec:BolKinetics}. We estimate the host-galaxy reddening in \textsection\ref{sec:HostExtinction} and use Gaussian Process Regression to derive the bolometric light curve for the explosion in \textsection\ref{sec:Bolo}. \textsection\ref{sec:BolKinetics} is devoted to the explosion parameters of SN~2020oi, which are \textcolor{black}{estimated} using three different models of the event in the photospheric phase and compared to previous stripped-envelope explosions. Next, we constrain the mass-loss rate of the progenitor from our X-ray observations in \textsection\ref{sec:MassLoss}. We model our spectral sequence \textcolor{black}{near peak light} using a radiative transfer code to characterize the ejecta in \textsection\ref{sec:SpectralAnalysis}, and independently fit the unique early-time spectrum in \textsection\ref{sec:day3spec}. In \textsection\ref{sec:fluxexcess}, we consider physical interpretations for the early-time optical and UV excess. \textsection\ref{sec:IsochroneFits} is devoted to fitting the \textit{HST} pre-explosion photometry of the stellar cluster coincident with the explosion \textcolor{black}{(see \textsection \ref{subsec:HST} for details)}. We then analyze the stellar population within SN~2020oi's local environment using MUSE \textcolor{black}{IFU} spectroscopy in \textsection\ref{sec:HostProperties}, and derive a final age for the SN progenitor in \textsection\ref{sec:ProgenitorConstraints}. We conclude by summarizing our major findings in \textsection\ref{sec:Discussion}.

\section{Observations}\label{Data}\label{sec:Data}
\subsection{\textit{HST} Pre- and Post-Explosion Observations}\label{subsec:HST}
\begin{figure}
    \centering
    \includegraphics[width=\linewidth]{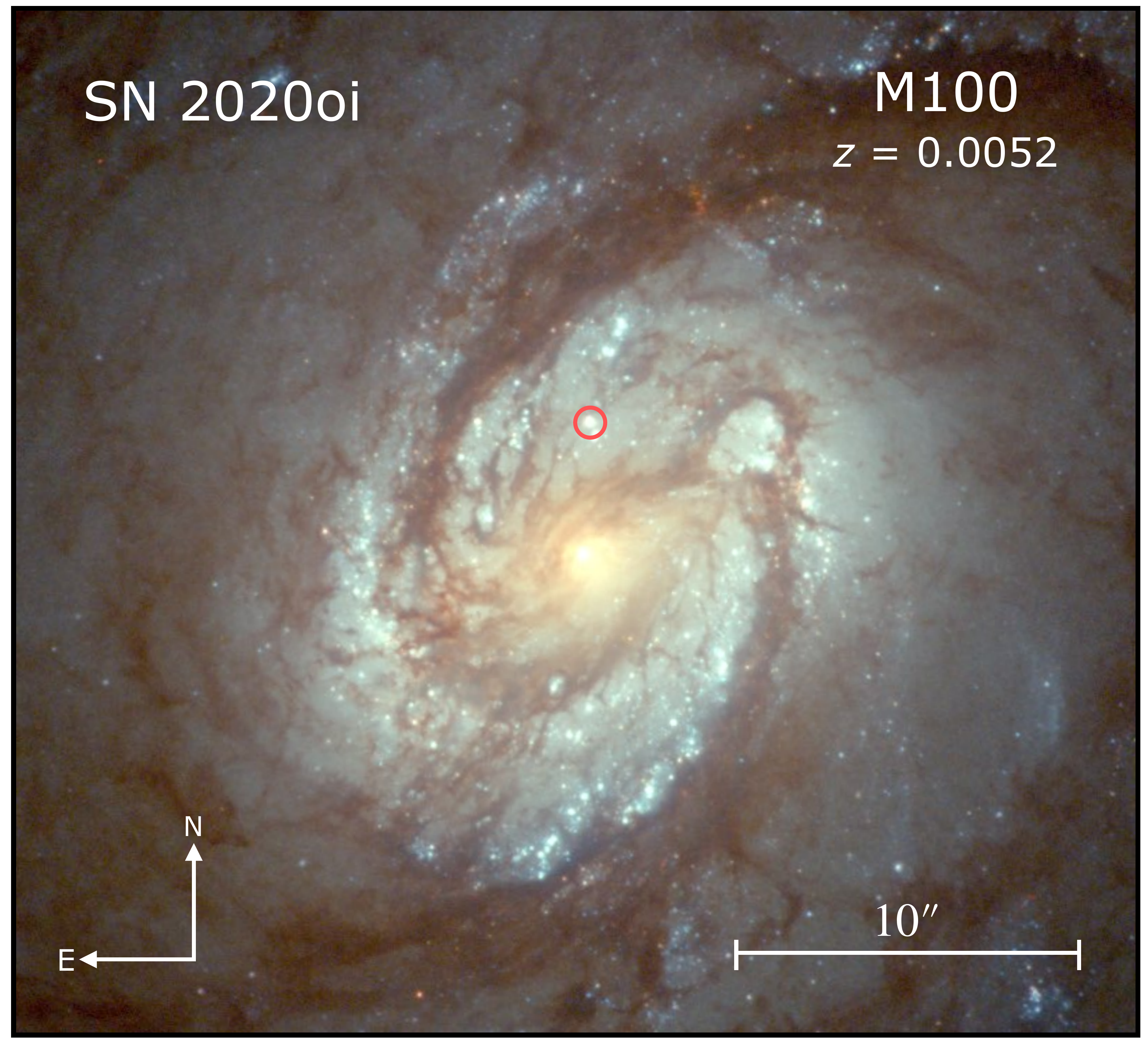}
    \caption{A false-color \textit{HST} image of the nucleus of M100 post-explosion. The location of SN~2020oi is circled, and the physical scale is given bottom right.}
    \label{fig:HST_SN2020oiExplosion}
\end{figure}

We obtained archival \textit{Hubble Space Telescope} (\textit{HST}) images of the central region of M100 using the Hubble Legacy Archive\footnote{\href{https://hla.stsci.edu/}{https://hla.stsci.edu/}} and the Mikulski Archive for Space Telescopes (MAST)\footnote{\href{https://archive.stsci.edu/}{https://archive.stsci.edu/}}. These observations span nearly three decades, beginning with the calibration of the Wide Field/Planetary Camera 2 (WFPC2) \citep{1992Brown_WFC2} for the \textit{Hubble Space Telescope}'s Key Project  \citep{1994Freedman,1995Hill} and ending  with a study (Proposal ID 16179; PI: Filippenko) into the host environments of nearby SNe.
We present a false-color composite of \textit{HST} pointings of M100 post-explosion in Figure \ref{fig:HST_SN2020oiExplosion}, in which we have marked the location of SN~2020oi. 
The diversity of studies involving M100, particularly concerning the dynamics and stellar populations immediately surrounding its nucleus, \textcolor{black}{provide ample context for studying} the pre-explosion environment. We present a detailed summary of the \textit{HST} observations in Table~\ref{tbl:hst_phot}. As in \citet{2018Kilpatrick_potentialProgenitor}, we use the \texttt{astrodrizzle} and \texttt{drizzlepac} packages \citep{2012Gonzaga} to reduce these archival images in the {\tt python}-based {\it HST} imaging pipeline {\tt hst123}\footnote{\url{https://github.com/charliekilpatrick/hst123}}.  
We performed all {\it HST} photometry using a circular aperture fixed to a 0.2\arcsec\ width and centered on the location of SN~2020oi as inferred from post-explosion F555W observations.  Using the {\tt python}-based {\tt photutils} package \citep{photutils}, we extracted an aperture in each drizzled frame and estimated the background contribution from the median value within an annulus with inner and outer radii of 0.4\arcsec and 0.8\arcsec, respectively, and centered on the circular aperture.  We derived the AB magnitude zero point within each frame from the {\tt PHOTPLAM} and {\tt PHOTFLAM} keywords in the original image headers\footnote{i.e., following the standard formula for WFPC2, ACS, and WFC3 zero points as in \url{https://www.stsci.edu/hst/instrumentation/acs/data-analysis/zeropoints}}.

Although no progenitor is immediately evident in the pre-explosion imaging, a marginally-extended brightness excess likely corresponding to a stellar cluster is nearly coincident with the SN explosion. We calculate the nominal offset between the cluster and the explosion in HST/WFC3 UVIS imaging to be $0.55$~px, corresponding to a physical separation of less than 2.3~parsecs. This cluster is also visible in the most recent HST images obtained (MJD~59267, $\delta t \approx 413$~days). We analyze the photometric properties of this source in \textsection\ref{sec:IsochroneFits}.

\definecolor{fu-blue}{RGB}{161, 101, 226} 
\newcommand{\shaderow}{\rowcolor{fu-blue!10}[2pt][2pt]}
\capstartfalse
 \begin{deluxetable*}{ccccccccccc}
 \tablecaption{\textit{HST} Pre-Explosion Cluster and SN~2020oi Photometry\label{tbl:hst_phot}}
 \tablecolumns{11}
 \tablewidth{\textwidth}
 \tablehead{
 \colhead{Date} &  \colhead{MJD} &  \colhead{Phase} &  \colhead{Instrument} & \colhead{Filter} & \colhead{Exposure} &  \colhead{Magnitude} &  \colhead{Uncertainty} &  \colhead{3$\sigma$ Limit} & \colhead{Proposal ID}& \colhead{PI} \\
 \colhead{(UT)} &  \colhead{} &  \colhead{(day)} &  \colhead{} & \colhead{} & \colhead{} &  \colhead{} &  \colhead{} &  \colhead{} & \colhead{}& \colhead{}
 }
\startdata
\shaderow 1993-12-31 & 49352.6     & \textcolor{black}{-9501.4} & WFPC2     & F555W  & 1800.0  & 19.419  & 0.062 & 26.242              & 5195  & Sparks \\
\shaderow 1993-12-31 & 49352.6     & \textcolor{black}{-9501.4} & WFPC2     & F439W  & 1920.0  & 19.460  & 0.114  & 24.857             & 5195  & Sparks \\
\shaderow 1993-12-31 & 49352.7     & \textcolor{black}{-9501.3} & WFPC2     & F702W  & 2400.0  & 19.657  & 0.057 & 25.990             & 5195  & Sparks \\
\shaderow 1994-01-07 & 49359.5     & \textcolor{black}{-9494.5} & WFPC2     & F555W  & 1668.5  & 19.443  & 0.048 & 26.228             & 5195 &  Sparks \\
\shaderow 1994-01-07 & 49359.5     & \textcolor{black}{-9494.5} & WFPC2     & F439W  & 1920.0  & 19.444  & 0.041 & 24.839             & 5195 & Sparks \\
\shaderow 1994-01-07 & 49359.6     & \textcolor{black}{-9494.4} & WFPC2     & F702W  & 2318.5  & 19.630  & 0.095 & 25.833             & 5195  & Sparks\\
\shaderow 1999-02-02 & 51212.0     & \textcolor{black}{-7642.0} & WFPC2     & F218W  & 1200.0  & 19.668  & 0.059 & 22.845             & 6358  & Colina\\
\shaderow 2004-05-30 & 53155.8     & \textcolor{black}{-5698.2} & ACS/HRC   & F814W  & 1200.0  & 19.837  & 0.005 & 25.430             & 9776  & Richstone \\
\shaderow 2004-05-30 & 53155.9     & \textcolor{black}{-5698.2} & ACS/HRC   & F555W  & 1200.0  & 19.432  & 0.005 & 25.886             & 9776 & Richstone \\
\shaderow 2006-01-26 & 53761.4     & \textcolor{black}{-5092.6} & ACS/HRC   & F330W  & 1200.0  & 19.271  & 0.005 & 25.728             & 10548 & Gonzalez-Delgado\\
\shaderow 2008-01-04 & 54469.8     & \textcolor{black}{-4384.2} & WFPC2     & F555W  & 2000.0  & 19.409  & 0.009 & 25.965             & 11171  & Crotts\\
\shaderow 2008-01-04 & 54469.9     & \textcolor{black}{-4384.1} & WFPC2     & F439W  & 1000.0  & 19.454  & 0.022 & 24.502             & 11171 & Crotts\\
\shaderow 2008-01-04 & 54469.9     & \textcolor{black}{-4384.1} & WFPC2     & F380W  & 1000.0  & 19.449  & 0.019 & 24.824             & 11171 & Crotts\\
\shaderow 2008-01-04 & 54469.9     & \textcolor{black}{-4384.1} & WFPC2     & F702W  & 1000.0  & 19.614  & 0.013 & 25.512              & 11171 & Crotts \\
\shaderow 2008-01-04 & 54470.0     & \textcolor{black}{-4384.1} & WFPC2     & F791W  & 1000.0  & 19.712  & 0.021 & 24.920              & 11171 & Crotts\\
\shaderow 2009-11-12 & 55147.1     & \textcolor{black}{-3707.0} & WFC3/UVIS & F775W  & 270.0   & 19.734  & 0.010 & 24.951              & 11646  & Crotts\\
\shaderow 2009-11-12 & 55147.1     & \textcolor{black}{-3706.9} & WFC3/UVIS & F475W  & 970.0   & 19.326  & 0.005 & 27.050              & 11646  & Crotts\\
\shaderow 2009-11-12 & 55147.1     & \textcolor{black}{-3706.9} & WFC3/UVIS & F555W  & 970.0   & 19.422  & 0.005 & 27.040              & 11646 & Crotts\\
\shaderow 2018-02-04 & 58153.7     & \textcolor{black}{-700.6} & WFC3/UVIS & F814W  & 500.0   & 19.849  & 0.007 & 25.252              & 15133 & Erwin \\
\shaderow 2018-02-04 & 58153.7     & \textcolor{black}{-700.3}  & WFC3/UVIS & F475W  & 700.0   & 19.328  & 0.005 & 26.500              & 15133  & Erwin\\
\shaderow 2018-02-04 & 58153.8     & \textcolor{black}{-700.3}  & WFC3/IR   & F160W  & 596.9   & 20.133  & 0.007 & 24.627              & 15133 & Erwin\\
\shaderow 2019-05-23 & 58626.8     & \textcolor{black}{-227.2}  & ACS/WFC   & F814W  & 2128.0  & 19.823  & 0.004 & 26.526              & 15645 & Sand \\
2020-01-29 & 58877.9     & \textcolor{black}{23.9}    & WFC3/UVIS & F814W  & 836.0   & 15.821  & 0.004 & 25.695             & 15654 & Lee\\
2020-01-29 & 58877.9     & \textcolor{black}{23.9}    & WFC3/UVIS & F438W  & 1050.0  & 16.783  & 0.004 & 26.150              & 15654 & Lee\\
2020-01-29 & 58877.9     & \textcolor{black}{23.9}    & WFC3/UVIS & F336W  & 1110.0  & 18.453  & 0.004 & 26.286              & 15654 & Lee\\
2020-01-29 & 58877.9     & \textcolor{black}{23.9}    & WFC3/UVIS & F275W  & 2190.0  & 19.114  & 0.005 & 26.535              & 15654 & Lee\\
2020-01-29 & 58877.9     & \textcolor{black}{23.9}    & WFC3/UVIS & F555W  & 670.0   & 16.421  & 0.004 & 26.426             & 15654 & Lee\\
2020-03-15 & 58923.6     & \textcolor{black}{69.6}    & WFC3/UVIS & F814W  & 836.0   & 16.943  & 0.004 & 25.672              & 15654 & Lee\\
2020-03-15 & 58923.6     & \textcolor{black}{69.6}    & WFC3/UVIS & F438W  & 1050.0  & 17.770  & 0.004 & 26.177              & 15654 & Lee\\
2020-03-15 & 58923.6     & \textcolor{black}{69.6}    & WFC3/UVIS & F336W  & 1110.0  & 18.928  & 0.005 & 26.150              & 15654 & Lee \\
2020-03-15 & 58923.6     & \textcolor{black}{69.6}    & WFC3/UVIS & F275W  & 2190.0  & 19.215  & 0.005 & 26.350              & 15654 &  Lee\\
2020-03-15 & 58923.6     & \textcolor{black}{69.6}    & WFC3/UVIS & F555W  & 670.0   & 17.372  & 0.004 & 26.283              & 15654 & Lee\\
2020-05-21 & 58990.6     & \textcolor{black}{136.6}   & WFC3/IR   & F110W  & 1211.8  & 19.853  & 0.005 & 25.873              & 16075 & Jacobson-Gal\'an\\
2020-05-21 & 58990.6     & \textcolor{black}{136.6}   & WFC3/IR   & F160W  & 1211.8  & 20.050  & 0.006 & 25.088              & 16075 & Jacobson-Gal\'an\\
2020-05-21 & 58990.7     & \textcolor{black}{136.6}   & WFC3/UVIS & F814W  & 900.0   & 18.984  & 0.005 & 25.943              & 16075 & Jacobson-Gal\'an\\
2020-05-21 & 58990.7     & \textcolor{black}{136.6}   & WFC3/UVIS & F555W  & 1500.0  & 19.027  & 0.004 & 27.071              & 16075 & Jacobson-Gal\'an\\
2021-02-21 & 59267.0     & \textcolor{black}{412.9}   & WFC3/UVIS & F625W  & 780.0   & 19.475  & 0.005 & 26.231              & 16179 & Filippenko\\
2021-02-21 & 59267.0     & \textcolor{black}{413.0}   & WFC3/UVIS & F438W  & 710.0   & 19.244  & 0.005 & 25.931              & 16179 & Filippenko
 \enddata
 \tablecomments{Apparent magnitudes are presented in the \textit{AB} photometric system and have not been corrected for host extinction. Rows corresponding to pre-explosion photometry are shaded violet. Phase is given relative to time of explosion (\textcolor{black}{MJD = 58854.0}). }
 \end{deluxetable*}
 \capstarttrue
 

\begin{figure*}
\includegraphics[width=\linewidth]{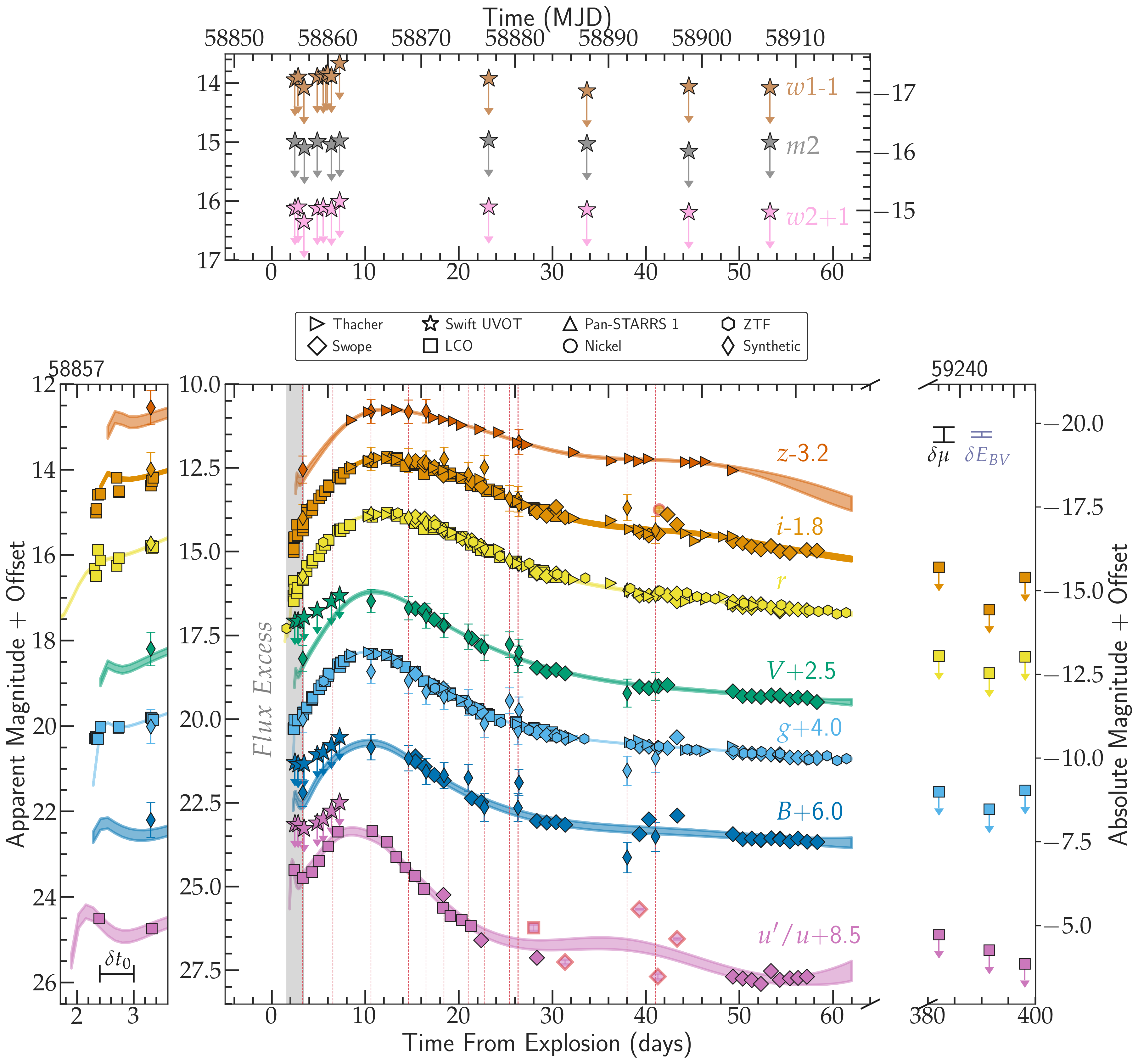}
\caption{Host-galaxy subtracted photometry for SN~2020oi relative to the calculated time of explosion \textcolor{black}{($MJD = 58854.0$)}. Markers are colored by filter and shapes indicate the instrument used to take the observation. Shaded light curves indicate the Gaussian-process fits integrated to construct the bolometric light curve for the event. Red dotted lines mark the phases where spectra were obtained and the grey shaded region spans the early-time optical and UV excess (shown in zoom at left). Late-time upper limits are shown at right. \textit{Swift} UV (optical) upper limits are plotted in the top (center) panel. 
The magnitude uncertainty from the reported distance to M100 is shown in black top-right, and the median uncertainty in magnitude due to uncertainty in host-galaxy extinction across all bands is shown \textcolor{black}{in purple}. Uncertainty in the time of explosion between our models is shown bottom left. Observations with photometric errors above 0.5~mag are not shown. 
Suspicious observations due to poor seeing and errors in background subtraction are outlined in red.}\label{fig:FullPhotometry}
\end{figure*}
\newpage
\subsection{Ground-Based Optical Photometry}\label{subsec:OpticalPhot}

We observed SN~2020oi with the Las Cumbres Observatory Global Telescope Network (LCO) 1m telescopes and LCO imagers from 8 Jan. to 5 Feb.\ 2020 in $g^{\prime}r^{\prime}i^{\prime}$ bands.  We downloaded the calibrated BANZAI \citep{BANZAI} frames from the Las Cumbres archive and re-aligned them using the command-line blind astrometry tool {\tt solve-field} \citep{astrometry.net}.  The images were also recalibrated using {\tt DoPhot} photometry \citep{Schechter93} and PS1 DR2 standard stars observed in the same field as SN~2020oi in $gri$ bands \citep{Flewelling+16}.  We then stacked $g^{\prime}r^{\prime}i^{\prime}$-band frames obtained from 31 Jan.\ to 7 Feb.\ 2021 as templates and reduced them following the same procedure using {\tt SWarp} \citep{swarp}.  The template images were subtracted from all science frames in {\tt hotpants} \citep{hotpants}, and finally we performed forced photometry of SN~2020oi on all subtracted frames using {\tt DoPhot} with a \textcolor{black}{point-spread function (PSF)} fixed to the instrumental PSF derived in each science frame.



SN~2020oi was also observed with the Nickel 1m telescope at Lick Observatory, Mt. Hamilton, California in conjunction with the Direct 2k $\times$ 2k camera ($6.8' \times 6.8'$) in $BVr^{\prime}i^{\prime}$ bands from 31 Jan. to 8 Aug.\ 2020.  All image-level calibrations and analysis were performed in {\tt photpipe} \citep{Rest05:photpipe,2018Kilpatrick_potentialProgenitor} using calibration frames obtained on the same night and in the same instrumental configuration.  We then aligned our images using 2MASS astrometric standards in the image frame, then each image was regridded to a corrected frame using {\tt SWarp} \citep{swarp} to remove geometric distortion.  All photometry was performed using a custom version of {\tt DoPhot} \citep{Schechter93} to construct an empirical \textcolor{black}{PSF} and perform photometry on all detected sources. We then calibrated each frame using PS1 DR1 sources  \citep{Flewelling+16} in $ri$ bands and transformed to $BV$ bands using transformations in \citet{Tonry12}.

Observations of SN~2020oi were also obtained with the Thacher 0.7m telescope located at Thacher Observatory, Ojai, California from 14 Jan. to 21 Dec.\ 2020 in $griz$ bands.  The imaging reductions followed the same procedure described above for our Nickel reductions and in \citet{Dimitriadis19:18oh_k2}.

We further observed SN~2020oi with the Swope 1m telescope at Las Campanas Observatory, Chile starting on 21 Jan.\ 2020 through 15 Mar.\ 2020 in $uBVgri$ bands.  Our reductions followed a procedure similar to the one outlined above for the Nickel telescope and described in further detail in \citet{Kilpatrick18:16cfr}.

In addition to the photometry listed above, we include observations obtained from the forced-photometry service \citep{2019Masci} of the Zwicky Transient Facility \citep[ZTF;][] {2019Bellm_ZTF,2019Graham_ZTF}. These data, which begin on 7 Jan.\ 2020 ($\delta t$ = 2~days) and continue through 26 April 2020 ($\delta t =$ 111~days), were obtained using the Palomar 48-inch telescope and reduced according to the methods outlined in \citet{2019Bellm_ZTFAnalysis}.

\subsection{\textit{Swift} Ultraviolet Observations}
To obtain ultraviolet \textcolor{black}{(UV)} photometry for SN~2020oi, we leverage the extensive observations made of M100 by the Neil Gehrels \emph{Swift} Observatory \citep{Gehrels2004}. The earliest of these was obtained in November 2005. The follow\textcolor{black}{-}up campaigns of SN~2006X and SN~2019ehk, acquired with the Ultraviolet Optical Telescope \citep[UVOT;][]{Roming2005}, provide excellent UV and \textit{UBV}-band template images for SN~2020oi, spanning a total of 22 pre-explosion epochs. Indeed, the first 2 post-explosion UVOT epochs come from the follow-up campaign of SN~2019ehk, which serendipitously observed SN~2020oi only \textcolor{black}{$\sim2.45$~days} after explosion. Observations were collected for SN~2020oi from 2 to 53 days post-explosion.

We performed aperture photometry with the \texttt{uvotsource} task within the \textsc{HEAsoft} v6.22\footnote{We used the calibration database (CALDB) version 20201008.}, following the guidelines in \citet{Brown2009} and using an aperture of 3\arcsec{}. Using the 22 pre-explosion epochs obtained, we have estimated the level of contamination from the host-galaxy flux. In doing so, we assume that excess flux contributions from the progenitor system (as in the case of outbursts or flares), if present, are negligible. This assumption is supported by our measurements of a consistent flux at the location of SN~2020oi across all pre-explosion observations. As a result, we averaged the photon count-rate across the 22 epochs for each filter and then subtracted this from the count rates in the post-explosion images, following the prescriptions in \citet{Brown2014}. 

To further constrain the host-galaxy contamination within our UVOT images, we perform the same aperture photometry described above at three other locations along the star-forming ring of M100 and equidistant from the nucleus. After host-galaxy subtraction, we find an unexplained flux increase at the same post-explosion epoch across all apertures. It is likely that this is a systematic effect in the \textit{Swift} instrumentation, but at present we are unable to validate this hypothesis. To eliminate the possibility of contaminating our photometry with systematics at other epochs, and because of the strong UV contamination from the M100 nucleus, we have replaced our \textit{Swift} photometry with upper limits derived prior to host subtraction.

We present our complete optical and ultraviolet light curve for the explosion in Figure \ref{fig:FullPhotometry}, where we have removed all observations with photometric uncertainties above 0.5~mag. Our full photometric dataset is listed in Table~\ref{tbl:optical_phot}.



\newpage
\subsection{Chandra X-ray Observations}\label{SubSec:Xraydata}
We obtained deep X-ray observations of SN\,2020oi with the Advanced CCD imaging spectrometer (ACIS) instrument onboard the \textit{Chandra X-ray Observatory} (\textit{CXO}) on February 15, 2020  and March 13, 2020, 40 and 67 days since explosion, respectively (PI Stroh, IDs 23140, 23141) under an approved DDT program 21508712. The exposure time of each of the two observations was 9.95\,ks, for a total exposure time of 19.9\,ks. These data were then reduced with the \texttt{CIAO} software package \textcolor{black}{\citep[version 4.13;][]{2006Fruscione_CIAO}}, using the latest calibration database CALDB version 4.9.4. As part of this reduction, we have applied standard ACIS data filtering.  

We do not find evidence for statistically significant X-ray emission at the location of the SN in either observations or in the co-added exposure. Using Poissonian statistics we infer a 3$\sigma$ count-rate limit of $\sim4.02\times10^{-4}\,\rm{c\,s^{-1}}$ and $\sim5.02\times10^{-4}\,\rm{c\,s^{-1}}$ for the two epochs of CXO observation (0.5 - 8 keV). The Galactic neutral hydrogen column density in the direction of the transient is $NH_{MW}=1.97\times 10^{20}\,\rm{cm^{-2}}$ \citep{Kalberla05}. Assuming a power-law spectral model with spectral photon index $\Gamma=2$, the above count-rate limits translate to 0.3-10~keV unabsorbed flux limits of $F_x<6.3\times 10^{-15}\,\rm{erg\,s^{-1}cm^{-2}}$ (first epoch), and $F_x<7.9\times 10^{-15}\,\rm{erg\,s^{-1}cm^{-2}}$ (second epoch). 
We note the presence of diffuse soft X-ray emission from the host galaxy at the SN site, which prevents us from achieving deeper limits on the X-ray emission of the explosion. 

\subsection{Optical Spectroscopy}\label{subsec:Spectroscopy}
\begin{figure*}[h]
    \centering
    \includegraphics[width=\linewidth]{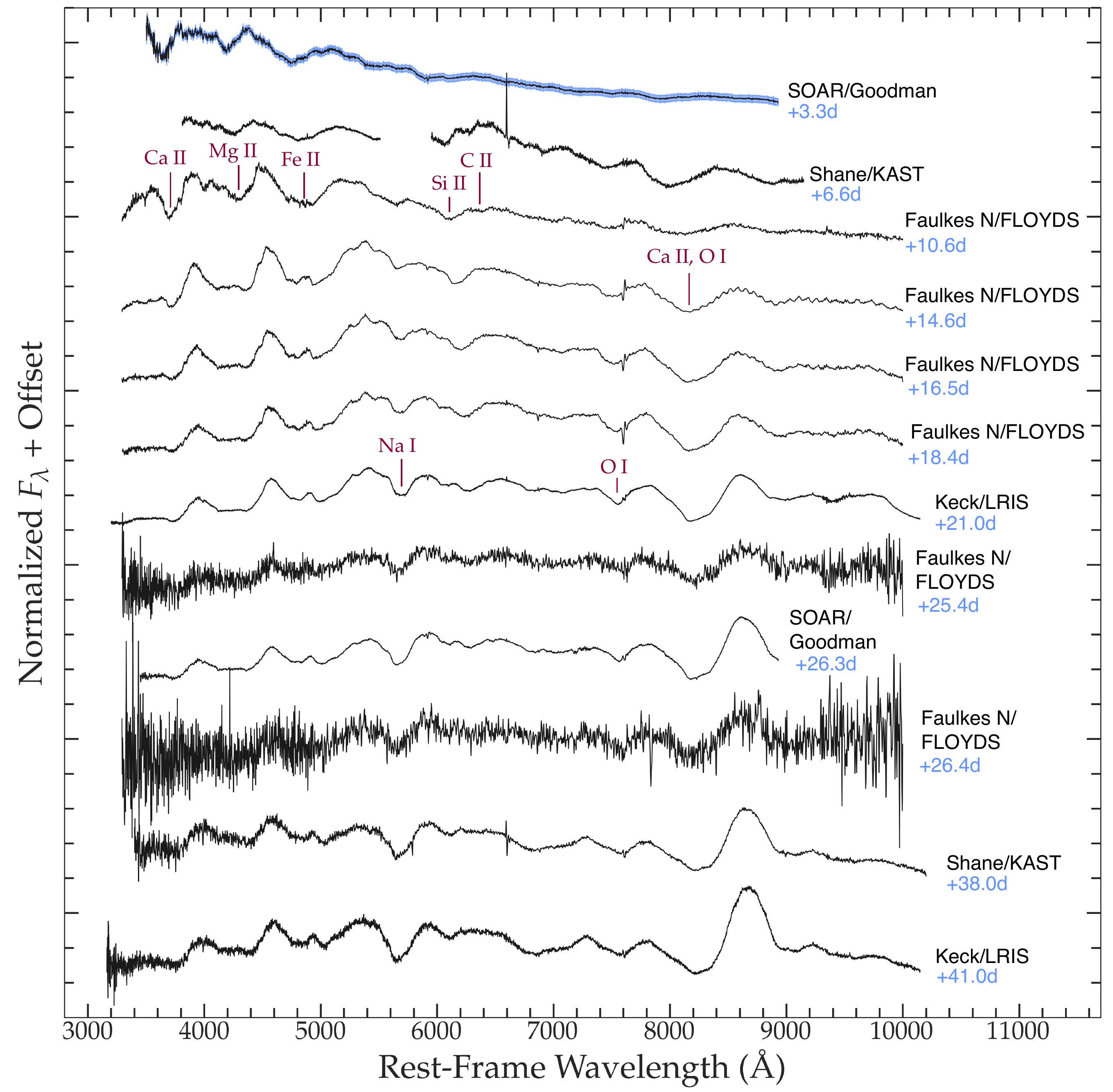}
    \caption{The full sequence of optical spectra obtained for SN~2020oi. Spectrographs used for observations are listed at right, along with the phase of the spectr\textcolor{black}{a} relative to the time of explosion. Prominent absorption lines are listed in red and the spectrum taken within one day of the early-time photometric excess is highlighted in blue at top.}
    \label{fig:SpectralSequence}
\end{figure*}

We have obtained 12 spectra from $\delta t \approx 3.3$ to $\delta t \approx 41.0$~days post-explosion. Two spectra, including the classification spectrum ($\delta t \approx 3.3$~days from explosion), were obtained with the Goodman High Throughput Spectrograph  \citep{Clemens04SPIE} at the Southern Astrophysical Research (SOAR) Telescope. Six were obtained with the FLOYDS spectrograph on the Faulkes 2m telescopes of the Las Cumbres Observatory Global Telescope Network \citep[LCOGT;][]{LCOGT13PASP}, two with the Low-Resolution Imaging Spectrometer \citep[LRIS;][]{LRIS} on the Keck I telescope\textcolor{black}{,} and two with the Kast spectrograph \citep{KAST} on the 3m Shane telescope at Lick Observatory. The FLOYDS spectra were reduced using a dedicated spectral reduction pipeline\footnote{\url{https://github.com/LCOGT/floyds_pipeline}} and the remaining ones with standard \textsc{iraf/pyraf}\footnote{\textsc{iraf} is distributed by the National Optical Astronomy Observatories, which are operated by the Association of Universities for Research in Astronomy, Inc., under cooperative agreement with the National Science Foundation.} and python routines \citep{Siebert2020ApJ}. All of the spectral images were bias/overscan-subtracted and flat-fielded, with the wavelength solution derived using arc lamps and the final flux calibration and telluric line removal performed using spectro-photometric standard star spectra \citep{Silverman2012MNRAS}. We provide a summary of our full spectral sequence, which spans 38 days of explosion, in Table~\ref{tbl:optical_spec}, and plot each obtained spectrum in Figure \ref{fig:SpectralSequence}. 

In addition to those described above, two optical spectra were obtained that did not contain obvious SN emission. The first was obtained using the Keck Observatory's Low Resolution Imaging Spectrometer (LRIS) on December 10th, 2020, $\sim336$~days from the explosion's maximum brightness in $r$ band. After reducing the data, it was determined that the spectrum was dominated by galaxy light. The second spectrum, which was obtained with the FLOYDS spectrograph on the Faulkes 2m telescope of the LCOGT in Siding Springs, Australia, was affected by poor seeing.

\capstartfalse
 \begin{deluxetable*}{cccccc}
 \tablecaption{Optical Spectroscopic Observations of SN~2020oi\label{tbl:optical_spec}}
 \tablecolumns{6}
 \tablewidth{\textwidth}
 \tablehead{
 \colhead{Date} &  \colhead{MJD} &  \colhead{Phase} &  \colhead{Telescope} & \colhead{Instrument} &  \colhead{Wavelength Range} \\
 \colhead{(UT)} & \colhead{}   & \colhead{(days)} & \colhead{} & \colhead{} & \colhead{(\AA)}
 }
 \startdata
2020-01-09 & 58857.3	& \textcolor{black}{+3.3}      &  SOAR  & Goodman   & 4000–9000  \\
2020-01-12 & 58892.0	& \textcolor{black}{+6.6}   & Shane    & KAST   & 3800–9100  \\
2020-01-16	& 58864.6	& \textcolor{black}{+10.6}  & Faulkes North     & FLOYDS  & 4800–10000  \\
2020-01-20	& 58868.6	& \textcolor{black}{+14.6}      & Faulkes North      & FLOYDS  & 4800–10000   \\
2020-01-22 & 58870.5	& \textcolor{black}{+16.5}    & Faulkes North    & FLOYDS  & 4800–10000   \\
2020-01-24 & 58872.4	& \textcolor{black}{+18.4}    & Faulkes North     & FLOYDS   & 4800–10000  \\
2020-01-27 & 58875.0	& \textcolor{black}{+21.0}   & Keck I     & LRIS   & 3200–10800 \\
2020-01-31  & 58879.4	&  \textcolor{black}{+25.4} & Faulkes North    & FLOYDS  & 4800–10000 \\
2020-02-01 & 58880.3	& \textcolor{black}{+26.3}    & SOAR      & Goodman   & 4000–9000\\
2020-02-01 & 58880.4	& \textcolor{black}{+26.4}   & Faulkes North  & FLOYDS   & 4800–10000\\
2020-02-13 & 58892.0	& \textcolor{black}{+38.0}   & Shane    & KAST   & 3500–11000  \\
2020-02-16 & 58895.0	& \textcolor{black}{+41.0}   & Keck I & LRIS   & 3200–10800  \\
 \enddata
 \tablecomments{Phase is given relative to time of explosion (MJD = \textcolor{black}{58854.0}).}
 \end{deluxetable*}

\section{Host Galaxy Extinction}\label{sec:HostExtinction}
We estimate the host-galaxy extinction along the line of sight to the SN first using the empirical relation between the reddening and the equivalent width of the spectrum's Na $\lambda\lambda5889,5895$ doublet \citep{2012Poznanski_Doublet}. Using our de-redshifted high resolution Keck/LRIS spectrum obtained on January 27, a pseudo-continuum is defined as a line at the edges of the absorption feature and the spectrum is then normalized at the feature's position. We then fit the sodium doublet, which we approximate as two Gaussians with their widths forced to be the same and their relative strengths constrained according to their oscillator strengths (obtained from the National Nuclear Data Center\footnote{\url{https://www.nndc.bnl.gov/}}). This process is repeated 10,000 times for different choices of the pseudo-continuum. We estimate a combined equivalent width of $0.88\,\pm\,0.05$~\AA, corresponding to a host reddening of $E(B-V)=0.15\,\pm\,0.03$~mag \citep[using equation 9 from ][]{2012Poznanski_Doublet}. This is comparable to the value provided by \citet{2020Horesh_oi}, who estimates $E(B-V)=0.14\,\pm\,0.05$~mag of reddening using this procedure. Assuming $R_V = 3.1$, this corresponds to a $V$ band extinction of $A_V\approx0.47$.

We additionally estimate the line-of-sight host-galaxy reddening by comparing the observed color evolution of SN~2020oi during the first 20 days following peak luminosity to the type-Ic color templates provided in \citet{2018Stritzinger}. First, we sample a range of $A_V$ and $R_V$ values across a uniformly\textcolor{black}{-}spaced grid spanning [0.0, 1.0]~mag and [1.0, 6.0], respectively. By interpolating the spectra spanning this range in phase, we obtain extinction corrections for each photometric band and calculate the $\chi^2$ value of our corrected color curve. Because we find $R_v$ to be poorly constrained from our photometry, we choose $A_V$ to be the value with the smallest $\chi^2$ value for a fixed $R_V = 3.1$ (corresponding to a Galactic extinction curve).  We note that infrared observations of SN~2020oi are needed to conclusively determine $R_V$ \citep{2018Stritzinger}.

We find a best-fit host-galaxy extinction of $A_V = 0.35$~mag for $R_V = 3.1$, corresponding to a reddening of \mbox{$E(B-V) = 0.11$~mag}. We estimate the error on $A_V$ to be $0.03$ mag by calculating the standard deviation of the best-fit values across each of our sampled $R_V$ values. We adopt this value as our host-galaxy extinction instead of the value derived from the Na doublet fitting due to the large dispersion associated with the latter relationship. A slightly higher \textcolor{black}{host} reddening of \mbox{$E(B-V) = 0.13$~mag} was adopted by \citet{2020Horesh_oi} from a comparison of the same color templates as we have used. We also report a Galactic reddening value of \mbox{$E(B-V)=0.0227\,\pm\,0.0002$~mag} in the direction of the SN based upon the maps of \citet{2011Schlafly_Dust}, leading to a combined reddening of $E(B-V)=0.133\,\pm\,0.03$~mag. This is consistent with the $0.153$~mag of \textcolor{black}{total} reddening reported by \citet{2020Horesh_oi}, who find a comparable Galactic value of $E(B-V)=0.023$~mag in the direction of M100. In the following sections, we adopt a combined reddening of $E(B-V)=0.133$.

\section{Bolometric Light curve Fitting}\label{sec:Bolo}

\begin{figure*}[h]
\includegraphics[width=\linewidth]{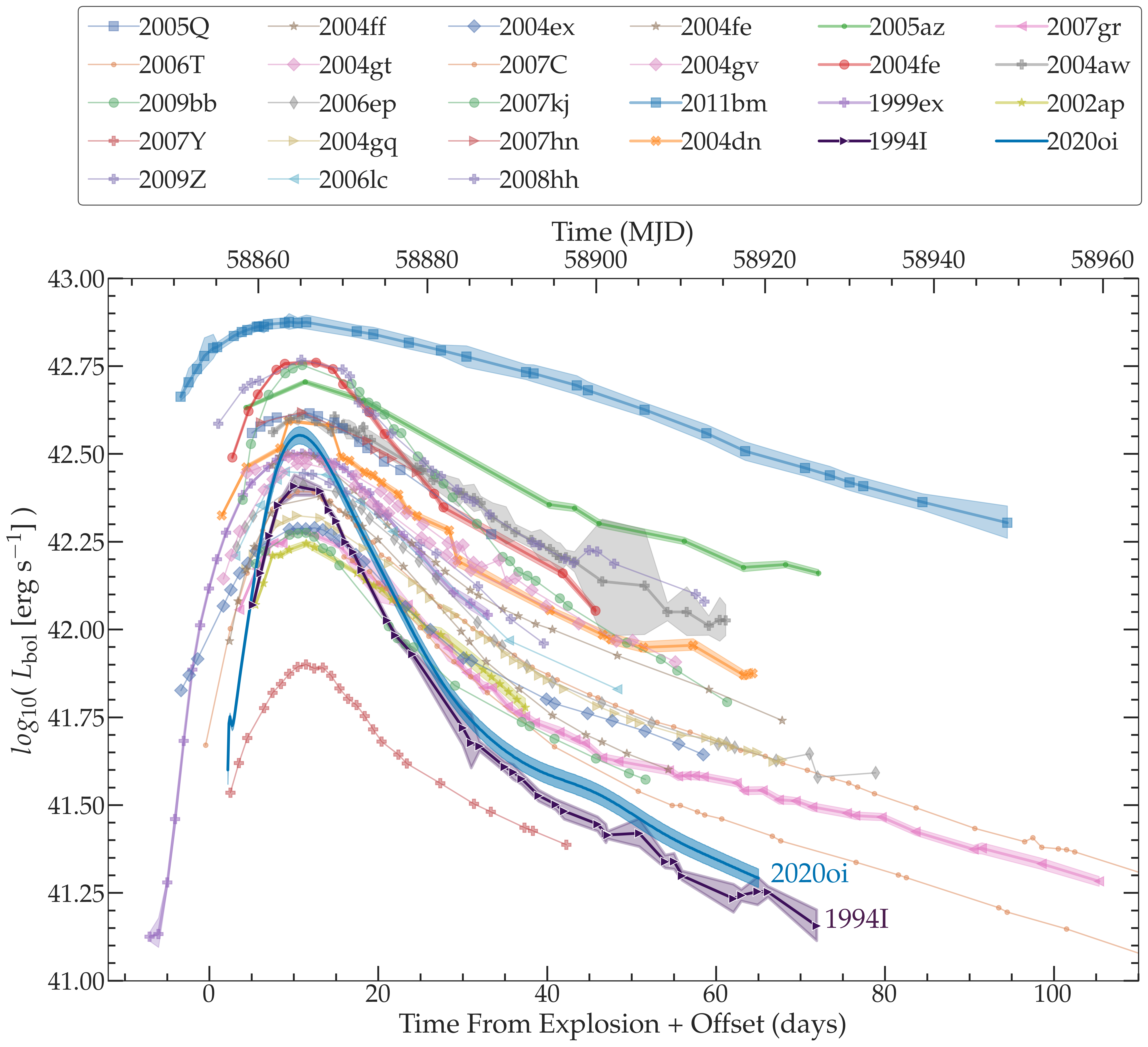}
\caption{The bolometric light curve for SN~2020oi (blue), plotted alongside the type-Ic/Ic-BL SN sample\textcolor{black}{s} from \citet{2016Lyman_bolo} \textcolor{black}{and \citet{2018Taddia}}. Light curves have been aligned at peak and the shaded regions correspond to 1-$\sigma$ confidence intervals \textcolor{black}{for the SNe in \citet{2016Lyman_bolo}}, which incorporate only uncertainty in the bolometric corrections for each event. Uncertainties in distance modulus and extinction along each line of sight are not shown and may affect this comparison. \textcolor{black}{For clarity, we plot only SNe with pre-maximum observations.} The rise and decline rate of SN~2020oi is similar to that of the characteristic type-Ic event SN~1994I (shown in violet), which is identified as a rapidly declining event in \citet{2016Lyman_bolo}. SN~2020oi appears more luminous than SN~1994I, but unaccounted-for extinction toward SN~1994I may also account for this difference \citep{1996Richmond_1994I}. The bolometric contribution from the SN~2020oi early-time bump can be seen in the first day of observations.}\label{fig:lcbolo}
\end{figure*}

To consolidate our panchromatic observations obtained at different epochs into a consistent bolometric light curve, we seek to construct a non-parametric model for the photometric evolution of the explosion in each filter using Gaussian Process Regression \citep[GPR;][]{2006Rasmussen_gaussianprocesses}. GPR is an approach to functional approximation that assumes that observations are realizations sampled from a latent function \textcolor{black}{with} Gaussian noise. The model is constrained by a kernel function that \textcolor{black}{describes} the similarity between observations using a length scale over which our observations are correlated. By conditioning a chosen kernel, which characterizes our prior, on the observations, we can generate a posterior distribution for a class of functions that describe the data. This procedure can additionally consider a mean model for the observations, and this further conditions the subsequent model predictions. We use the GPR implementation in \texttt{George} \citep{2014Ambikasaran_George}.


The mean model we construct for our light curve in each band must be sensitive to the early-time bump observed within the first five days, but insensitive to late-time galactic contamination from the bright nucleus. We use \texttt{Scipy}'s \texttt{splrep} function, which determines a basis (B-) spline representing a one-dimensional function, to construct this model
\citep{dierckx1995curve}. The basis calculated by this method is determined both by the degree of the spline fit and the weights imposed on each observation. The observations with highest relative weighting most tightly constrain the B-spline, allowing us to determine the light-curve features captured in the mean model and those smoothed in it. 

First, we calculate a B-spline for our $r$-band photometry with polynomial order five. Observations taken before MJD 58858.0 ($\delta t \approx 4.0$ days) are given a weight of 60, those within 3~days of the $r$-band peak are given a weight of 50, and all other points are given a weight of 10. We then construct a mean model for our GPR according to the following equation:
\begin{equation}
      \bar{m}(t) =
  \begin{cases}
    B(t+\alpha) + \beta + \gamma & \text{for } t < 58858.0 \\
    B(t+\alpha) + \beta, & \text{for } t \geq 58858.0
  \end{cases}
\end{equation}

Where $t$ is the time in MJD, $B$ is the $r$-band B-spline interpolation function, $\alpha$ is a parameter that shifts the entire curve forward in phase, $\beta$ is a parameter that shifts the model in magnitude, and $\gamma$ is a parameter that determines the height of the early-time brightness excess relative to the rest of the light curve. Although this model was constructed from only our $r$-band photometry, it serves as the mean model for all our passbands. The parameters described above allow the model to account for the difference in light curve properties between $r$ and the other fitted bands.

These three free parameters, in addition to a fourth to account for the intrinsic photometric dispersion, are then fit in each band independently using an exp-sine-squared kernel of length-scale $\Gamma = 0.9$ and period $\textrm{ln}(P) = 5$ to smoothly predict the rise and decay of the luminosity. The period was chosen to be approximately twice the duration of the photometry ($\sim70$~days), ensuring that the rise and fall to the light curve corresponds to the first half-wavelength of the model. The value for $\Gamma$ was determined empirically; larger $\Gamma$ values \textcolor{black}{resulted in a mean model that overfit the observations} and preserve\textcolor{black}{d} small-scale correlations. This results in a set of 500 interpolated observations in $UBgVriz$ bands spanning MJD 5885\textcolor{black}{4}-58919 (corresponding to the first $65$~days of the explosion). We present the posterior distributions obtained from this method in Figure \ref{fig:FullPhotometry}. 

Next, we use the \texttt{Superbol} package\footnote{\url{https://github.com/mnicholl/superbol}} \citep{2018Nicholl_Superbol} to calculate the integrated bolometric luminosity of SN~2020oi. After shifting to the rest frame and correcting for the combined Milky Way and host-galaxy extinction, we model the explosion at each epoch as a black-body (a good approximation during the photospheric phase owing to the optically thick ejecta) and use the \verb|curve_fit| routine within the Python package \texttt{Scipy} to determine the photospheric temperature and radius that best describe each interpolated observation. These curves are then integrated to account for the unobserved far-UV and near-IR flux from the event and calculate the bolometric luminosity at each epoch. We present the final bolometric light curve in Figure \ref{fig:lcbolo}, along with those  \textcolor{black}{reported by \citet{2016Lyman_bolo} and \citet{2018Taddia}} for previous SN~Ic and SN~Ic-BL events.

 We find SN~2020oi to be less luminous than nearly all previous SNe~Ic from \citet{2016Lyman_bolo} for the majority of its evolution. Roughly 10 days before peak, the explosion is the \textcolor{black}{second-}dimmest type-Ic event for which data are available. Similarly, at the end of the photospheric phase ($t\approx30$ days, after which point the SN ejecta can no longer be approximated as a black-body due to its decreasing opacity as it expands), SN~2020oi is dimmer than all but \textcolor{black}{two} SN~Ic reported \citep[the tail of the SN~1994I is slightly less luminous, although the extinction in the direction of SN~1994I remains highly uncertain; see][]{2006Sauer_1994I}. Interestingly, although we find the event to be dimmer than most other SNe~Ic at early- and late-times, at peak SN~2020oi rises to within less than half a standard deviation of the mean peak luminosity for the SN~Ic sample. 
 
\begin{figure}
    \centering
    \includegraphics[width=\linewidth]{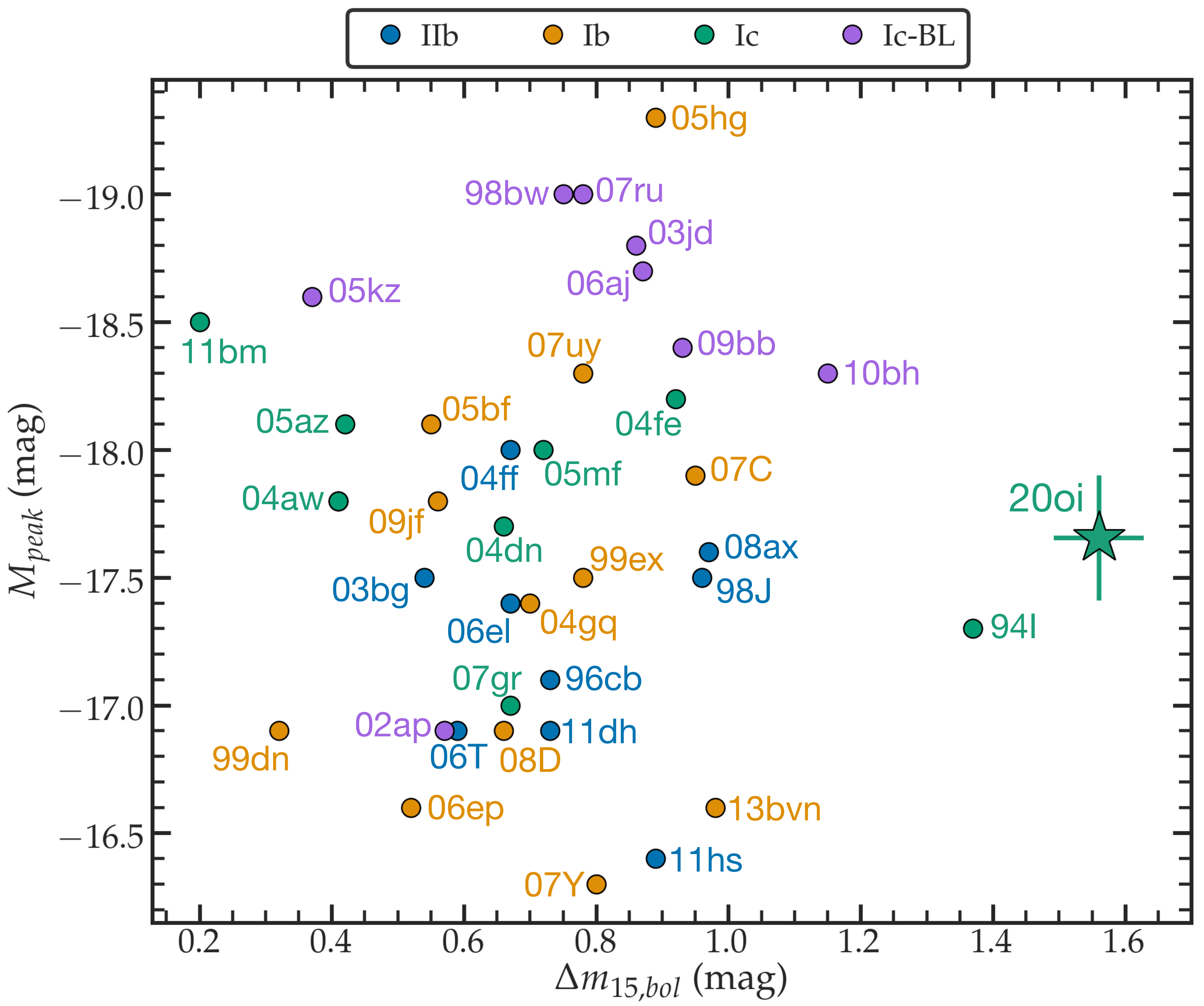}
    \caption{Light curve absolute magnitude at peak ($M_{\rm peak}$) and linear decline rates ($\Delta m_{15, bol}$) for stripped-envelope SNe from \citet{2016Lyman_bolo}. The location of SN~2020oi is denoted by a star at right. Error in the value of $\Delta m_{15, bol}$ is propagated from photometric uncertainties, and error in $M_{\rm peak}$ combines uncertainty in event photometry, distance, and extinction. The decline rate for SN~2020oi from peak to 15~days following is $\sim1$~mag higher than the median for type-Ic events shown and $\sim0.3$~mag higher than the decline rate of the next-closest type-Ic event, SN~1994I \citep[although the decline of SN~1994I may have been higher than is shown here due to uncertainty in extinction estimates in the direction of the SN; see][]{1996Richmond_1994I}. Figure adapted from \citet{2016Lyman_bolo}.}
    \label{fig:dm15}
\end{figure}

An event with a lower luminosity pre- and post-maximum but reaching comparable brightness at peak to other type-Ic explosions must necessarily exhibit rise and decline rates greater than other type-Ic events. Indeed, as is reported in \citet{2020Horesh_oi} and visible in Fig.~\ref{fig:lcbolo}, the slope of the bolometric luminosity of SN~2020oi after maximum is steeper than most previously observed SNe~Ic. The overall bolometric evolution can be seen to roughly match that of SN~1994I.
We can parameterize the decline rate of SN~2020oi by $\Delta m_{15, \rm bol}$, the difference in the absolute bolometric magnitude from peak brightness to 15~days following peak. We find a value of $\Delta m_{\rm 15,bol} \approx 1.63\,\pm\,0.14$, higher than any other stripped-envelope SN reported by \textcolor{black}{either} \citet{2016Lyman_bolo} \textcolor{black}{or \citet{2018Taddia}} and $\sim1$ mag higher than the median for type-Ic events. The wide scatter in SN~Ic decline rates was reported in \citet{2011Li_SNDeclines} in a study of eight events, although of the SNe considered only the decline of SN~1994I was characterized as rapid. Larger type-Ic samples are needed to determine whether these rapidly-declining events are intrinsically rare. We present the peak absolute magnitudes and $\Delta m_{\rm 15, bol}$ values for SN~2020oi and other stripped-envelope events in Figure \ref{fig:dm15}.




\subsection{Photospheric Properties of SN~2020\lowercase{oi}}\label{sec:PhotosphericProperties}

\begin{figure}
    \centering
    \includegraphics[width=\linewidth]{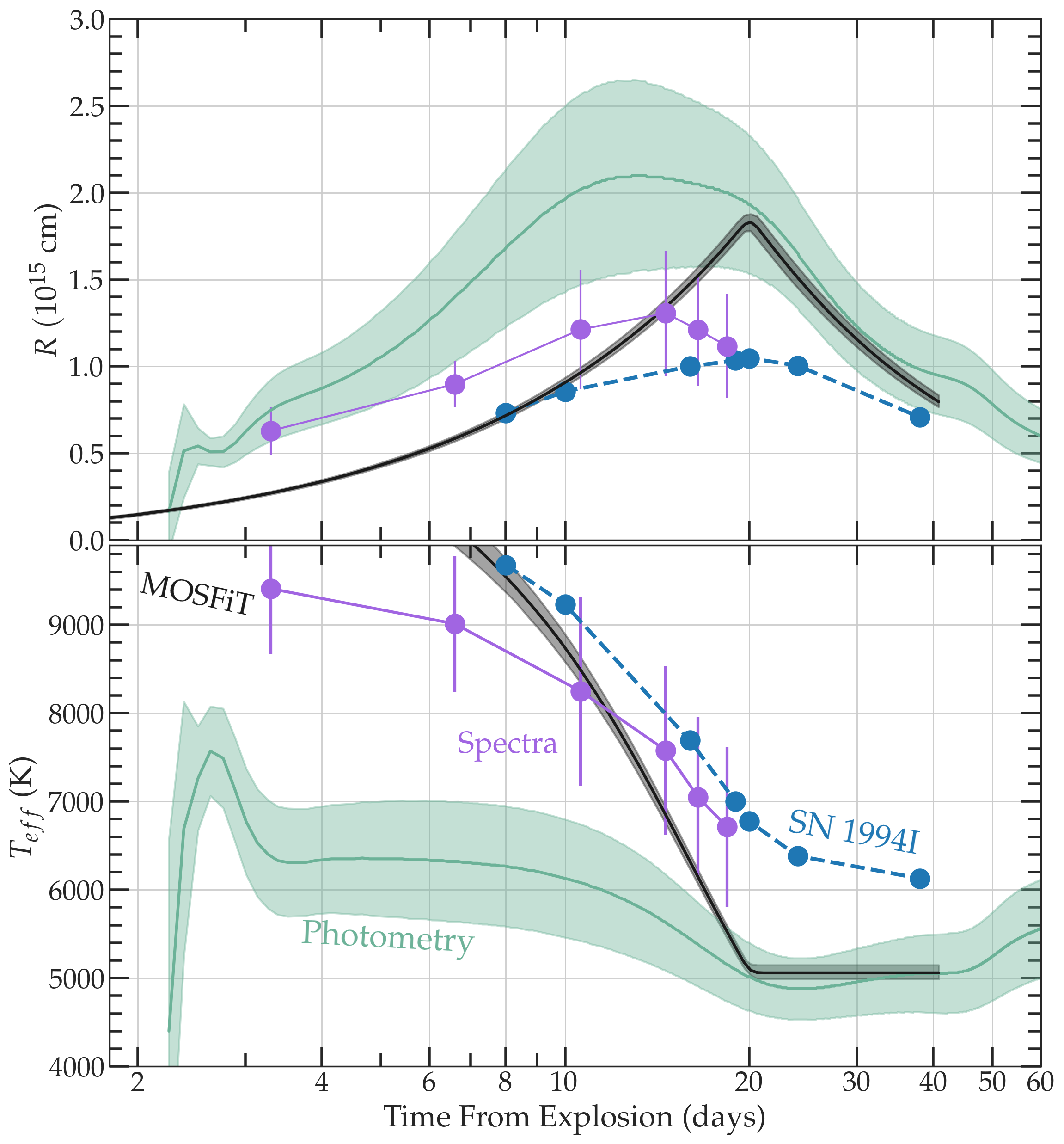}
    \caption{\textcolor{black}{T}emperature and radius estimates of the SN photosphere from black-body fits to the photometry at each interpolated epoch \textcolor{black}{in green}.  Spectra-derived black-body values for SN~2020oi are shown in violet. The spectra-derived photospher\textcolor{black}{ic properties of} the type-Ic SN~1994I \citep{2006Sauer_1994I} are shown as blue points. The photospheric properties of the best-fit \texttt{MOSFiT} model (described in \textsection\ref{subsec:MOSFiT}) are given in black. \textcolor{black}{Shaded regions denote 1-$\sigma$ confidence intervals.} The difference between spectroscopic and photometric estimates of these properties are not physical, but instead reflect the approximate nature of \textcolor{black}{each} indicator.}
     \label{fig:PhotosphericProperties}
\end{figure}

We now leverage our best-fit black-body \textcolor{black}{model} from \texttt{Superbol} to \textcolor{black}{estimate} the radius and temperature of the photosphere of the SN as it explodes, which we present in Figure \ref{fig:PhotosphericProperties} for the first 60 days of the explosion. Due to the unknown nature of the flux excess, we consider only the radius and temperature estimates following $\delta t\sim 3$~days from the time of explosion.

Following the observed flux excess, the first photospheric radius observed is $R = 5.1\,\pm\,0.9 \times 10^{14}$~cm at $\delta t \approx 3$~days from explosion. The corresponding effective temperature at this epoch is $T_{eff} = 6300\,\pm\,600$~K. As the ejecta expands, \textcolor{black}{the temperature of the ejecta probed by the photosphere decreases} and so does its scattering opacity. At $\delta t = 11.7$~days, the photosphere radius reaches a maximum of $2.1\,\pm\,0.6\,\times\,10^{15}$~cm, or 140~AU, and a temperature of $T_{eff} = 5800\,\pm\,600$~K. The opacity of the external layers of expanding ejecta has now decreased sufficiently to allow central radiation to escape, causing the photosphere to recede inward. Past $\delta t = 11.7$~days, the radius of the photosphere decreases gradually until $\delta t = 20$~days (10~days following peak luminosity) and then remains roughly constant for the following 30~days considered.

For comparison, we have also estimated the photospheric properties of the explosion as derived from the classification spectrum and the \textcolor{black}{five spectra proceeding it}. At each spectral epoch, we obtain an upper limit on the photospheric velocity from the minimum of the \ion{Si}{2} $\lambda$6355 line. Assuming homologous expansion, the radius is then estimated as $R(\delta t) = v_{\rm exp} \delta t$. We caution that this estimate is highly sensitive to our estimated time of explosion. The effective temperature is calculated from the bolometric luminosity as \begin{equation} 
T_{eff}(\delta t) =  \left(\frac{L_{bol}(\delta t)}{4 \pi R_{ph} (\delta t)^2 \sigma_{SB}}\right)^{1/4}
\end{equation}

where $\sigma_{SB}$ is the Stefann-Boltzmann constant. We find systematically higher temperatures and lower radii using the spectroscopic indicators for the epochs studied, although \textcolor{black}{the overall evolution is similar}. 

We also plot the best-fit spectroscopic estimates of the photospheric temperature and radius for the similar type-Ic SN~1994I in Fig.~\ref{fig:PhotosphericProperties}. \textcolor{black}{W}e note a more gradual temperature evolution for SN~2020oi compared to SN~1994I \textcolor{black}{when derived from the event phometry; this difference is less prominent in the spectroscopic indicators, and may be a reflection of the method used rather than an intrinsic difference in the two explosions.}

The evolution provided by the black-body fits excluding the early excess closely mimics that of the \textcolor{black}{stripped-envelope} SNe considered in \textcolor{black}{both} \citet{2019Prentice} \textcolor{black}{and \citet{2018Taddia}}. \textcolor{black}{The maximum photospheric radius for SN~2020oi is in agreement with the range reported by \citet{2018Taddia} of $0.6-2.4\times10^{15}\;\rm{cm}^{-1}$, and the SNe in both samples also} exhibit a maximum photospheric temperature of $T = 4000 - 8000\;\rm{K}$ followed by a cooling phase to roughly $\sim 5000\;\rm{K}$ $10$~days following maximum light. \textcolor{black}{We similarly note} an apparent increase in temperature following this leveling off, as is reported in \citet{2019Prentice} \textcolor{black}{and \citet{2018Taddia}}; however, this is most likely nonphysical and instead a consequence of non-thermal effects following the photospheric phase of the explosion (as we mention above, the explosion is not well-characterized by a black-body following $\delta t \approx$ 30 days).


\section{Explosion Kinetics from Bolometric Fitting}\label{sec:BolKinetics}

\textcolor{black}{The rapid brightening of the explosion as observed in Fig. \ref{fig:lcbolo} indicates a short diffusion time for photons produced by the radioactive decay of synthesized $^{56}$Ni and $^{56}$Co.} We derive \textcolor{black}{this timescale along with other} explosion parameters for the SN using \textcolor{black}{three} independent methodologies\textcolor{black}{, which we describe and compare below.}

\subsection{The Arnett (1982) Model Applied to the Bolometric Light Curve of SN~2020oi}\label{subsec:ArnettFitting}
In this section, we use \textcolor{black}{a modified} one-component \textcolor{black}{Arnett} model \citep{Arnett1982a} to \textcolor{black}{constrain $M_{\rm Ni56}$, the mass of $^{56}$Ni synthesized in the explosion; $t_{\rm exp}$, the time of explosion; and $t_d$, the diffusion timescale. We further derive The total kinetic energy $E_k$ and the total mass ejected in the explosion $M_{\rm ej}$ from these estimates.} The Arnett model contains a number of assumptions which are applicable during the photospheric phase of most standard SN explosions ($t\,\lesssim\,30$~days): that the ejecta undergo homologous expansion and \textcolor{black}{are} both optically thick and radiation-pressure dominated; that the \textcolor{black}{energy density of the ejecta} is most concentrated at \textcolor{black}{its center;} and that the explosion exhibits spherical symmetry. \textcolor{black}{This formalism has proven valuable for characterizing the bolometric evolution of both type-Ia SNe and stripped-envelope events \citep[see e.g., ][]{2007Phillips_2005hk, 2009Foley_2008ha, 2010Scalzo_2007if, 2011Drout, 2016Lyman_bolo, 2018Sahu_2014ad, 2020Barbarino_PalomarIcSNe}. In this work, we adopt the modified Arnett model developed by \citet{2008_Valenti2003jd} in which} the emission of the SN is assumed to be dominated by the radioactive decay of $^{56}$Ni into $^{56}$Co early in the explosion and from $^{56}$Co to $^{56}$Fe at late times. 

\textcolor{black}{W}e iteratively fit our bolometric light curve excluding the early-time flux excess, first for $t_d$ and next for $t_{\rm exp}$. This procedure requires an estimate of the ejecta velocity at peak bolometric luminosity, which we estimate spectroscopically using the Si line to be $v_{\rm exp} = -12750\,\pm\,250\;\rm{km\,s^{-1}}$. We limit our search for $t_{\rm exp}$ to within 5~days of our earliest observation \textcolor{black}{but no later than MJD 58855.54 (the date of the first explosion detection)} and our search for $t_d$ to (0, 20)~days. 
We also assume an optical opacity of $\kappa_{\rm opt} = 0.07$ $\rm cm^{2} \, g^{-1}$ as is typically adopted for hydrogen-poor CCSNe \citep{2016Taddia}. Using this routine, we find a diffusion timescale for the event of \textcolor{black}{$t_d =8.41 \,\pm\, 0.28$~days} and a predicted time of explosion of $t_{\rm exp} = 58855.4\,\pm\,0.2$ (MJD). The uncertainties reported are propagated from our photometric and spectroscopic uncertainties, and do not include uncertainty in the host-galaxy extinction or the distance to the SN. From this procedure, we further derive a total kinetic energy of $E_k = 0.97\,\pm\,0.13 \times 10^{51}$~erg, \textcolor{black}{comparable to the estimate of $1\times10^{51}$~erg} provided in \citet{2020Rho}.

\textcolor{black}{Because the unusual early-time photometric evolution of the explosion can bias the Arnett estimates for $t_0$ toward later epochs, we derive the time of explosion by fitting the SN rise (excluding epochs of optical and UV excess) to an expanding-fireball model. We elaborate on this model in \textsection \ref{sec:fluxexcess}. We impose an assumption of zero flux at MJD 58852.55 corresponding to the epoch of the last $r$-band non-detection from ZTF. From our best-fit model, we predict an explosion time of MJD 58854.0 $\pm$ 0.3. We adopt this value throughout this work. We note that this estimate is consistent with the MJD date of $58854.0\pm1.5$ estimated by \citet{2020Rho} and that of $58854.50\,\pm\,1.46$ predicted by \citet{2020Horesh_oi}. Further, taking the mean between the last ZTF non-detection and the time of the first explosion detection (on MJD 58855.54), we obtain a comparable MJD date of 58854.05.} 

\textcolor{black}{\subsection{Constraining the Ejecta Mass of SN~2020\lowercase{oi} Using the Khatami and Kasen (2019) Formalism}\label{subsec:K&KFitting}}
In addition to the Arnett \textcolor{black}{prescription}, we use the model described in \citet{2019KhatamiKasen_Ni56} to constrain $M_{\rm ej}$ and $M_{\rm Ni56}$. \textcolor{black}{Although the Arnett model provides an estimate for the mass of synthesized $^{56}$Ni, the model assumes that the peak luminosity of the event is equal to the heating rate at peak. This ignores radiative diffusion originating from the central engine and extending to the surface of the ejecta, which can lead to the true peak luminosity being underestimated if the heating source is centrally concentrated and overestimated if the heating source is highly mixed. For stripped-envelope supernovae such as the one considered here, this can have a large effect on the reported $^{56}$Ni mass \citep{2019KhatamiKasen_Ni56}. By parameterizing the degree of mixing for different classes of explosions with a factor $\beta$, the Khatami \& Kasen model attempts to account for this diffusion and provide a more accurate estimate of the nickel mass.} 

\textcolor{black}{With an estimate for the peak luminosity of the event $L_{\rm peak}$,  the time of peak light $t_{\rm peak}$, and the mixing parameter $\beta$, $M_{\rm Ni56}$ can be determined by re-arranging equation A.13 from \citet{2019KhatamiKasen_Ni56}:}
\begin{dmath}
    M_{\rm Ni56} = \frac{L_{\rm peak} \beta^2 t_{\rm peak}^2 }{2 \varepsilon_{\rm Ni} t_{\rm Ni}^2} \left( \left(1 - \frac{\varepsilon_{\rm Co}}{\varepsilon_{\rm Ni}} \right)\times \\ (1 - 
    (1 + \beta t_{\rm peak}/t_{\rm Ni})e^{-\beta t_{\rm peak}/t_{\rm Ni}}) + \\ \frac{\varepsilon_{\rm Co} t_{\rm Co}^2}{\varepsilon_{\rm Ni} t_{\rm Ni}^2} \left(1 - (1+\beta t_{\rm peak}/t_{\rm Co} )e^{-\beta t_{\rm peak}/t_{\rm Co}} \right)  \right)^{-1}
\end{dmath}

\textcolor{black}{where $t_{\rm Ni} = 8.77$ days
 is the timescale for the radioactive decay of $^{56}$Ni into $^{56}$Co, $t_{\rm Co} = 111.3$ days is the timescale for the radioactive decay of $^{56}$Co into $^{56}$Fe, and $\varepsilon_{\rm Ni}$ and $\varepsilon_{\rm Co}$ are the amount of energy per unit mass released from these decays.} We adopt \textcolor{black}{a} value of $\beta = 0.9$ that has been empirically calibrated from a sample of well-studied SNe Ic \citep{2020Afsariardchi_IcBeta}. \textcolor{black}{The diffusion timescale $t_{d}$ can be calculated from the rise time by numerically solving the equation }
 \begin{equation}
     \frac{t_{\rm peak}}{t_d} = 0.11\; \textrm{ln}\left(1 + \frac{9t_s}{t_d} \right) + 0.36
 \end{equation}
\textcolor{black}{and, from the diffusion timescale, the total ejecta mass is then found by}
\begin{equation}
M_{\rm ej} = t_d^2 \; v_{\rm ej} \; \frac{c}{\kappa_{\rm opt}}
\end{equation}

\textcolor{black}{As in the Arnett treatment, we derive the kinetic energy from the ejected mass:}
\begin{equation}
    E_{k} = \frac{3}{10} M_{\rm ej} v_{\rm exp}^2
\end{equation}
where $v_{\rm exp}$ is the velocity of the explosion at peak \textcolor{black}{(found} spectroscopically with the Si line\textcolor{black}{)}.


We report a synthesized $^{56}$Ni mass of $M_{\rm Ni56} = 0.08\pm0.02\;M_{\odot}$ and a total ejecta mass of $M_{\rm ej} = 0.81\pm0.03\;M_{\odot}$ \textcolor{black}{from this method}. These estimates are slightly higher than the $M_{\rm 56Ni} = 0.07\;M_{\odot}$ and $M_{\rm ej} = 0.7\;M_{\odot}$ values reported by \citet{2020Rho}, \textcolor{black}{where they are estimated} by comparing photometric observations of the event to a library of explosion models. \textcolor{black}{The best-fit values from our one-component Arnett model are $M_{\rm Ni56} = 0.16\,\pm\,0.02\;M_{\odot}$ and $M_{\rm ej} = 1.00\,\pm\,0.08\;M_{\odot}$.} The \textcolor{black}{larger} $M_{\rm ej}$ values \textcolor{black}{derived with the Arnett model} is \textcolor{black}{a direct consequence of their distinct treatments of the diffusion timescale; by considering the additional contributions from radiative diffusion, the timescale calculated using the \citet{2019KhatamiKasen_Ni56} method is significantly higher than is found usin the Arnett method. \citet{2019KhatamiKasen_Ni56} also note that the Arnett model yields less-accurate parameter estimates for lower values of $M_{\rm 56Ni}$, in some cases deviating from the true mass by a factor of two (as is shown for the type-II SN~1987A relative to the value determined from late-time light curve fits). Because of the limitations of the Arnett model, we adopt the \citet{2019KhatamiKasen_Ni56} estimates for the nickel and ejecta masses\textcolor{black}{, as well as the kinetic energy} of the explosion.} 

\begin{figure}
    \centering
    \includegraphics[width=\linewidth]{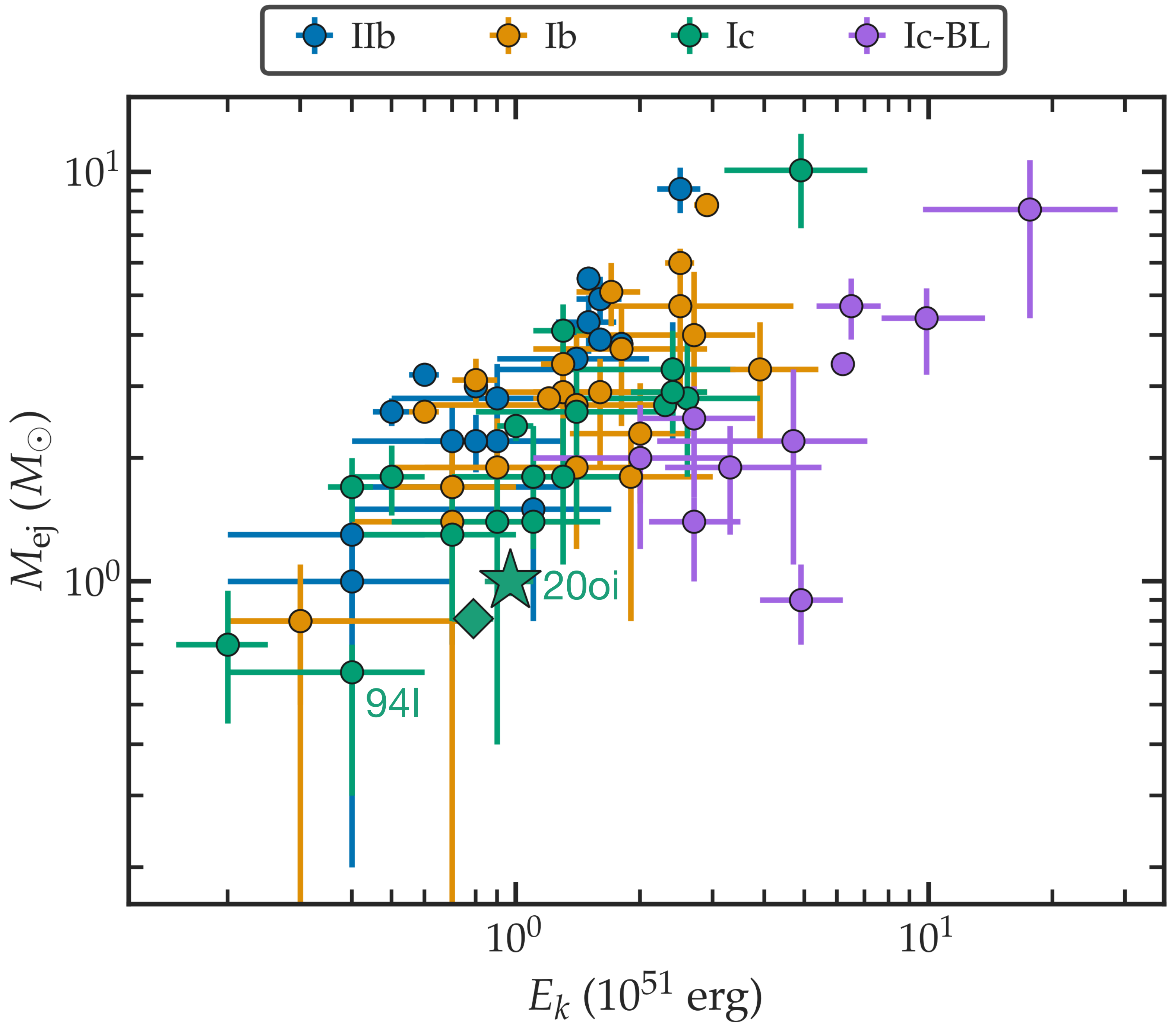}
    \caption{The \textcolor{black}{total ejecta masses and explosion energies for the stripped-envelope SNe in \citet{2016Lyman_bolo} and \citet{2018Taddia} derived using a semi-analytic Arnett model \citep{Arnett1982a}. The star denotes the parameter values derived for SN~2020oi in this study using an Arnett model, whereas the diamond denotes the values adopted from the \citet{2019KhatamiKasen_Ni56} prescription (see text for details). The location of SN~1994I is labeled bottom left. SN~2020oi was more energetic than the well-studied SN~1994I but the two explosions ejected similar masses of material.}}\label{fig:supernovaEnergies}
\end{figure}

The ejected nickel mass estimated for SN~2020oi \textcolor{black}{from both the Arnett and the Khatami \& Kasen formalisms} is lower than the median for SNe~Ic presented in \citet{2019Anderson_Ni56}. Similarly, \citet{2018Taddia} suggest that the mass of nickel synthesized in SN~Ic events is \textcolor{black}{$\sim0.09-0.17\;M_{\odot}$, and our estimates occupy the lower end of this distribution}. Because the radioactive decays $^{56}$Ni$\rightarrow^{56}$Co and $^{56}$Co$\rightarrow^{56}$Fe are the dominant energy sources powering the emission at early and late times, respectively, this finding is consistent with the low luminosity of the bolometric light curve observed in Fig.~\ref{fig:lcbolo}. 
The estimated mass of synthesized $^{56}$Ni is comparable to the $0.07\;M_{\odot}$ value reported for SN~1994I \citep{1994Iwamoto_1994I}, explaining their similar bolometric evolution. We compare the best-fit explosion parameters for SN~2020oi to other \textcolor{black}{stripped-envelope} SNe in Figure \ref{fig:supernovaEnergies}, and report the derived explosion properties in Table~\ref{tbl:explosionParams}.



\newpage
\subsection{The MOSFiT type-Ic Model Applied to the Optical/UV Photometry of SN~2020oi}\label{subsec:MOSFiT}
In addition to estimating the properties of SN~2020oi from the bolometric light curve, we use the SN~Ic model within the Modular Open Source Fitter for Transients \citep[\texttt{MOSFiT};][]{ 2018AGuillochon_MOSFiT} to validate the \textcolor{black}{SN} explosion parameters and constrain the photospheric properties of SN~2020oi. \textcolor{black}{In this framework, a forward model for the emission of an explosive transient is constructed by specifying its central engine and emission SED. In the default SN~Ic model, energy from $^{56}$Ni decay is deposited following the rates provided in \citet{1994Nadyozhin_NiDecay}. This produces black-body radiation that diffuses from the SN ejecta according to \citet{Arnett1982a}. \texttt{MOSFiT} is implemented using a Bayesian framework for iteratively sampling the SN parameter space and approximating the solution with maximum likelihood.} As in the models described \textcolor{black}{in previous sections}, \texttt{MOSFiT} \textcolor{black}{constrains $M_{\rm ej}$ and $M_{\rm Ni56}$ (parameterized by the fraction of $M_{\rm ej}$ comprised of nickel, $f_{\rm Ni}$)} and assumes homologous expansion of the ejecta. \textcolor{black}{We additionally solve for the $\gamma$-ray opacity $\kappa_{\lambda}$ of the ejecta, which controls the degree of trapping of $\gamma$-rays generated from $^{56}$Ni and $^{56}$Co decay; as well as $T_{\rm min}$, the temperature floor of the model photosphere. We exclude photometry after $\delta t > 30$ days from our fit.} We use the dynamic nesting sampling method in \texttt{dynesty} \citep{2020Speagle_dynesty}, with a burn-in phase of 500 and a chain length of 2000, to sample our parameter space. We have verified that we \textcolor{black}{obtain} comparable results using MCMC sampling \textcolor{black}{with} \texttt{emcee}\textcolor{black}{, a Python-based application of an affine invariant Monte Carlo Markov Chain (MCMC) with an ensemble sampler \citep{2013Foreman_MCMC}}.

\begin{figure}
    \centering
    \includegraphics[width=1.0\linewidth]{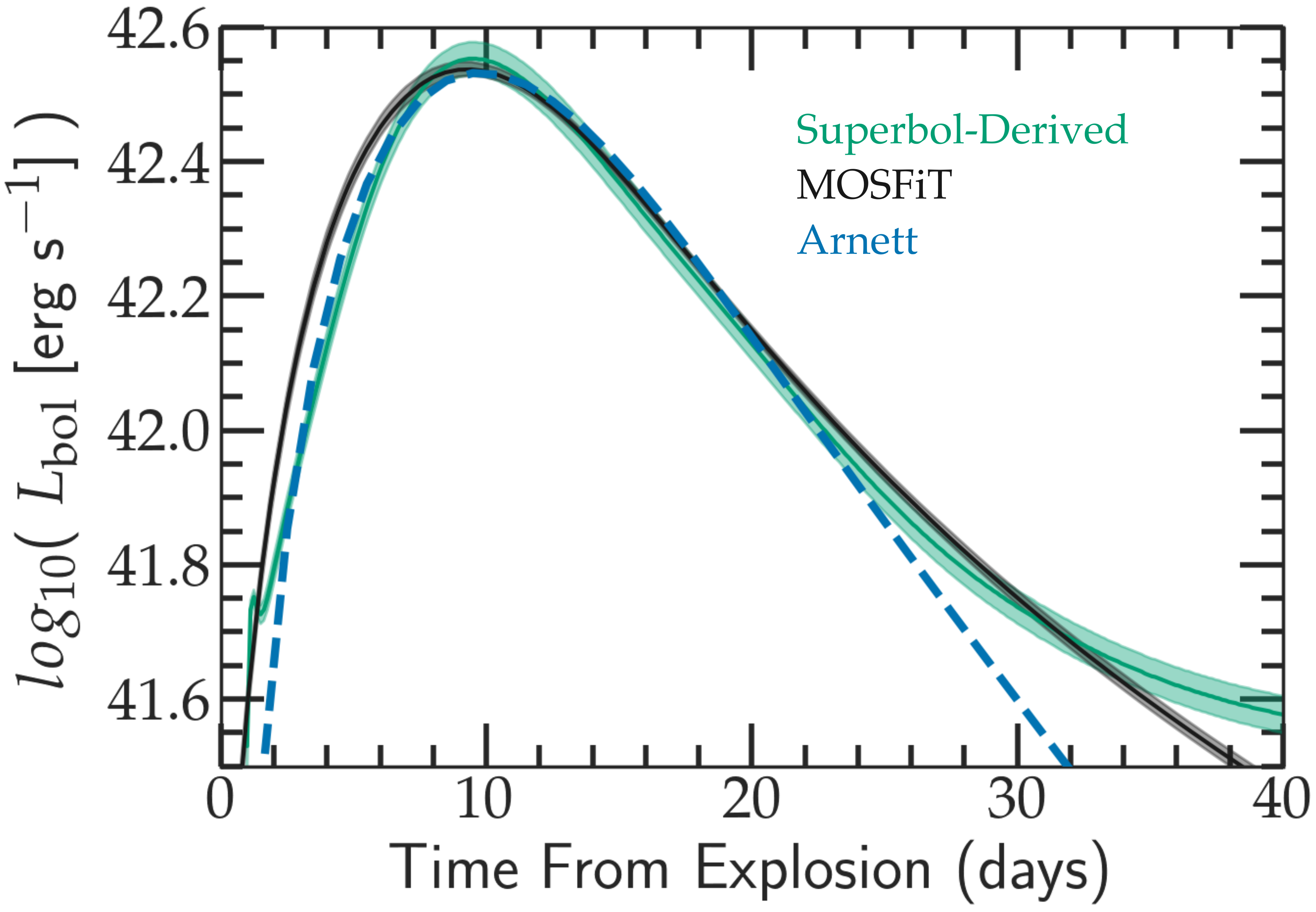}
    \caption{\textcolor{black}{Best-fit bolometric light curve models for the photospheric phase of SN~2020oi. The Arnett model calculated in \textsection \ref{subsec:ArnettFitting} is shown as a blue dashed line, and the black shaded region describes the fit determined using the code \texttt{MOSFiT} (\textsection \ref{subsec:MOSFiT}). The derived bolometric light curve for SN~2020oi is shown in green. Derived parameters are presented in Table \ref{tbl:explosionParams}. The bolometric luminosity of the explosion is not well-described by either model 30 days after explosion due to the rapidly decreasing opacity of the ejecta.} }
    \label{fig:LBolModels}
\end{figure}

\begin{figure*}
    \centering
    \includegraphics[width=\linewidth]{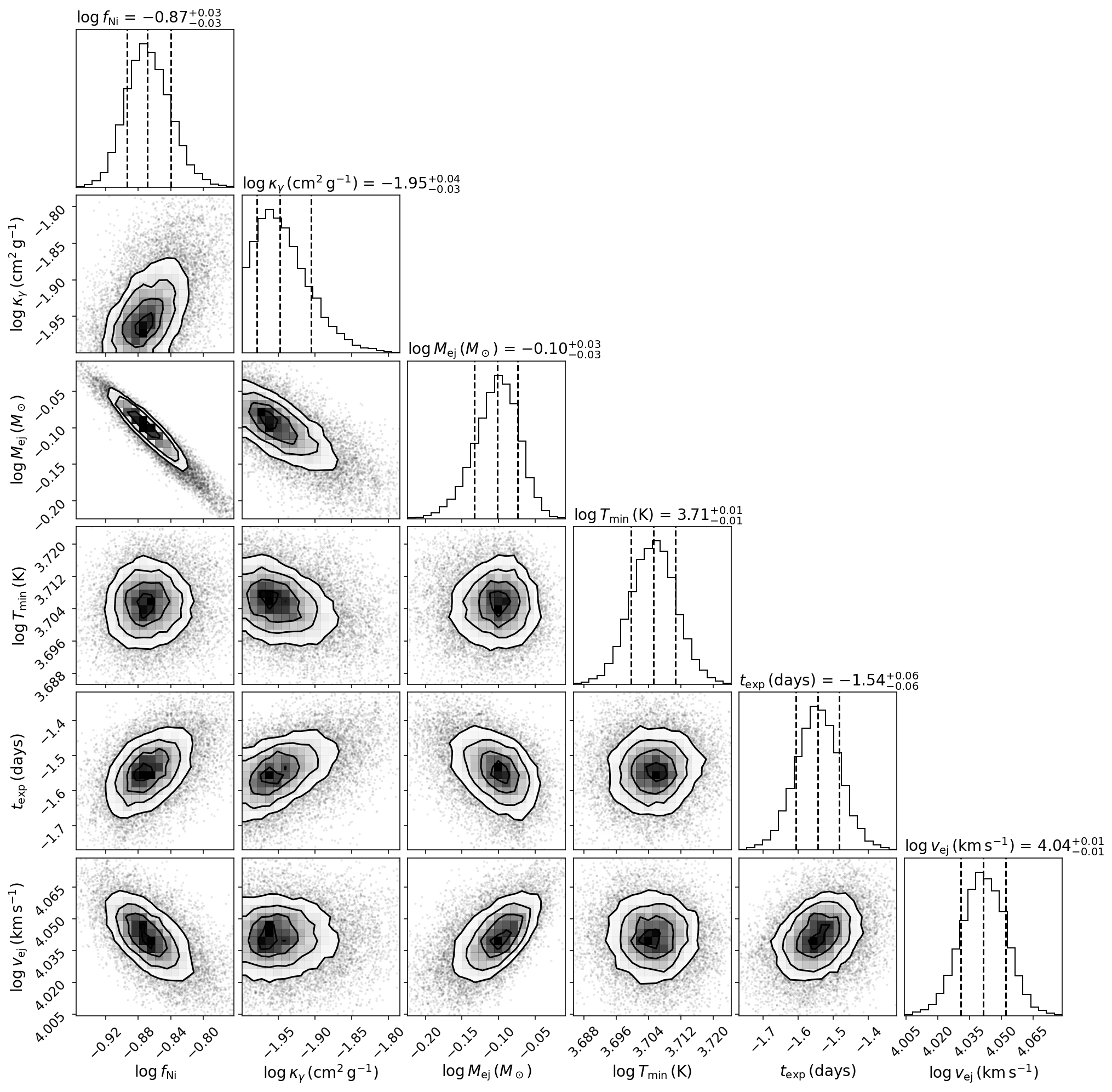}
    \caption{Corner plot of the model parameters for the SN~2020oi explosion found using the nested-sampler implementation in \texttt{MOSFiT}. Marginal distributions from the nested chains are shown at top along with the median parameter values \textcolor{black}{and their 1-$\sigma$ uncertainties}. The parameter $t_{\rm exp}$ indicates the date of explosion relative to the first ZTF observation at MJD 58855.54.}
    \label{fig:MOSFiT_results}
\end{figure*}

We \textcolor{black}{list} \textcolor{black}{our best-fit \texttt{MOSFiT} parameters} in Table~\ref{tbl:explosionParams}. We also \textcolor{black}{compare the bolometric light curves associated with our MOSFiT and Arnett models in Fig.~\ref{fig:LBolModels}, and} present the corner plot \textcolor{black}{from our \texttt{MOSFiT} run} in Fig.~\ref{fig:MOSFiT_results}. We have found during this analysis that, by fitting the model band-by-band under the assumption of black-body radiation (as opposed to our Arnett fit to the bolometric light-curve), the \texttt{MOSFiT} model is more sensitive to deviations from black-body. This was particularly evident later in the event's evolution, where the inclusion of photometry $>30$ days from explosion resulted in a best-fit \texttt{MOSFiT} model whose bolometric light-curve was under-luminous relative to that of SN~2020oi.


We can now compare the photospheric evolution of our \texttt{MOSFiT} model to that derived photometrically and spectroscopically. We plot the black-body radius and temperature for the first 60 days of the model in Fig.~\ref{fig:PhotosphericProperties}. The temperatures predicted by the model within the first $\sim6$~days are higher than those derived from photometry and spectra, but the plateau starting \textcolor{black}{20}~days following explosion is \textcolor{black}{consistent}. The photospheric radius suggested by the model is \textcolor{black}{lower than} the photometric estimates \textcolor{black}{before 20 days and consistent thereafter}.

\capstartfalse
 \begin{deluxetable*}{cccccc}\label{tbl:explosionParams}
 \tablecaption{\textcolor{black}{SN~2020oi Explosion Parameters Derived Using Multiple Models$^1$ }}
 \tablecolumns{6}
 \tablewidth{\textwidth}
 \tablehead{
\colhead{Method} &  \colhead{$M_{\rm Ni}$} &  \colhead{$M_{\rm ej}$ } &  \colhead{ $t_{\rm exp}$} & \colhead{$t_{\rm d}$ } & \colhead{$E_{k}$ }  \\
\colhead{} & \colhead{$(M_{\odot}$)} &  \colhead{$(M_{\odot}$)} &  \colhead{(MJD)} & \colhead{(days)}&  \colhead{$10^{51}$~erg} 
 } 

  \startdata
  \vspace*{1mm}
 \citet{Arnett1982a}                       & $0.16^{+0.02}_{-0.02}$    & $1.00^{+0.08}_{-0.08}$    & $58855.4^{+0.2}_{-0.2}$  & $8.41^{+0.28}_{-0.28}$      & $0.97^{+0.13}_{-0.13}$ \\ \vspace*{1mm}
 \citet{2019KhatamiKasen_Ni56}             & $0.08^{+0.02}_{-0.02}$        & $0.81^{+0.03}_{-0.03}$   & --                 & $19.88^{+0.36}_{-0.36}$                 &    $0.79^{+0.09}_{-0.09}$       \\ \vspace*{1mm}
 \citet[MOSFiT;][]{2018AGuillochon_MOSFiT} & $0.107^{+0.003}_{-0.003}$      & $0.79^{+0.06}_{-0.07}$  & $58853.99^{+0.08}_{-0.07}$  & $8.08_{-0.29}^{+0.23}$ & $0.77_{-0.10}^{+0.10}$ \\ \hline \vspace*{1mm}
\textcolor{black}{Final Values} & $0.08^{+0.02}_{-0.02}$      & $0.81^{+0.03}_{-0.03}$  & $58854.0^{+0.3}_{-0.3}$  & $19.88_{-0.36}^{+0.36}$ & $0.79_{-0.09}^{+0.09}$ 
 \enddata
 \tablenotetext{1}{\textcolor{black}{The adopted explosion time was determined by fitting the early-time rise to a fireball explosion model.}}
 \end{deluxetable*}
 \capstarttrue

\newpage
\section{Inferences on the Pre-Explosion Mass-Loss History}\label{sec:MassLoss}
The X-ray emission from H-stripped SNe exploding in low-density environments is dominated by Inverse Compton (IC) radiation for $\delta t\lesssim 40$~days \citep[e.g.,][]{Chevalier06}. In this scenario, the X-ray emission is generated by the upscattering of seed optical  photospheric  photons by  a  population  of  relativistic  electrons  that have been accelerated  at  the  SN  forward shock. We followed the IC formalism by \citet{Margutti12} modified for a massive stellar progenitor density profile as in \citet{Margutti14b}. Specifically, we assumed a wind-like environment density profile $\rho_{CSM}\propto r^{-s}$ with $s=2$ as appropriate for massive stars \citep{2018Chandra_CSM}, an energy spectrum of the accelerated electrons $N_e(\gamma)\propto \gamma^{-p}$ with $p=3$ as commonly found from radio observations of Ib/c SNe \citep[e.g.,][]{Soderberg06,Soderberg06b,Soderberg06c,Soderberg10} and as observed at late times in SN\,2020oi \citep{2020Horesh_oi}, and a fraction of postshock energy into relativistic electrons  $\epsilon_e=0.1$. We further adopted the explosion parameters $M_{\rm ej} = 0.81\;M_{\odot}$  and \textcolor{black}{$E_k=0.79\times 10^{51}$~erg} inferred from the modeling of the bolometric light curve in \S\ref{sec:BolKinetics}. Under these assumptions, our deep X-ray upper limits from \S\ref{SubSec:Xraydata} lead to a mass-loss rate limit of \textcolor{black}{$\dot M\approx1.5\times 10^{-4}\,\rm{M_{\sun}yr^{-1}}$}  for a wind velocity of $v_w=1000\,\rm{km\,s^{-1}}$.

In an earlier analysis on SN~2020oi by \citet{2020Horesh_oi}, radio observations obtained with the Karl G. Jansky Very Large Array (VLA) beginning on day~5 of the explosion \citep{2020Horesh_FirstRadio} were explained as radiation originating from a shock-wave interaction between the SN ejecta and surrounding circumstellar material. These data were then modeled using the synchrotron self-absorption (SSA) formalism derived in \citet{1998Chevalier_SSA}. In this model, the microphysics of the interaction are parameterized by the ratio between $\epsilon_{e}$, the fraction of energy from the shock-wave injected into relativistic electrons; and $\epsilon_B$, the fraction of energy converted to magnetic fields. The best-fit model found by \citet{2020Horesh_oi} suggests a strong departure from equipartition, with $\frac{\epsilon_{e}}{\epsilon_{B}}\,\geq\,200$. Further, \citet{2020Horesh_oi} predict an X-ray emission from Inverse Compton of $L_x \approx 1.2\times 10^{39}$ erg s$^{-1}$. This corresponds to a flux of $F_x \approx 5.1 \times 10^{-14}\,\rm{erg\,s^{-1}cm^{-2}}$ for their estimated distance of 14~Mpc. 

We find no evidence for statistically significant X-ray emission using Chandra and infer a 0.3-10~keV unabsorbed flux limit of $F_x<6.3\times 10^{-15}\,\rm{erg\,s^{-1}cm^{-2}}$ at $\delta t = 40$~days (see \textsection\ref{SubSec:Xraydata}). Their derived progenitor mass-loss rate of $\dot{M} =1.4 \times 10^{-4}\;M_{\odot}\; \rm{yr}^{-1}$ \textcolor{black}{is comparable to the value calculated in this work; however,} our deeper flux limit indicates either different microphysical parameters ($\epsilon_e$ and $\epsilon_B$) than the ones adopted by \citet{2020Horesh_oi} or suppression of the X-ray emission due to photoelectric absorption by a thick neutral medium.



\section{Spectral Analysis}\label{sec:SpectralAnalysis}
We have used the 1D Monte Carlo radiative transfer code \texttt{TARDIS}\footnote{https://tardis-sn.github.io/tardis/index.html} \citep{Kerzendorf2014,Kerzendorf2018} to estimate the composition of the SN ejecta from the obtained spectra. This requires us to assume a density distribution for the SN ejecta and a bolometric luminosity for each spectrum. For the bolometric luminosities corresponding to each spectral epoch, we have evaluated the bolometric light curve derived in \textsection \ref{sec:Bolo}. Given the similarity of the explosion to SN~1994I, we have adopted the density distribution model corresponding to a carbon-oxygen core of mass $M_f  = 2.1$~M$_{\odot}$ immediately before explosion \citep[CO21, ][]{Nomoto1994,Iwamoto1994}. 
Each spectrum has been computed within a given range of velocities, in which we have assumed the ejecta \textcolor{black}{undergo} homologous expansion. The minimum ejecta velocity for each spectrum was derived from the P-Cygni profile associated with its primary absorption features. Elemental abundances are assumed to be uniform within the velocity range considered. We concentrate our analysis on the four spectra measured closest to peak luminosity ($\delta t = 10.6$, 14.6, 16.5, and 18.4~days from explosion).

\begin{figure}
\centering
\includegraphics[width=\linewidth]{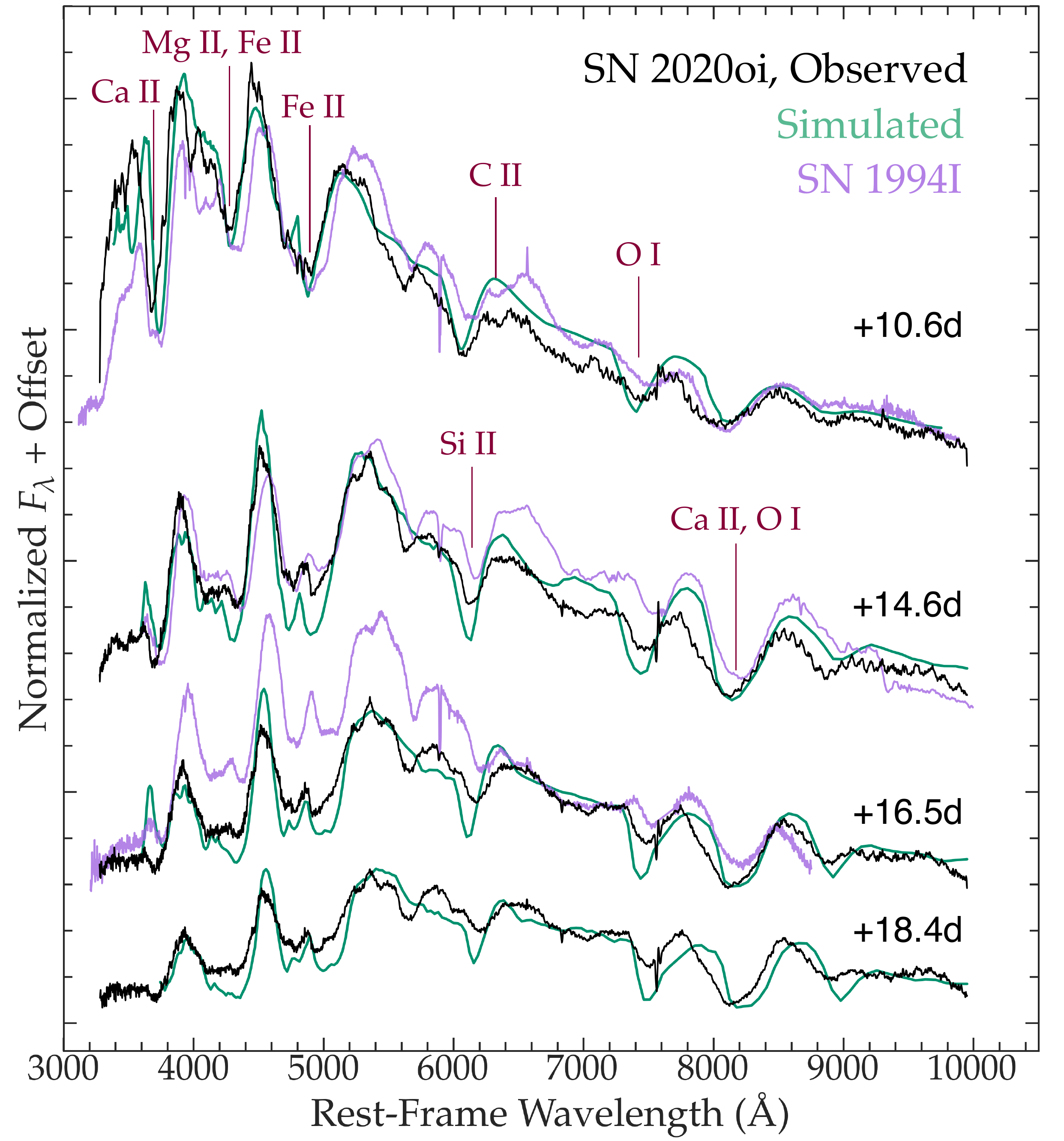}
\caption{The spectra observed at $\delta t \approx$ 11, 15, 17, and 18~days from explosion (black), along with the corresponding best-fit models (green). The spectra of SN~1994I are shown in violet for comparison. Mutual features associated with the presence of Ca, Mg, Fe, Si, and C are shown. The similarity between spectral sequences suggests similar ejecta composition and photospheric evolution for the two SNe.}\label{Fig:SpecFits}
\end{figure}

\begin{figure}[h]
\includegraphics[width=\linewidth]{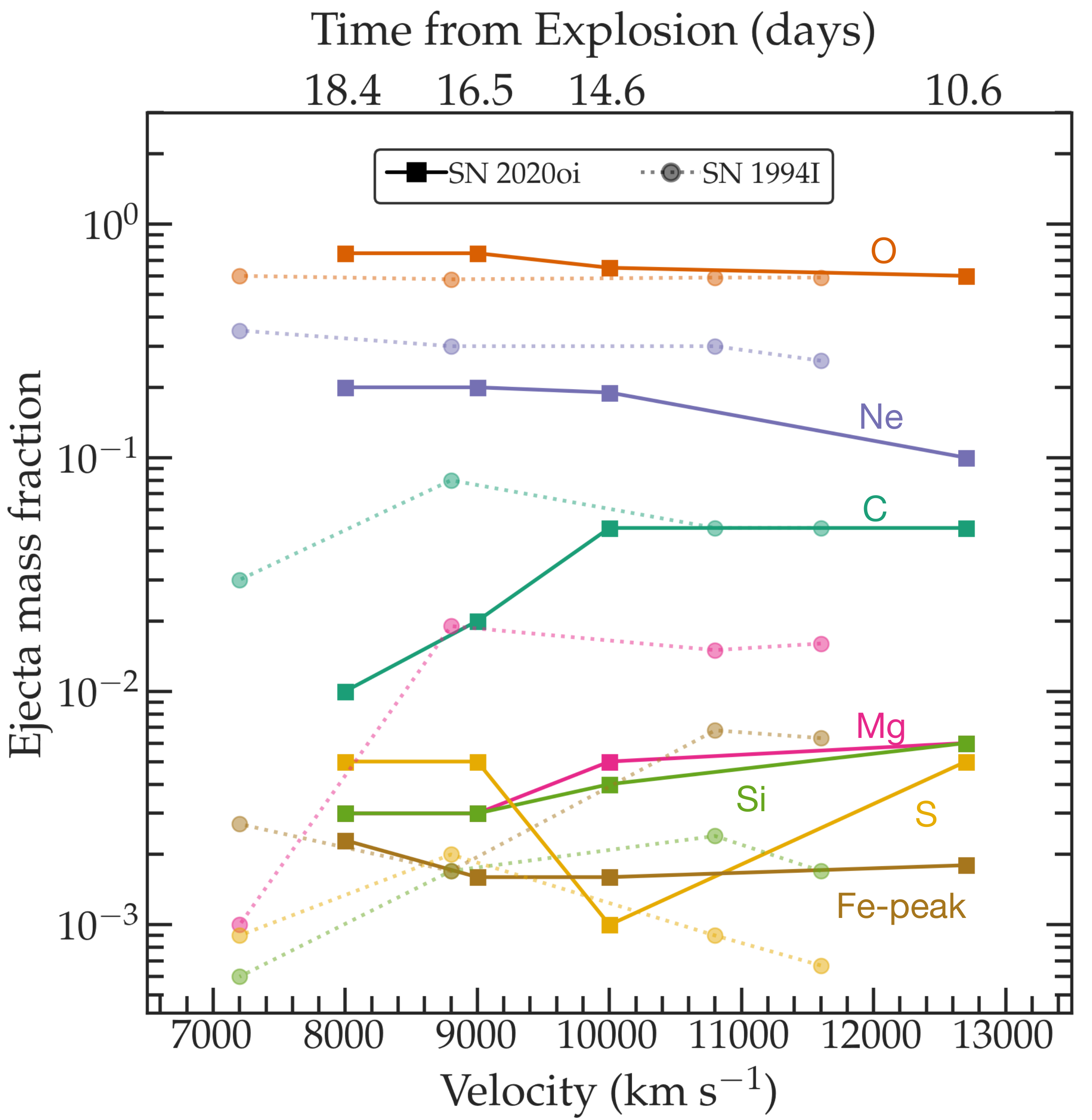}
\caption{The best-fit ejecta composition for the four epochs corresponding to the modeled peak spectra. The epochs relative to the explosion date are listed at top, and the velocity values adopted for each epoch are shown at bottom. The best-fit composition for the SN~1994I ejecta at similar epochs \citep{2006Sauer_1994I} is also shown. Although the composition of the SN~2020oi ejecta \textcolor{black}{varies} between epochs, the comparable abundances of O and Ne across the 8-day evolution indicates partial ejecta mixing.}\label{fig:EjectaComposition}
\end{figure}

Our models are able to reproduce the dominant features identified in the observed spectra: we replicate the profiles of the \ion{Si}{2} $\lambda$6355 \textcolor{black}{feature}, the near-IR \ion{Ca}{2} triplet, the \ion{Fe}{2} contributions and the \ion{Mg}{2} $\lambda$4481 lines. Some discrepancies remain; for example, the simulated \ion{O}{1} line predicts a slightly larger absorption than the observed line \citep[similar to what is shown in][]{2021Williamson_Hiddenhelium}. We have identified the \ion{C}{2} $\lambda$6540 line in the day 10.6 spectrum, and in order to simulate this feature, we have increased the abundance of carbon in the corresponding velocity regime.
We have also included a non-negligible sodium abundance to reproduce the absorption observed around 5600 \AA. This feature may include some contribution from \ion{He}{1} $\lambda$5876, which is excited by non-thermal processes originating \textcolor{black}{from} the decay of \textcolor{black}{n}ickel generated in the explosion \citep{Lucy1991}. A similar line of reasoning applies for the \ion{C}{2} $\lambda$6540 \textcolor{black}{feature}, which can be contaminated by residual absorption from \ion{He}{1} $\lambda$6678.  We do not identify clear He {\sc I} features in our spectral series, such as the triplet 2p-3s transition \ion{He}{1} $\lambda$7065 that is usually observed in \textcolor{black}{the spectra of} type-Ib SNe. The other optical \ion{He}{1} $\lambda$4471 \textcolor{black}{feature} is located in a region contaminated by other absorptions, mainly from Mg and Fe. Unfortunately, our spectral data do not cover the near-IR range where the bright lines \ion{He}{1} $\lambda \lambda$10830, 20580 are visible from the 2s-2p singlet/triplet transitions, and as a result we are unable to conclusively verify contributions \textcolor{black}{from helium}. 

To further investigate the presence of a non-negligible helium abundance, we have also used the \texttt{recomb-nlte} option \textcolor{black}{in \texttt{TARDIS}}. For the day 11 spectrum, we have considered an amount of $\sim0.01\;M_{\odot}$ of helium in our simulated ejecta, and we obtain slightly stronger agreement with the observed spectrum. Nevertheless, we are unable to unambiguously confirm the presence of helium in the SN~2020oi ejecta. We note that the potential \textcolor{black}{presence} of helium was also considered in the case of the type-Ic SN~1994I \citep[see e.g.,][]{Filippenko1995,Baron1999} and previously for SN~2020oi \citep{2020Rho}. 



In Figure \ref{Fig:SpecFits}, we show the spectral series obtained near peak with the FLOYDS spectrograph along with the results of our spectral synthesis simulations. As an additional comparison, we plot three spectra corresponding to the type-Ic SN~1994I at comparable epochs in its explosion \citep{Filippenko1995}. The two events show notable similarities in their evolution and in the presence of \ion{Ca}{2}, \ion{Mg}{2}, \ion{Fe}{2}, \ion{Si}{2}, and \ion{O}{1} features. SN~2020oi shows slightly higher ejecta velocities than SN~1994I \citep{1999Millard_1994I}, as estimated from the minima of the P-Cygni absorptions lines (in particular, from the \ion{Si}{2} $\lambda$6355 transition). This result is also consistent with the higher kinetic energy found for this SN (see \textsection\ref{sec:BolKinetics}) compared to SN~1994I (\textcolor{black}{\textcolor{black}{$0.6 - 0.8\times 10^{51}$~erg}}; see \citealt{1999Millard_1994I}), and also its higher bolometric peak in Figure \ref{fig:lcbolo}. 

The dominant species recovered from the \texttt{TARDIS} \citep{Kerzendorf2014,Kerzendorf2018} simulations of the peak spectra are shown in Figure \ref{fig:EjectaComposition}, and the full abundance pattern found for each spectrum is presented in Table~\ref{tbl:abundances}. The abundance pattern varies only marginally across the epochs that we have simulated and within the velocity range considered, suggesting mixing within the ejecta. Our simulated composition is also similar to that reported for other type-Ic SNe for which element mixing has been discussed \citep{2006Sauer_1994I}. \textcolor{black}{A more detailed analysis of these spectra considering a stratified abundance distribution is planned for an upcoming work, allowing us to further investigate mixing signatures.}

\capstartfalse
 \begin{deluxetable*}{ccccccccccccccccc}\label{tbl:abundances}
 \tablecaption{Abundance Patterns for Simulated SN~2020oi Spectra}
 \tablecolumns{16}
 \tablewidth{\textwidth}
 \tablehead{
 \colhead{Phase} &   \colhead{$X_{He}$} &  \colhead{$X_{C}$} & \colhead{$X_{O}$} &  \colhead{$X_{Ne}$}&  \colhead{$X_{Na}$} & \colhead{$X_{Mg}$} & \colhead{$X_{Si}$} & \colhead{$X_{S}$} & \colhead{$X_{\rm Ca}$} & \colhead{$X_{Ni}$} & \colhead{$X_{Fe}$} & \colhead{$X_{Co}$} & \colhead{$X_{Cr}$}  & \colhead{$X_{Ti}$}  & \colhead{$X_{Ar}$} }
 \startdata
 \textcolor{black}{+3.3d}   & 0.65 & 0.10 & 0.168  & 0.00 & 0.000 & 0.030 & 0.040 & 0.000 & 0.0050 & 0.0001 & 0.0010 & 0.0000  & 0.00001 & 0.00001 & 0.00 \\
 \textcolor{black}{+10.6d}  & 0.14 & 0.05 & 0.600  & 0.10 & 0.001 & 0.006 & 0.006 & 0.005 & 0.0005 & 0.0005 & 0.0010 & 0.0001  & 0.00010 & 0.00010 & 0.02 \\
 \textcolor{black}{+14.6d}  & 0.01 & 0.05 & 0.650  & 0.19 & 0.010 & 0.005 & 0.004 & 0.001 & 0.0005 & 0.0005 & 0.0010 & 0.0000  & 0.00005 & 0.00005 & 0.04 \\
 \textcolor{black}{+16.5d}  & 0.00 & 0.02 & 0.750  & 0.20 & 0.010 & 0.003 & 0.003 & 0.005 & 0.0005 & 0.0005 & 0.0010 & 0.0000  & 0.00005 & 0.00005 & 0.02 \\
 \textcolor{black}{+18.4d}  & 0.00 & 0.01 & 0.750  & 0.20 & 0.010 & 0.003 & 0.003 & 0.005 & 0.0010 & 0.0007 & 0.0015 & 0.0000  & 0.00006 & 0.00006 & 0.03
 \enddata
 \tablecomments{Values listed are fractional abundances.}
 \end{deluxetable*}
 

\section{The Very Early Spectrum of SN~2020\lowercase{oi}}\label{sec:day3spec}
We now consider the peculiar features of the SN~2020oi spectrum obtained at $\delta t = 3.3$~days. This spectrum is one of the earliest obtained for a type-Ic SN. 

This spectrum shows considerable absorption features from Si-burning elements, including \ion{Si}{2} $\lambda$6355 and the Ca II NIR triplet jointly expanding at a velocity of $v_{\rm{exp}}=-24$,000$\,\pm\,500\;\rm{km\,s^{-1}}$. At the same velocity, we have identified the feature at $\sim4500$~\AA\ as \ion{Fe}{2} (multiplet 42), although this feature is likely blended with other fainter absorptions of Fe-peak elements \citep[e.g., $\lambda=4508.29$\AA; see][]{2017Aleo_IronAbundance}. The lack of a substantial absorption from \ion{O}{1} $\lambda$7773 indicates that the line-forming region of this spectrum is located in the most external layers of the ejecta, where the abundance pattern is enriched in lighter elements such as carbon and helium. Indeed, we find evidence for \ion{He}{1} $\lambda$5876 and \ion{C}{2} $\lambda$6580, \textcolor{black}{and cannot rule out a potential contribution} from \ion{He}{1} $\lambda$6678. Unfortunately, our spectrum does not cover the near-IR region where the the \ion{He}{1} $\lambda$10830 line is typically prominent in the presence of a helium-rich gas. 

\begin{figure}
    \centering
    \includegraphics[width=\linewidth]{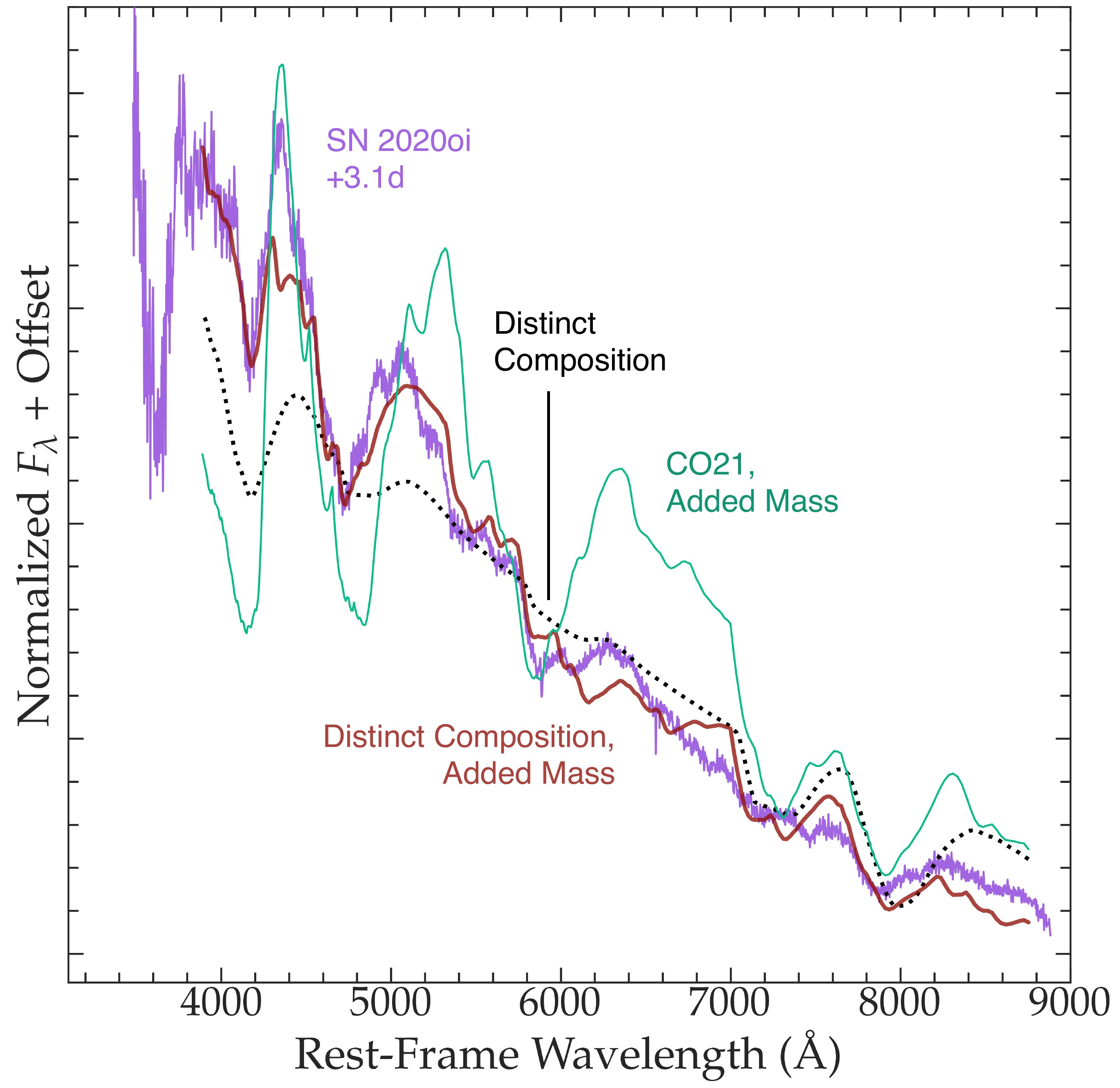}
    \caption{\textcolor{black}{The early-time spectrum for SN~2020oi (violet) compared to three composition models: One in which a best-fit composition distinct from the CO21 model was used (black, dashed); one in which additional mass was added at the highest velocities with composition matching that of CO21 model (green); and one in which high-velocity mass was added with a composition distinct from CO21 (red). The model with a distinct composition (Table \ref{tbl:abundances}) of additional mass at high velocities provides the best fit to the day 3 spectrum.}}
    \label{fig:EarlySpectrum}
\end{figure}

To characterize this early spectrum, we undertake the same composition modeling using \texttt{TARDIS} as was done for the peak spectra. However, we are unable to reproduce the observed spectrum using the same SN~1994I CO21 density distribution \citep{Nomoto1994,Iwamoto1994} that was adopted for the peak spectra; in particular, we cannot reproduce the blue excess observed at wavelengths $\leq$ 5000 \AA. Consequently, we have considered deviations from the pure CO21 model for this spectrum caused by the presence of a gas excess at larger radii. We note that a similar approach has been recently adopted in \citet{2021Williamson_Hiddenhelium} in an analysis of SN~1994I. We find that our observed spectrum can be reproduced by an excess of $ \sim 0.2 \; M_{\odot}$ of material composed of a large amount of carbon, helium, oxygen, and traces of heavy element signatures (Ca, Si, S, Fe) at the highest velocities ($v_{\rm{exp}}\approx-24000\,\rm{km\,s^{-1}}$), roughly corresponding to $\sim${}$10^{14}$ cm at the time the spectrum was obtained \textcolor{black}{(}assuming homologous expansion\textcolor{black}{)}. We show this best-fit spectrum\textcolor{black}{, as well as those predicted by the CO21 composition and density models,} in Fig.~\ref{fig:EarlySpectrum}. \textcolor{black}{\textit{Our fits suggest that the blue excess of the day 3.3 spectrum can be explained by an additional-mass component with a composition distinct from the ejecta near peak.}} However, we note that our \textcolor{black}{final} simulation does not 
precisely reproduce the continuum at bluer wavelengths (e.g. $\lambda << $ 5000 \AA. 

\textcolor{black}{If the blue excess observed in the day 3.3 spectrum is the result of emission from material present at the highest explosion velocities, any additional signatures within the day 6.6 spectrum will better constrain its mass and composition. This analysis is beyond the scope of this work but is planned for a separate publication.}

\section{Characterizing the Early-Time Optical and UV Emission of SN~2020\lowercase{oi}}\label{sec:fluxexcess}

\subsection{Evidence for Flux in Excess of an Expanding-Fireball Explosion Model}\label{subsec:fluxExcess}
We now consider the evidence for a bump in the photometry at day $\delta t \approx 2.5$ in excess of the emission expected for traditional SN explosions. 

\begin{figure}
    \centering
    \includegraphics[width=\linewidth]{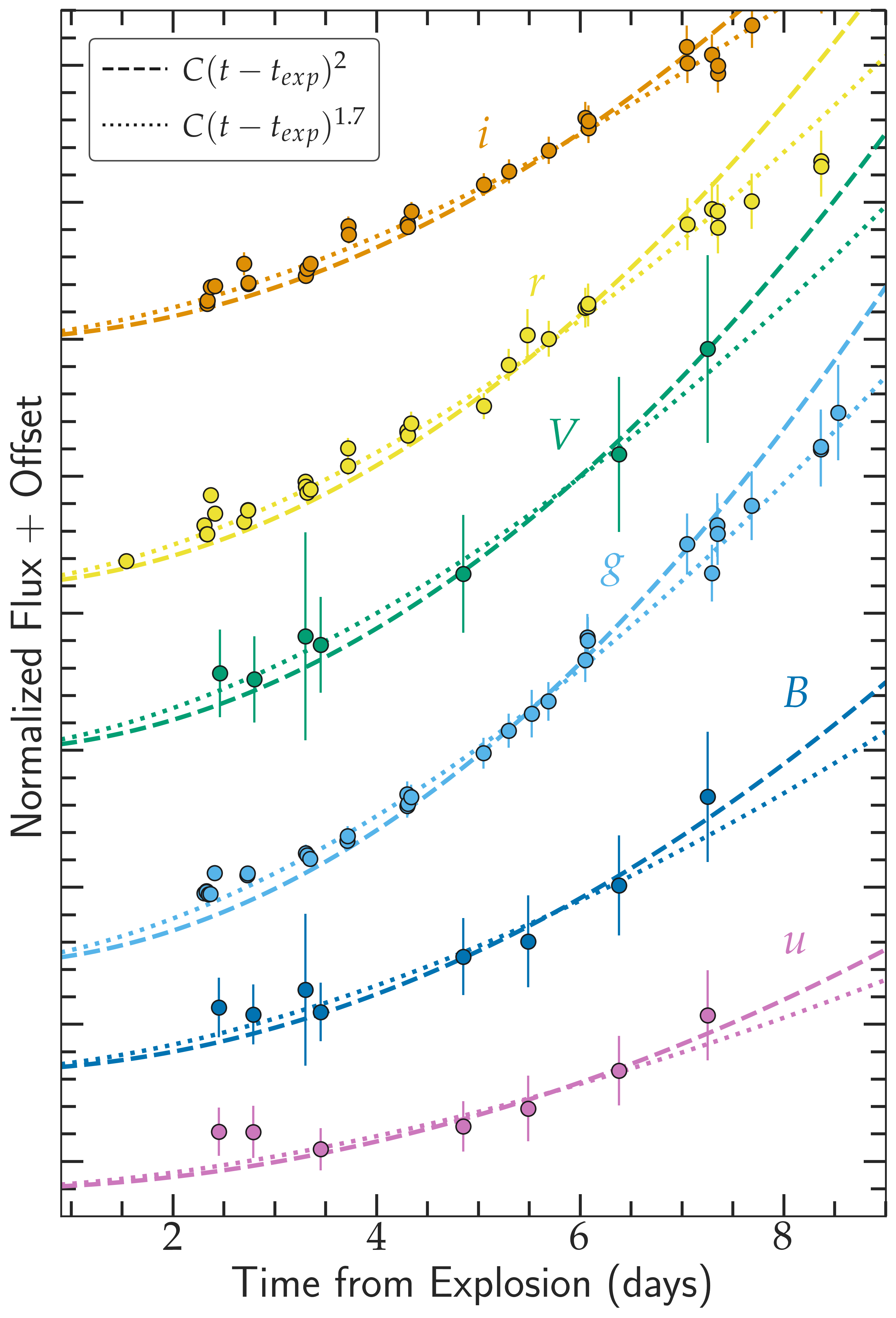}
    \caption{The early-time normalized flux of SN~2020oi. The dashed line corresponds to a canonical expanding-fireball model $f\propto (t-t_{\rm exp})^2$ applied to days $3-10$ of the photometry in each band, while the dotted line describes a model with $f\propto (t-t_{\rm exp})^{1.7}$ to more accurately capture the photometry in $gri$ bands following 7~days. Flux in excess of that predicted by both models can be seen at $\delta t \approx$ 2.5~days from explosion for the majority of bands.}
    \label{fig:FluxExcess}
\end{figure}

We fit the extinction-corrected flux spanning 3 to 10~days post-explosion in each band (excluding the early-time bump) to a canonical expanding-fireball model ($f \propto \left(t - t_{\rm exp} \right)^2$, where $t_{\rm exp}$ is the time of explosion). 
We have also fit a $\left(t-t_{\rm exp}\right)^n$ model where we allow $n$ to vary between $1.0$ and $3.0$, finding reasonable agreement with the rise across all bands for $n=1.7$. We present both models in Fig.~\ref{fig:FluxExcess} along with the associated photometry. 
Although neither model perfectly captures the early-time rise of the SN due to their simplicity, the $n=1.7$ model more accurately describes the gradual increase in explosion flux past $\delta t \approx 6$ days. The models most closely fit the data between 4 and 6~days, which is unsurprising given the higher photometric uncertainties for data obtained at later epochs. We calculate the reduced-$\chi^2_\nu$ Goodness-of-Fit across all bands for our analytic fireball models, where $\nu$ quantifies the degrees of freedom in our early-time dataset.

We find a $\chi_\nu^2$ value of \textcolor{black}{1.9} for the $n=2$ model and \textcolor{black}{0.5} for the $n=1.7$ model. Next, we calculate the reduced-$\chi^2$ across all bands for the values between 2.2 and 2.7 days (comprising the early bump). We find a $\chi_\nu^2$ value of \textcolor{black}{15.0} for the $n=2$ model and \textcolor{black}{4.7} for the $n=1.7$ model, indicating significantly worse fits for these observations than for the rest of the data composing the rise.
Further, the consistency of the flux in excess of the best-fit models between bands (which is not captured by our $\chi_\nu^2$ metric) and within photometry taken at multiple observatories indicate\textcolor{black}{s} a physical origin. We investigate potential explanations for this excess in the following sections.

\subsection{Emission from Shock Cooling}\label{subsec:shockcooling}
To characterize the excess flux observed in the pre-maximum UV and optical photometry, we first consider four distinct shock-cooling models. In the first two models, we apply the \citet{2017Sapir_SBO} treatment using two values for the polytropic indices of the progenitor star. These models assume a progenitor composed of a uniform density core of mass $M_c$ and a polytropic envelope in hydrostatic equilibrium. Immediately following shock breakout, the emission is assumed to be dominated by the outermost layers of the envelope; in subsequent epochs, the emission from successively deeper layers dominate. We adopt polytropic indices of $n=3/2$ and $n=3$, appropriate for a red super-giant with a convective envelope and a blue super-giant with a radiative envelope, respectively. Although these \textcolor{black}{extended hydrogen} envelopes have been stripped in the case of SNe~Ic such as SN~2020oi, this is one of the only shock-cooling treatments \textcolor{black}{in the literature} that attempts to account for the density profile of the progenitor (by the ability to change the polytropic index of its envelope). As a result, it remains a valuable probe of the shock breakout kinetics of stripped-envelope events.

 For the third model, we consider the one-zone analytic solution \textcolor{black}{described} in \citet{2015Piro_SBO}. This model considers shock-cooling from surrounding circumstellar material and is independent of the chemical composition and density profile of the material. The fourth model uses a revised treatment for this emission \textcolor{black}{from} \citet{2021Piro}, \textcolor{black}{which differs from the original formalism with the addition of a power-law dependence of the luminosity with time during the rise of the early emission.}
 
 
\begin{figure*}[h]
    \centering
    \includegraphics[width=\linewidth]{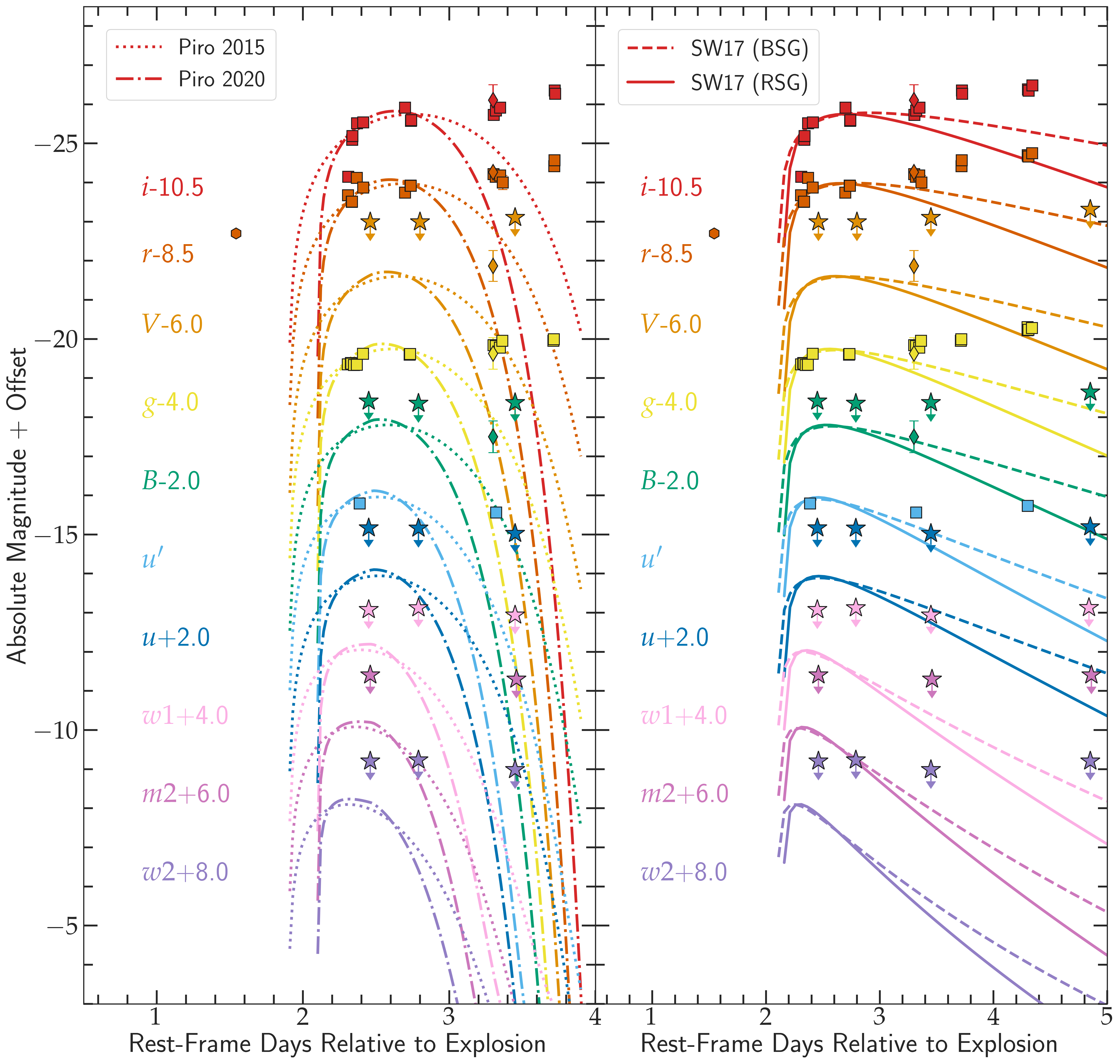}
   \caption{The best-fit shock-cooling models for SN~2020oi excluding the first ZTF observation in $r$ band, shown along with the optical and UV photometry corresponding to the first \textcolor{black}{five} days of explosion. Four analytic fits were considered to characterize the early-time observations: \citet{2015Piro_SBO}; \citet{2021Piro}; and \citet{2017Sapir_SBO} using polytropic indices of $n=3/2$ and $n=3$.} 
    \label{fig:shockCooling}
\end{figure*}

Each of these models allows us to constrain the mass ($M_{\rm env}$) and the radius ($R_{\rm env}$) of extended material surrounding the progenitor; the shock velocity $v_s$; and the \textcolor{black}{time between the early excess and} the time of explosion $t_{\rm exp}$. As in \citet{2020Jacobson-Galan}, we use the package \texttt{emcee} to sample our model parameter space and obtain the fit with the smallest $\chi^2$ value. 

Adopting the procedure outlined above, none of the four models successfully converged to a solution that accurately characterized the early-time photometry. The reason for this lies in the first photometric observation for the event (see Fig.~\ref{fig:FluxExcess}) in $r$ band, which was originally reported in the ZTF alert stream \citep{2019Bellm_ZTFAnalysis}. If the explosion occurred within an environment free of surrounding material, the emission during \textcolor{black}{shock-breakout} of the progenitor's photosphere should be the earliest optical emission observed. The initial $r$ band observation occurs $>0.5$ days earlier than the rest of the photometry and agrees with the continuum predicted by the analytic rise models outlined in the previous section. This suggests that shock breakout from the stellar surface occurred earlier than the optical excess at $\delta t \approx 2.5$ days, and \textcolor{black}{the models considered are} unable to reconcile these two phases of early-time photometry. \textcolor{black}{T}he timescale of these observations disfavors shock-cooling of surrounding material as the cause of the flux excess\textcolor{black}{; nevertheless, we caution that these simplified models have been validated against prominent early emission signatures and may be unsuitable for more subtle excesses.} 

To account for the possibility that the first ZTF observation was not caused by the explosion, we manually fit our shock-cooling models to the early-time bump excluding this point to estimate the properties of the resulting progenitor photosphere. Both these parameters and those corresponding to the full MCMC fit are presented in Table~\ref{tbl:shocktable}. From the manual fits, which are shown in Fig.~\ref{fig:shockCooling}, we derive $M_{\rm env} \approx 0.5-70 \times 10^{-2} \; M_{\odot}$, $R_{\rm env} \approx 4-14 \ R_{\odot}$, and $v_{\rm env} \approx 2-4\times 10^4\; \rm{km\,s^{-1}}$. Although the range in shock velocities found is consistent with the value of $2.4\,\pm\,0.2\;\times\;10^4\; \rm{km\,s^{-1}}$ estimated spectroscopically for the photosophere at $\delta t \approx 3.5$~days, \textcolor{black}{binary evolution models from \citet{Yoon+10_BinaryProg} (Fig.~12) predict larger radii for a progenitor of final mass $2.1\;M_{\odot}$ as is suggested by} the spectroscopic analysis detailed in Section~\ref{sec:SpectralAnalysis}. Although these results suggest that only a small amount of mass \textcolor{black}{located at} the photosphere of the progenitor is needed to explain this emission, \textcolor{black}{additional analysis is required to reconcile the characteristics of the observed bump with the initial ZTF detection.}

Although shock-heating of dense CSM has been proposed to explain the VLA radio observations of SN~2020oi \citep{2020Horesh_oi}, the first radio emission was detected at $\delta t=4.9$ days. This is $\sim 2.5$~days later than the early-time optical and UV excess. If both emission is caused by shock-heated media, the radio-emitting material must either exist at significantly higher radii than the optically-emitting material or the same material must be dense enough to explain the delay (in which case the material would likely be optically thick to the radio emission in the first place). This suggests that the SN~2020oi radio observations are uncorrelated \textcolor{black}{with} the optical excess, \textcolor{black}{and that} the two signatures are probing distinct environments. Without radio observations closer to the epoch of the photometric bump, we are unable to use the VLA data to verify the presence of nearby CSM. 

 \begin{deluxetable*}{cccccccc}\label{tbl:shocktable}
 \tablecaption{Shock Cooling Models}
 \tablecolumns{7}
 \tablewidth{\textwidth}
 \tablehead{
 \colhead{Model} & \colhead{$R_{\rm env}$} & \colhead{$M_{\rm env}$} & \colhead{$v_{\rm env}$} & \colhead{$t_{\rm exp}$} & \colhead{$\chi_{\nu}^2$} & \colhead{DOF} \\
 \colhead{} & \colhead{$R_{\odot}$} & \colhead{$[\times 10^{-2}$] \ $M_{\odot}$} & \colhead{$[\times 10^4]$ km s$^{-1}$} & \colhead{MJD} & \colhead{days} & \colhead{} & \colhead{}
 }
 \startdata
 P15 & $\sim 6$ & $\sim1.5$ & $\sim2.2$ & -- & -- & --\\
 P15$^a$ & $7.23^{+2.33}_{-0.45}$ & $0.82^{+0.02}_{-0.03}$ & $2.45^{+0.10}_{-0.20}$ & $58855.9^{+0.08}_{-0.03}$ & 51.7 & 21\\
 P20 & $\sim 18$ & $\sim0.9$ & $\sim3.6$ & -- & -- & --\\
 P20$^a$ & $13.6^{+1.31}_{-1.24}$ & $0.47^{+0.02}_{-0.02}$ & $4.03^{+0.10}_{-0.10}$ & $58856.1^{+0.01}_{-0.01}$ & 53.7 & 20\\
 SW17 [n=3/2] & $\sim7.1$ & $\sim6$ & $\sim1.6$ & -- \\
   SW17$^a$ [n=3/2] & $4.3^{+0.4}_{-0.3}$ & $2.4^{+0.3}_{-0.2}$ & $2.36^{+0.11}_{-0.11}$ & $58856.2^{+0.1}_{-0.1}$ & 50.6 & 20 \\
 SW17 [n=3] & $\sim7.2$ & $\sim 70$ & $\sim1.8$ & -- \\
  SW17$^a$ [n=3] & $5.7^{+1.1}_{-1.1}$ & $67.0^{+1.5}_{-2.3}$ & $2.04^{+0.20}_{-0.21}$ & $58856.1^{+0.1}_{-0.1}$ & 52.9 & 20 \\
 \enddata
 \tablenotetext{a}{Fitting only the flux excess i.e., $2 < t < 3$~days after explosion.}
 \end{deluxetable*}
 \capstarttrue

\subsection{Emission from Companion Interaction}\label{subsec:companion_interaction}
The \textcolor{black}{ejecta mass derived in} \textsection\ref{sec:BolKinetics} and the \textcolor{black}{agreement of the CO21} composition model \textcolor{black}{with peak spectra in} \textsection\ref{sec:SpectralAnalysis} both \textcolor{black}{suggest} that SN~2020oi originated in a binary system. \textcolor{black}{For systems with low binary separations, the explosion of the primary star will affect the secondary, and} it has been theorized that the presence of a companion can be deduced by the signature it imprints on the earliest moments of an SN explosion. 

The study by \citet{2010Kasen_BinaryInteraction} in connection with SNe~Ia is illustrative. In the conceptual framework presented, the presence of the companion blocks the expansion of the explosion ejecta and carves out a cavity behind it. Thermal diffusion from the heated ejecta, which is typically unable to escape at early times because of the high optical depths involved, then leaks into this rarefied space as radiation. This emission, which varies in intensity based on the binary separation $a$ and the viewing angle $\theta$, can be observed as an optical and UV excess at $\delta t < 8$ days above the broad continuum dominated by synthesized $^{56}$Ni. 

 For the type-Ia simulated by Kasen, the emission timescale associated with companion interaction varies from $\sim 2$~days for highly inclined viewing angles to $\sim 8$~days for an interaction along the line of sight. The lower end of this timescale range agrees more with the inclusion of the early ZTF observation than the timescales associated with \textcolor{black}{the} shock-cooling \textcolor{black}{models in the previous section}, although we caution that this range may differ for SN~Ic progenitor interactions. In addition, as is detailed in \textsection\ref{sec:day3spec}, interaction with material at $\geq 10^{14}$ cm \textcolor{black}{can explain} the blue excess in the day 3.3 spectrum. 

Interaction of the explosion with a binary companion, proceeding in a manner similar to that outlined in \citet{2010Kasen_BinaryInteraction}, should produce additional \textcolor{black}{early-time} signatures. When the initial SN shock collides with the surface of the companion, the post-shock energy is released as an X-ray burst spanning the first few hours of the event in advance of the UV/optical emission. Further, because the SN ejecta are distorted by the presence of the companion, the subsequent emission should show polarization indicative of ejecta asymmetries. Observations of SN~2020oi taken using the WIRC+Pol instrument at Palomar Observatory \citep{2021Tinyanont} near peak found a broadband polarization of $p= 0.37\,\pm\,0.09$\%, low enough to be explained by interstellar dust scattering and not asymmetry within the explosion itself. Because the flux-excess timescale \textcolor{black}{agrees more closely with the highly-inclined interactions simulated in \citet{2010Kasen_BinaryInteraction}}, and the polarization measurements were taken long after any \textcolor{black}{potential} interaction, early asymmetry may be difficult to detect; further, the polarization signature of companion interaction at peak light (or lack thereof) remains unconstrained \textcolor{black}{in the literature}. 

Nevertheless, the question remains as to whether the interaction of a type-Ic explosion with a binary companion would produce a similar flux excess to that predicted for SNe~Ia. The analysis in \citet{2010Kasen_BinaryInteraction} considered a low-mass companion with radius between $10^{11}$~cm (for an evolved sub-giant) and $10^{13}$~cm (for a red giant). In contrast, most companions of stripped-envelope supernovae should reside on or near the ZAMS \citep{2017Zapartas}, and so the signatures of binary interaction should be relatively faint \citep{2015Lui_CompanionInteraction} except for rare close-binary systems \citep{2016Rimoldi_BinarySims}. \textcolor{black}{The stellar cluster coincident with SN~2020oi limits our ability to constrain the brightness of a companion and derive its physical properties.} The majority of binary evolution models in \texttt{BPASS} that agree with our derived ejecta mass (see \textsection\ref{sec:ProgenitorConstraints}) \textcolor{black}{feature} a companion with radius immediately pre-explosion below $2\times10^{11}$~cm and an orbital separation below $10^{12}$~cm. 80\% of these systems feature radial separations higher than the close-binary systems considered in \citet{2016Rimoldi_BinarySims}. Further, the optical bump occurs $\sim0.7$~days after the first ZTF detection. Estimating the ejecta velocity as $-23$,000~km~s$^{-1}$ at early times, this corresponds to a distance of $\sim10^{14}$~cm. As a result, the likely binary separation for this system \textcolor{black}{is lower than suggested by} the timescale of the excess if caused by companion interaction and \textcolor{black}{higher than} the necessary separation for a bright signature. 




\subsection{\textcolor{black}{Emission from Hydrodynamical Interaction of the Ejecta with Circumstellar Material}}\label{subsec:CSMinteraction}
\textcolor{black}{The rapidly-expanding shock wave from an SN is followed by its more slowly-moving ejecta. For progenitor systems surrounded by CSM, the collision of the ejecta with this material creates a high-temperature interface whose multi-wavelength emission is re-processed and re-emitted. Although many stripped-envelope supernovae (SE SNe) for which CSM interaction has been proposed have been SNe~IIb  \citep[e.g., 1993J and ZTF18aalrxas;][]{1993Schmidt_1993J, 2019Fremling_ZTF18aalrxas}, there is increasing evidence that this process can also occur in SNe~Ib/c \citep{2015Milisavljevic_2014C, 2018De_SESNe, 2020Sollerman_CSM}.}

\textcolor{black}{The presence of local CSM as inferred from an early-time signature indicates a mass-loss episode concurrent with or immediately preceding the explosion. It has been recently realised that SNe can occur even for the fraction of stripped stars that are stably transferring mass onto a binary companion \citep{Laplace+20}, potentially providing fresh CSM with which the ejecta could collide.}

Current models \citep{Laplace+20,Gotberg+20,Mandel+21} indicate that significant expansion of the progenitor star occurs only at sub-solar metallicity, $\sim50$~kyr before the explosion and once again a few kyr before the explosion \textcolor{black}{(although different progenitor mass-loss histories may allow for expansion at higher metallicities, as is suggested by \citealt{2019Gilkis_Winds})}. During the \textcolor{black}{latter} interaction phase, the radius of the SN progenitor exceeds several $R_{\odot}$, thus creating a CSM cloud of at least 10$^{13}$~cm. Much less mass ($< 0.1\;M_{\odot}$) is shed during this secondary pre-explosion interaction relative to the first. Given that the envelope mass will be continuously ejected over the few kyr before the SN, and assuming a characteristic ejection velocity of $100\,\rm{km\,s^{-1}}$ (comparable to the orbital velocity at such separations), one may realistically expect a tenuous cloud extending up to $\rm 10^{17.5}\rm cm$ around the system by the explosion time. Such clouds are sufficient to produce an early excess \citep{Chevalier82}. Because the density of this material strongly decreases with radius, a flux excess \textcolor{black}{from CSM interaction} would originate in the inner layers ($10^{14}-10^{15}\rm cm$) of the cloud and the \textcolor{black}{collision} shock would accelerate as it expanded into the outer\textcolor{black}{most} low-density media. This distance is consistent with the timescale for the optical bump observed. The SN energy, in turn, would decrease due to the mass loss in the preceding binary interactions but remain comparable to typical type-Ic SN \textcolor{black}{energies} as an upper limit \citep{Zenati21}.

The main prediction of this scenario is that the event must have originated in a location with sub-solar metallicity, which supports the findings both from \textit{HST} photometry in \textsection\ref{sec:IsochroneFits} and from MUSE spectroscopy in \textsection\ref{sec:HostProperties}. Further, the explosion of the progenitor into CSM composed of its own lost envelope should lead to early-time spectroscopic signatures of the light elements shed, as is strongly suggested by the spectroscopic analysis in \textsection\ref{sec:day3spec}. \textcolor{black}{Radiative diffusion through asymmetrically-distributed or clumpy CSM may also explain the offset of the excess relative to the initial ZTF observation.}

Another interesting line of evidence that may indicate CSM interaction lies in the rising $K$ band continuum found by \citet{2020Rho} 63 days from MJD 58854, which can be attributed to infrared emission from dust. \citet{2020Rho} suggest that this signature may be produced by dust condensing directly from the SN ejecta, pre-existing CSM dust heated by SN radiation, newly formed dust from CSM interactions with the explosion, or an infrared echo from dust in the galaxy's interstellar medium. A dusty pre-existing CSM shell heated by the SN shock at the time of explosion should be located at a distance of $10^{16}-10^{17}$~cm to generate infrared (IR) emission $\sim60$~days post-explosion. This distance is in general agreement with the limits placed on the sizes of previously-observed dust shells \citep{2013Fox_IInCSM}, but it remains unclear whether any CSM surrounding SN~2020oi at these radii would be dense enough to produce the day-63 IR emission. Additional analysis is therefore necessary to determine whether the most-likely CSM density structure created by type-Ic SNe undergoing Roche-lobe overflow could be responsible for both optical and IR signatures. 





\subsection{Emission caused by Asymmetric $^{56}$Ni}\label{subsec:asymmetric_Ni}
It is possible that the presence of decaying $^{56}$Ni in the outer layers of the SN ejecta is the source of the early flux excess, as has been proposed for stripped-envelope events with multiple light curve peaks \citep{2016Drout_2013ge} \textcolor{black}{and other type-I events with less-prominent photometric excesses \citep{2020Magee_NiShells}}. An asymmetric \textcolor{black}{or shallow} distribution, in comparison to \textcolor{black}{the} centrally concentrated $^{56}$Ni ejecta assumed by the Arnett model, would power an event that is blue at early times and red at late times as the outer layers are locally heated \citep{2018Magee}. We do not find \textcolor{black}{significant} evidence for this trend in our spectral sequence relative to that for SN~1994I in \textsection\ref{sec:SpectralAnalysis}. 

It is possible that a jet can deposit $^{56}$Ni into the outermost, high-velocity ejecta of an SN, as was proposed for the type-Ib SN~2008D \citep{2013Bersten_2008D}; however, we have detected no X-ray emission associated with SN~2020oi as would be expected for a jet. In theory, the mass of nickel-rich material needed to explain our early-time emission is \textcolor{black}{likely} small \citep{2020Magee_NiShells}.  However, as we note in earlier sections, significant asymmetry in the ejecta is at odds with the negligible polarization at peak light observed by \citet{2021Tinyanont}. Further, we have found in \textsection\ref{sec:day3spec} that by including C and He at significantly higher radii than the rest of the ejecta, we are able to reproduce the day 3.3 spectrum more faithfully than by considering an excess contribution of Ni and Fe. 
\subsection{Conclusions on the Photometric Excess}\label{subsec:FluxExcessConclusions}
\begin{figure}
    \centering
    \includegraphics[width=\linewidth]{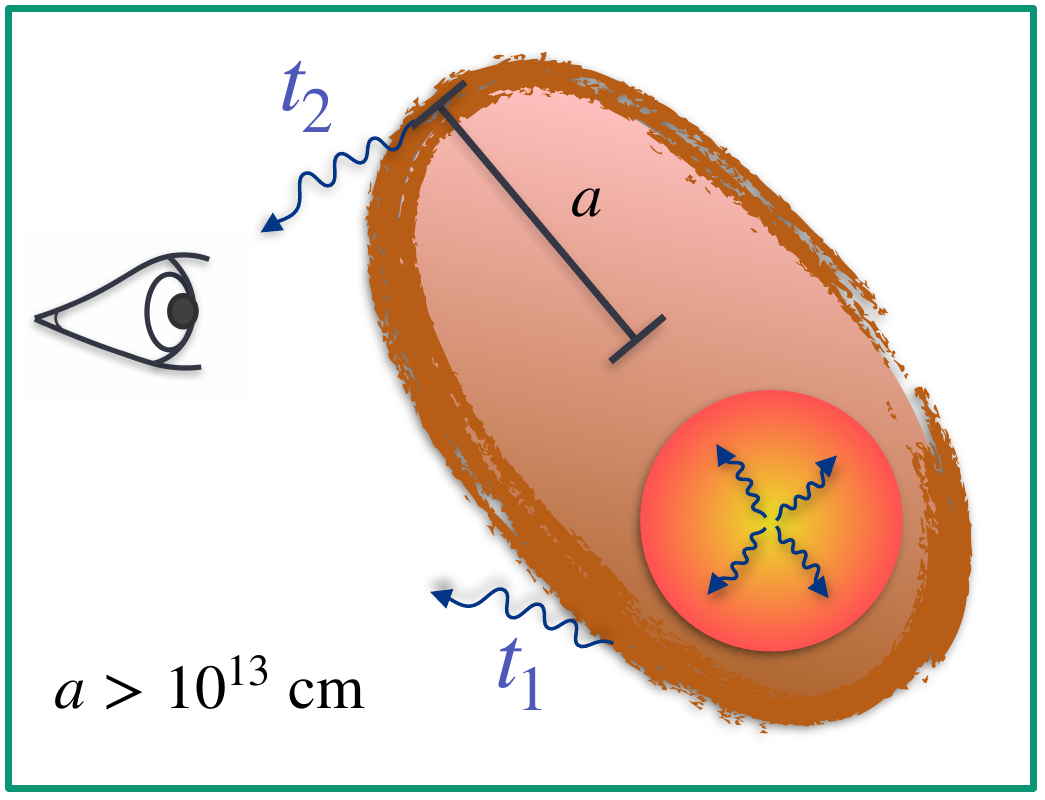}
    \caption{Diagram illustrating flux excess \textcolor{black}{from ejecta interaction with asymmetric CSM}. In this scenario, the SN ejecta collide with an asymmetric cloud and the thermal emission of the material as it cools is observed at \textcolor{black}{distinct epochs} ($t_1$ and $t_2$ corresponding to the epoch of the first ZTF point and the epoch of the photometric bump, respectively) based on its optical depth. From simulations of similar binary systems, the minimum semi-major axis of the cloud predicted  is $\geq 10^{14}$ cm, which would agree with the presence of material as is inferred in \S\ref{sec:day3spec}. 
    }
    \label{fig:PhysicalCartoon}
\end{figure}

The above considerations lead us to the conclusion that the early-time flux excess \textcolor{black}{may be the} emission from ejecta interaction with CSM at \textcolor{black}{large} radii. We illustrate this scenario in Fig.~\ref{fig:PhysicalCartoon}. We note that the interpretation of CSM interaction is not inconsistent with the absence of \textcolor{black}{narrow photoionization} features in the day 3.3 spectrum from \textsection\ref{sec:day3spec}, as these may have been detectable at earlier epochs \citep{2016Khazov_FlashSpec}. \textcolor{black}{We caution that given the limited number of predictive excess models available in the literature for stripped-envelope events, other interpretations are possible. Further, at present we are unable to \textcolor{black}{constrain} whether CSM surrounding the SN is the result of late-stage Roche-lobe overflow, the tenuous remnant of a previous mass-transfer episode, or an eruptive mass-loss event \citep[e.g.,][]{2014Shiode_WaveDrivenMassLoss}.} The viability of \textcolor{black}{late-stage Roche-lobe overflow} from theoretical simulations of this explosion will be the focus of a subsequent paper.



\section{Properties of the Stellar Cluster Coincident with 2020\lowercase{oi} from Pre-Explosion Photometry}\label{sec:IsochroneFits}
In this section, we derive the properties of the stellar cluster associated with SN~2020oi from pre-explosion photometry obtained with the \textit{Hubble Space Telescope} (\textit{HST}). 


We use the code \texttt{Prospector} \citep{2017Leja} to generate synthetic integrated Spectral Energy Distributions (SEDs) corresponding to a series of Simple Stellar Populations (SSPs, which are assumed to be created instantaneously). The \texttt{Prospector} package allows for both MCMC sampling in \texttt{emcee} \citep{2013Foreman_MCMC} and dynamic nested sampling in \texttt{Dynesty} \citep{2020Speagle_dynesty} to generate posterior estimates for a set of model parameters. In addition, it provides an interpolation scheme for generating SEDs spanning an arbitrarily fine grid in parameter space. 

To characterize the stellar cluster associated with the SN, we first calculate \textcolor{black}{its} extinction-corrected flux in each \textit{HST} filter prior to explosion. We then develop an SED model in \texttt{Prospector} parameterized by the age of the stellar cluster $t_{\rm Clust}$; the cluster metal mass fraction log$_{10}(Z/Z_{\odot})$; and the cluster mass $M_{\rm Clust}$. We implement top-hat priors for log$_{10}(Z/Z_{\odot})$ and $t_{\rm Clust}$ spanning [-2, 0.2] and [0.1, 300] Myr, respectively, informed both by our later MUSE analysis and the stellar populations predicted in \citet{2006Allard}. For our prior on $M_{\rm Clust}$, we impose a log-uniform distribution spanning [$10^4$, $10^{11}$]$\;M_{\odot}$. We then sample the posterior distribution of each SED model marginalized by our \textit{HST} observations using \texttt{emcee}, where we have chosen 128 walkers for two rounds of burn-in of length 25 and 50, respectively, and a run length of 1000 iterations. 

For comparison, we have additionally calculated the results obtained using \texttt{dynesty} and from a targeted brute-force grid search of the parameter space, in which we have sampled 200 values each of $M_{\rm Clust}$, log$_{10}(Z/Z_{\odot})$, and $t_{\rm Clust}$ within [$10^{4.5}, 10^{6.5}$]$\;M_{\odot}$, [-2, 0], and [1, 100] Myr, respectively. For each of our SSPs, we assume the Chabrier log-normal stellar IMF \citep{2003Chabrier_IMF} and a Milky-Way curve for extinction of starlight from dust surrounding old stars \citep{1989Cardelli_Av}. We have verified that the use of the \citet{2000Calzetti_Av} extinction law does not alter our results.

\begin{figure*}[htp]
\centering
\includegraphics[width=\linewidth]{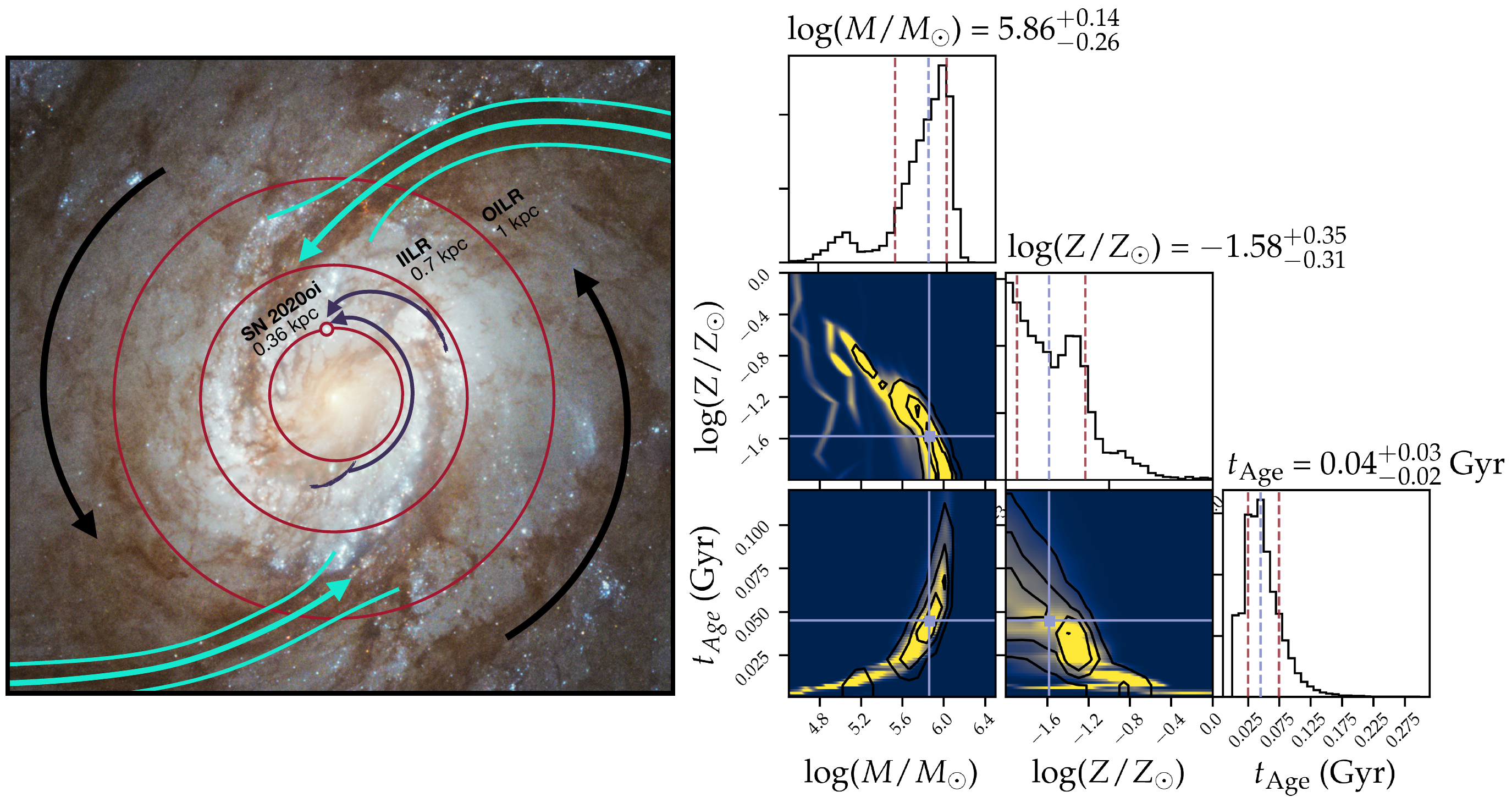}
\caption{\textbf{Left:} Schematic diagram illustrating the \textcolor{black}{proposed} star formation mechanism associated with M100's nuclear ring. \textcolor{black}{Cold gas flows inward} along the spiral arms \textcolor{black}{(turquoise channels)} and collects between the Outer Inner and Inner Inner Lindblad resonances (OILR and IILR, indicated as red circles at $\sim 1$~kpc and $\sim 0.7$~kpc, respectively), and then sinks toward the nucleus from gravity. This material then sweeps past the spiral arm shock \textcolor{black}{fronts} in its rotation and collapses\textcolor{black}{, forming} new stars. Two possible paths for the SN~2020oi progenitor \textcolor{black}{from formation to explosion} are shown in violet \textcolor{black}{ and used to provide an independent estimate for the age of the system (see \textsection \ref{LindbladResonances}).} \textcolor{black}{T}he \textcolor{black}{innermost red} circle marks the \textcolor{black}{radial offset} of SN~2020oi. \textbf{Right:} Corner plot corresponding to our best-fit parameters \textcolor{black}{for} the \textit{HST} pre-explosion photometry of the stellar cluster associated with SN~2020oi. \texttt{Emcee} results are shown in black contours and posterior probabilities derived from a manual grid search is shown in color (where yellow corresponds to the highest-probability parameters and blue corresponds to the lowest). Marginal histograms are plotted at top, with median posterior values \textcolor{black}{marked by light blue lines} and first and third quartiles \textcolor{black}{marked by red lines}.}\label{Fig:IsoFit}
\end{figure*}

We present a corner plot of our posterior estimates from both \texttt{emcee} and grid search in the right panel of Figure \ref{Fig:IsoFit}. Both methods predict a best-fit median cluster mass of $\textrm{log}\left( M_{\rm Clust}\right) = 5.86^{+0.14}_{-0.26} \; M_{\odot}$, a cluster metallicity of $\textrm{log}(Z/Z_{\odot}) = -1.58^{+0.35}_{-0.31}$, and \textcolor{black}{a cluster} age of $t_{\rm Age} = 40^{+30}_{-20}\;\rm{Myr}$. Our \texttt{dynesty} values are \textcolor{black}{consistent with} these estimates. 


\citet{1995Knapen} undertakes a similar analysis in the innermost region of M100 by fitting spatially-averaged optical and IR \textcolor{black}{observations} of dominant star-forming regions to stellar population models. For the region coincident with SN~2020oi, the authors find a best-fit model composed of multiple stellar populations but dominated by stars of age $\sim 40$~Myr, in close agreement with our estimate. A further study by \citet{2006Allard} derived an age of $10-30$~Myr for the stellar population associated with \textcolor{black}{SN~2020oi}. These studies, coupled with our \texttt{Prospector} results from above, suggest that the SN progenitor is coincident with a young ($\sim 40$~Myr) stellar \textcolor{black}{cluster}. Although we do not find evidence for multiple populations of stars as a direct consequence of our simplified SSP treatment, we do not have the wavelength coverage to constrain a more complex star formation history. 


\section{Host-Galaxy Properties from MUSE Spectroscopy}\label{sec:HostProperties}

\begin{figure}
    \centering
    \includegraphics[width=\linewidth]{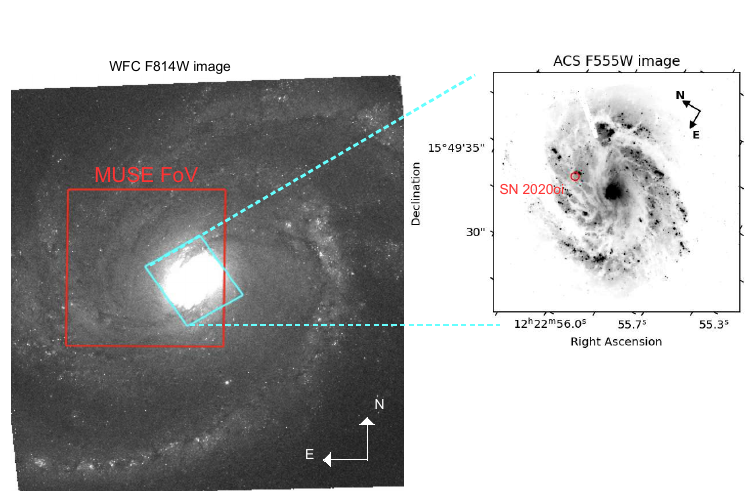}
    \caption{\textit{Hubble Space Telescope} images of the host galaxy of SN~2020oi. The left image corresponds to a Wide-Field Camera observation in the F814W filter, which covers almost the entire galaxy, while the right panel shows the inner region of M100 as observed by the Advanced Camera for Surveys \textcolor{black}{(ACS)} in the F555W filter. \textcolor{black}{T}he position of SN~2020oi is shown as a red circle. On the WFC image, the corresponding field of view of the ACS, as well as the region covered by MUSE observations, are over-plotted with cyan and red squares, respectively.}
    \label{fig:mosaic_HST}
\end{figure}

The inner region of NGC 4321/M100 was observed with the European Southern Observatory Very Large Telescope \citep{2003Henault_MUSE} with the Multi Unit Spectroscopic Explorer (MUSE) in the wide-field mode with adaptive optics configuration (WFM-AO) on April 28, 2019 (Prog. ID 1100.B-0651, PI: Schinnerer). Using the code described in \citet{2020Fusco_MUSEResolution} to reconstruct the atmospheric conditions at the epochs observed, we derive PSF FWHM values of 0.677$\arcsec$, 0.509$\arcsec$, and 0.375$\arcsec$, for 5000 \AA, 7000 \AA, and 9000 \AA, respectively, for our MUSE data. MUSE data have been reduced using standard \texttt{esorex} recipes that were embedded in a general python-based script. The final data cube covers $\sim 90\%$ of the \textit{HST/ACS} \textit{F814W} image, corresponding to the bright star-form\textcolor{black}{ing} ring surrounding the center of the galaxy as can be seen in Fig.~\ref{fig:mosaic_HST}.

To analyze the MUSE data cube, we have first corrected for the Galactic reddening in the direction of the galaxy and then reported each single spaxel in the rest-frame, assuming a redshift of $z = 0.0052$. Then, we have applied the Voronoi spatial binning method \citep{Cappellari2003} assuming a signal-to-noise value of 40 in a wavelength range characterized by an absence of spectral features ($\Delta \lambda = 5600-5700\;$\AA). After this binning, we use our analysis tools to study the properties of the underlying stellar component and nebular gaseous emission in each spectral bin. For each specific physical property we aim to study, we obtain a detailed spatially-resolved map across the full data cube and in the immediate surroundings of SN~2020oi.  

\newpage
\subsection{Stellar Populations within M100}\label{StellarPopulations}

\begin{figure*}
    \centering
    \includegraphics[width=\linewidth]{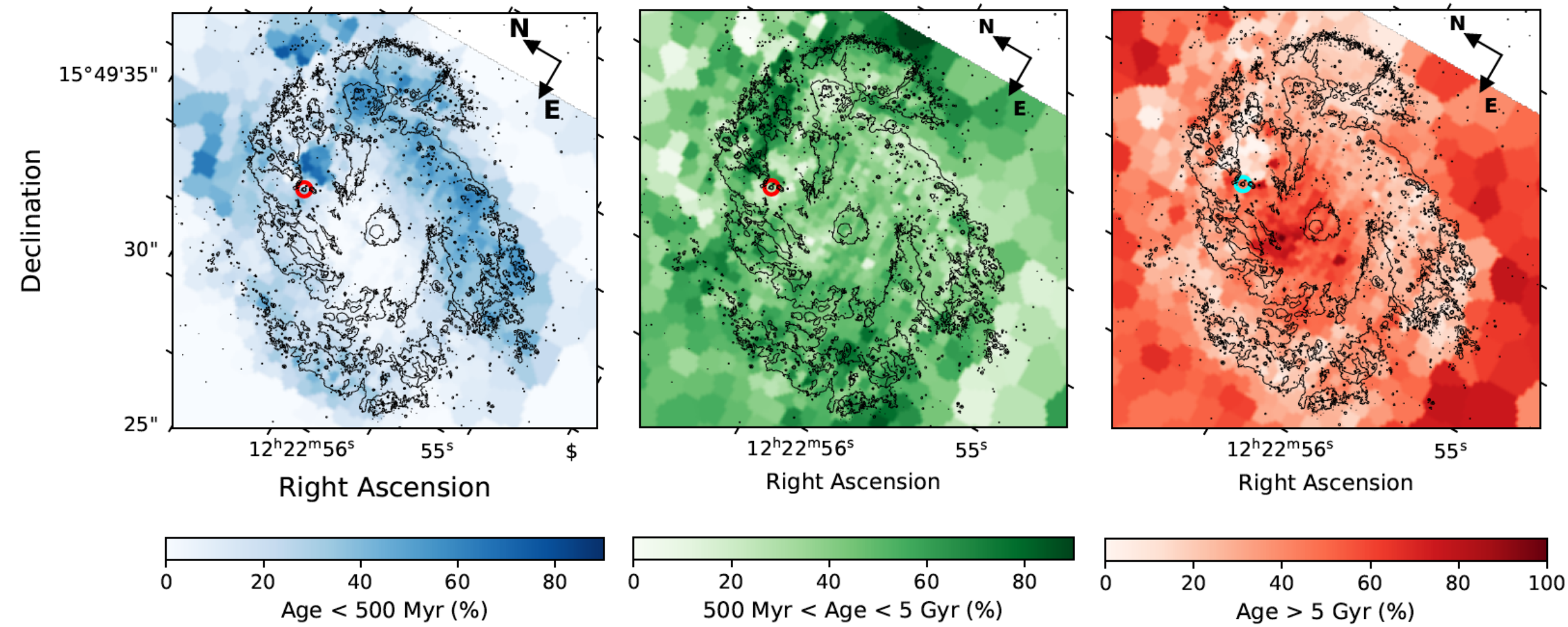}
    \caption{Light fraction contributions for three populations of stars within the nucleus of M100 as derived from MUSE spectroscopy. The location of SN~2020oi is marked with a circle in the upper left corner of each map. }
    \label{fig:StellarAges}
\end{figure*}

To distinguish the underlying stellar continuum from the gaseous emission, we have applied the stellar population synthesis code \texttt{STARLIGHT} \citep{2005CidFernandes_STARLIGHT} to each spectral bin. \texttt{STARLIGHT} allows us to fit an observed spectrum to a combination of template spectra, which can be composed of either individual stellar spectra or \textcolor{black}{distinct} stellar population models obtained from evolutionary codes. In the current work, we have used the stellar population synthesis models described in \citet{Bruzual2003}. This library consists of 150 stellar templates generated with a Chabrier initial mass function \citep{2003Chabrier_IMF} with ages varying between 10$^6$ yrs and 1.8 $\times$ 10$^{10}$ yrs, and with metallicity spanning from $Z = 0.0001$ to $Z = 0.05$ in six bins (where $Z_{\odot}$ = 0.02). This allows us to generate best-fit estimates for the age and metallicity distribution of M100, according to the input templates. A caveat is given by the wavelength range provided by MUSE: with a rest-frame range of 4675-9300 \AA, we miss the bluest region of stellar spectra where important indicators for the star formation history are present (e.g., Mg and Ca H\&K absorption lines). As a result, the values provided are mainly based on indicators available in the wavelength range covered by MUSE at $z = 0.0052$, e.g. the Ca II near-IR triplet.

We plot the light fraction contributions for young \\ ($t\,<\,500\;\rm{Myr}$), intermediate age ($500\;\rm{Myr}\,<\,t\,<\,5\;\rm{Gyr}$), and old ($t\,>\,5\;\rm{Gyr}$) stellar populations in Fig.~\ref{fig:StellarAges}. Most evident is an anti-correlation of old stellar light with the spiral arms that comprise the nuclear ring. This anti-correlation is not evident in either of the two other maps, suggesting that the nuclear ring is comprised primarily of a combination of young and intermediate-age stars. Light from all three of these populations can be seen near the location of the SN, and because of the limited resolution of the IFU data we are unable to definitively associate it with a single stellar population. 



\subsection{SN~2020oi as Evidence for Cold Gas Dynamics in M100}\label{LindbladResonances}

M100 has been extensively studied due to its close proximity and its active star formation sites \citep{Sakamoto1995,1998Garcia_GasM100,2007Castillo_M100,2016Azeez_KSLaw_M100,2018Elmegreen_IRPeaks_M100}. To date, seven SNe have been discovered within M100, but only SN~2020oi occurred within \textcolor{black}{its} central 5$\arcsec$. This makes it possible to leverage previous analyses to further characterize the progenitor system and its formation as a consequence of the dynamical evolution of its host galaxy.

SN~2020oi exploded within a ``nuclear ring'' of radius $\sim 5\arcsec$ where the majority of star formation within M100 occurs \citep{2001Ryder}. \citet{2005Allard} used SAURON IFU spectroscopy to probe the ring's H$\beta$ emission and gas dispersion. In their model of nuclear ring formation, cold gas is channeled inward along the dust lanes of the spiral arms under the gravitational influence of the central bar. This gas settles near the inner Lindblad resonances for the galaxy at the contact points between the nuclear ring and the innermost spiral arms. At the trailing edge of the spiral arms, where the velocity gradient is smaller than at the shock fronts, cold gas clumps and star formation is induced. These locations are predicted to contain the youngest stellar populations within the nuclear ring. The connection between core-collapse progenitors and the clumping of atomic gas by the motion of spiral arms has also been explored in the galaxy M74 \citep{2020Michaelowski_M74Accretion}. We illustrate this mechanism in the left panel of Fig.~\ref{Fig:IsoFit}.


Because the SN took place within the co-rotation radius for M100, the gas and dust at the radius of SN~2020oi is rotating more rapidly than the pattern speed of the spiral arms. If the SN~2020oi progenitor formed from the action of the spiral arms, we can obtain a rough estimate for its age from the time over which the newly-formed stellar cluster \textcolor{black}{underwent} roughly circular motion from within a spiral arm to its current location. We first use a PS1 $gri$-band composite pointing of M100 to estimate the coordinates of a point along the leading edge of each of the inner dust lanes, such that they are roughly the same distance from the nucleus as SN~2020oi ($\sim 4.5$~arcsec). Assuming the cluster undergoes circular rotation, we evaluate the rotation curve for M100 from \citet{2000Knapen} at 4.5$\arcsec$ (using both the H$\alpha$ and CO-derived estimates) and determine the differential speed between the matter at this radius and the pattern speed of the spiral arms from \citet{2005Hernandez_TremaineWeinberg}. We then calculate the length of the circular arc connecting SN~2020oi to each of the dust lanes, \textcolor{black}{accounting for an} extinction with respect to our line of sight of $i=30\degree$ \citep{2000Knapen}. From these estimates, we derive an upper limit to the age of the progenitor cluster of $t_{\rm Age} \approx 9-17$~Myr, if it formed from the passage of the nearest spiral arm; and $t_{\rm Age} \approx 14-26$~Myr if it formed from the furthest arm. The second age range overlaps both with our earlier stellar cluster age estimate and with the age provided by \citet{1995Knapen} (who estimates an age of $\sim 15$~Myr for the majority of stars in the star-forming region coincident with SN~2020oi). Although neither of these estimates alone is conclusive evidence for the age of the SN~2020oi progenitor (and earlier passes of the material through the spiral arms could have equally triggered star formation events), in conjunction with the cluster age estimates from \texttt{Prospector} they present a consistent picture for its formation.

Using population synthesis models, \citet{2006Allard} finds that the spectral emission from the nuclear ring is \textcolor{black}{equally well-}explained by two models. In the first, an initial period of star formation ($t\sim3$~Gyr ago) concludes and is followed only by the starburst event currently observed. In the second, the period of initial formation was followed by multiple continuous starburst events \textcolor{black}{occurring every} $\sim100$~Myr and starting $t\sim500$~Myr \textcolor{black}{ago}. \citet{2006Allard} favors the latter hypothesis, which is consistent with a continuous inflow of gas under the gravitational pumping action of the central bar. While we are unable to distinguish between these two scenarios, our estimate of $\sim40$~Myr for the age of the SN~2020oi cluster suggests that its formation corresponds to the most recent burst of star formation.

\subsection{Metallicity and Star Formation at the Supernova Site}\label{LocalProperties}
Because our IFU data span the inner region of M100, we can use traditional emission-line flux indicators to estimate the metallicity at the location of the SN. \textcolor{black}{W}e employ the empirical relations derived by \citet{Marino2013} to estimate the metallicity at the SN~2020oi spectral bin location based on the (\ion{O}{3} $\lambda$5007/H$\beta$)/(\ion{N}{2} $\lambda$6583/H$\alpha$) and (\ion{N}{2} $\lambda$6583)/(H$\alpha$) line ratios (the O3N2 and N2 indices, respectively), as is appropriate for low-redshift HII regions:
\begin{equation}
    12+\textrm{log(O/H)} = 8.743 + 0.462 \times \textrm{log(N2)}
\end{equation}\label{N2}

\begin{equation}
    12+\textrm{log(O/H)} = 8.753 - 0.214 \times \textrm{log(O3N2)}
\end{equation}\label{O3N2}

\begin{figure}
    \centering
    \includegraphics[width=\linewidth]{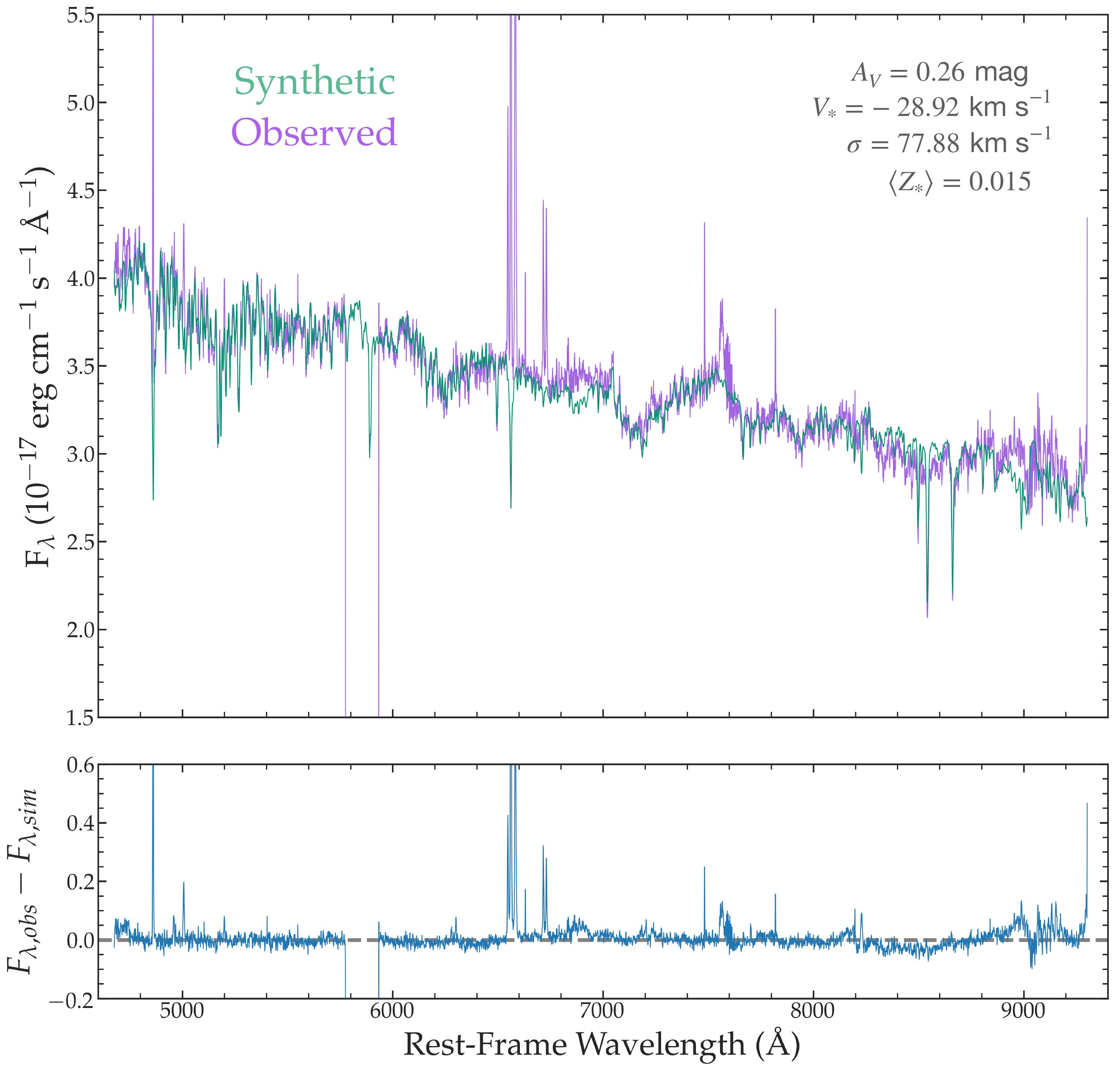}
    \caption{The spectrum at the bin of SN~2020oi as observed by MUSE (in violet). The composite stellar population spectrum obtained from \texttt{STARLIGHT} is shown in green. Emission-line fluxes have been measured from the residual spectrum, calculated \textcolor{black}{by subtracting} the synthetic spectrum by the observed spectrum. Host-galaxy extinction has been calculated internally from emission-line fluxes.}
    \label{fig:spectralbin}
\end{figure}

Line fluxes have been measured on the spectrum \textcolor{black}{obtained by} the subtraction of the composite stellar population best-fit spectrum obtained from \texttt{STARLIGHT} with the observed spectrum (see Fig.~\ref{fig:spectralbin}). Using the N2 and O3N2 indices, we find a metallicity at the location of SN~2020oi of $12+\textrm{log(O/H)} = 8.50\,\pm\,0.01$ ($\pm\;0.18$ sys), and $12+\textrm{log(O/H)} = 8.57\,\pm\,0.03$ ($\pm\;0.18$ sys), respectively. Averaging these, we find $12+\textrm{log(O/H)} = 8.55\,\pm\,0.03$. Assuming a value for solar metallicity of $12+\textrm{log(O/H)} = 8.69$ \citep{Asplund2009}, the metallicity at the position of SN~2020oi is found to be slightly sub-solar. Another estimate for the metallicity comes from the final results of the \texttt{STARLIGHT} fits, where we have averaged the metallicities of each stellar base with its corresponding stellar mass weighted by the eigenvalues of the results obtained. From the analysis of the spectral bin corresponding to the location of SN~2020oi we find $\left<Z\right> = 0.015$, where the Solar value is $Z_{\odot} = 0.02$. We conclude that the stellar metallicity inferred from the analysis of the stellar population underlying the SN is consistent with the metallicity obtained from the analysis of the nebular gas. Both values are also consistent with the average values for the gas-phase and stellar metallicities found from the analysis of a IFU-data sample of type-Ic SN host galaxies \citep{2016Galbany}.

We have also estimated the star-formation rate at the location of SN~2020oi using the method delineated in \citet{1998Kennicutt_SFR}, which is based on the luminosity of the extinction-corrected H$\alpha$ recombination line, $L_{H\alpha}$ = $6.9\,\pm\,1.4\;\times\;10^{37}$~erg s$^{-1}$: we obtain an effective star-formation rate of \\ $\rm{SFR} =6.0\,\pm\,1.2\;\times\;10^{-3}\;M_{\odot}\;\rm{yr}^{-1}\;\rm{kpc}^{-2}$. This value is lower than the average SFR value found in a systematic analysis of type-Ic SN local environments \citep{2018Galbany_PISCO}.  


\section{Deducing the Properties of the SN~2020\lowercase{oi} Progenitor}\label{sec:ProgenitorConstraints}

\begin{figure*}
    \centering
    \includegraphics[width=\linewidth]{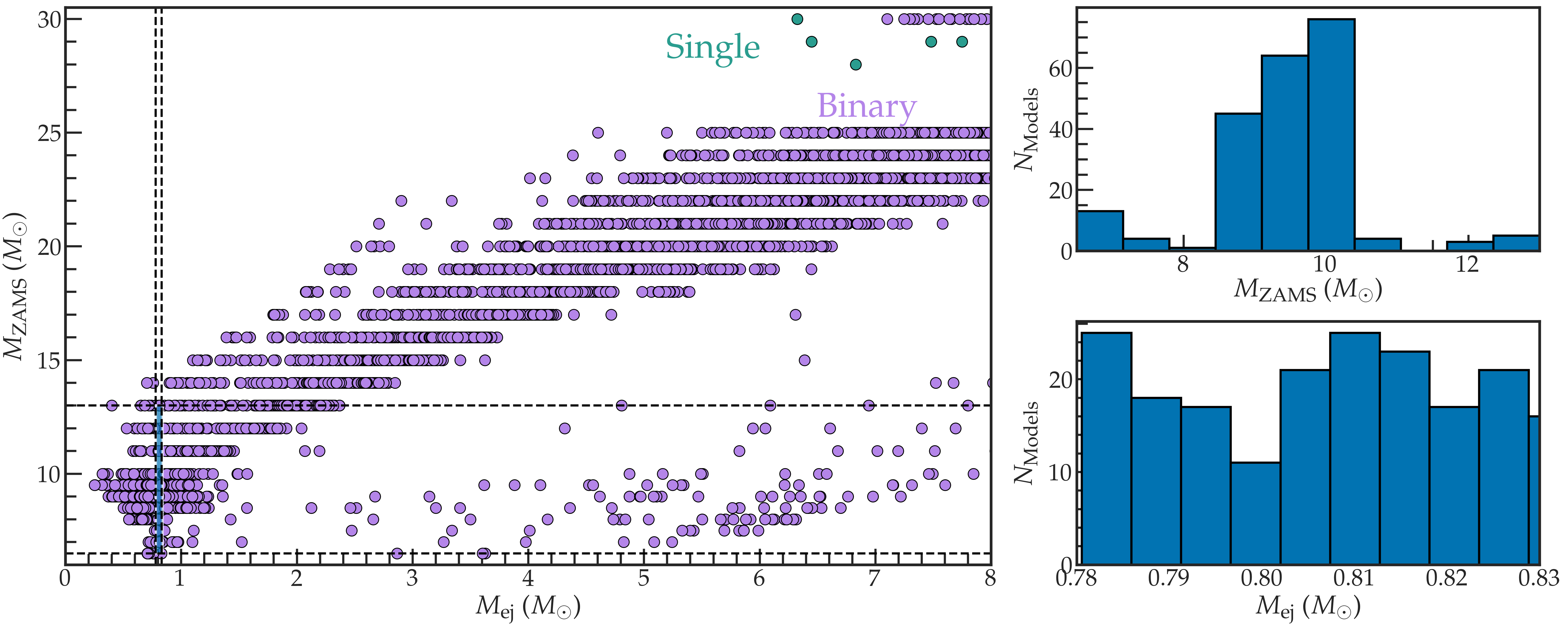}
    \caption{\textbf{Left:} The Zero-Age Main Sequence (ZAMS) mass of the progenitor star $M_{\rm ZAMS}$ versus the ejecta mass $M_{\rm ej}$ following explosion for single (green) and binary (violet) progenitor systems in \texttt{BPASS}. The blue shaded region captures the models in \texttt{BPASS} with predicted $M_{\rm ej}$ values within the range estimated for SN~2020oi. These models are weighted by the initial mass function with properties determined by \citet{2017Moe_PeriodMassBinaries} to reproduce observed binary populations. \textbf{Right Upper Panel:} The distribution of $M_{\rm ZAMS}$ values for the models within the blue shaded region at left. Considering only these models, the most likely mass for the SN progenitor is $9.5\;M_{\odot}$. \textbf{Right Lower Panel:} The \textcolor{black}{range of} $M_{\rm ej}$ values for the same set of models as above.}
    \label{fig:BPASS_Kinetics}
\end{figure*}


\begin{figure}
    \centering
    \includegraphics[width=\linewidth]{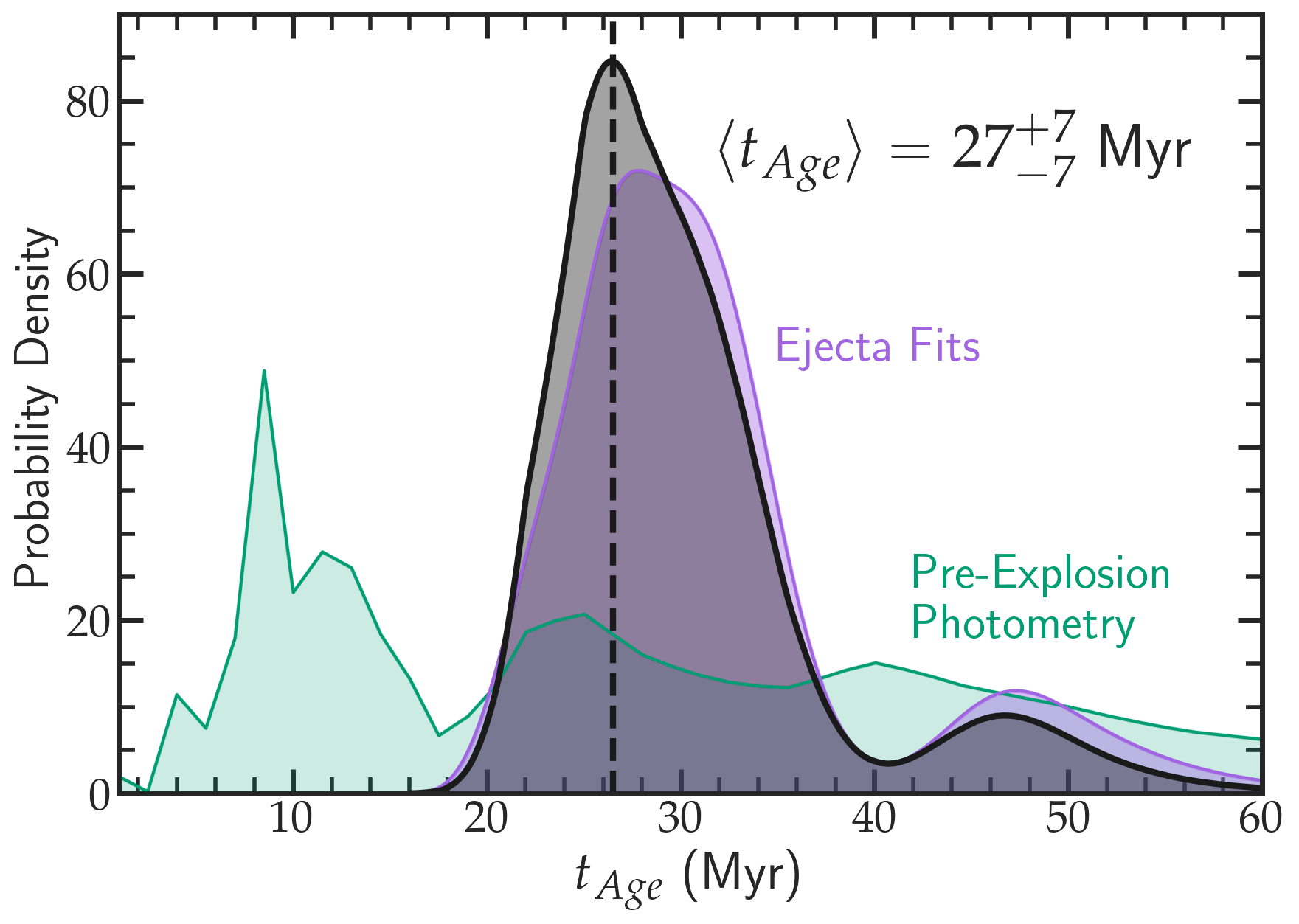}
    \caption{The probability density functions (PDFs) associated with the age of the SN~2020oi progenitor. The estimates derived from pre-explosion cluster photometry are given in green, and those derived from comparing explosion parameters to stellar evolution models are given in violet. The normalized probability density found by combining these two estimates are given as the black PDF at center, and the age with highest posterior probability is reported at right along with the standard deviation of the combined PDF.}
    \label{fig:2020oi_ProgenitorAge}
\end{figure}

The estimated mass ejected in the explosion has strong implications for its progenitor system. We evaluate these implications by comparing our results to events simulated using the binary and single-star models from v2.2 of the Binary Population And Spectral Synthesis (\texttt{BPASS}) code\footnote{\href{https://bpass.auckland.ac.nz/}{https://bpass.auckland.ac.nz/}}, which are described in \textcolor{black}{detail} in \citet{2017Elridge}. We constrain \texttt{BPASS} simulations to those \textcolor{black}{consisting of} a primary star with a CO-core mass greater than $1.38 \;M_{\odot}$ and a total mass greater than $1.5\;M_{\odot}$ immediately pre-explosion, as progenitors less massive than this are unlikely to undergo core-collapse \citep{2017Elridge}; and to only those systems containing a primary star with a hydrogen mass of less than $10^{-3}\;M_{\odot}$ immediately prior to explosion (the threshold reported in \texttt{BPASS} as corresponding to a stripped-envelope event). The resulting models span stellar metallicities from $Z=10^{-5}$ to $Z=0.04$. We plot the ejected mass for a fiducial SN explosion energy of $10^{51}$ erg (roughly corresponding to the energy of SN~2020oi) against the progenitor mass of the system at the beginning of the simulation in Figure \ref{fig:BPASS_Kinetics}. We find the $M_{\rm ej}$ value estimated for SN~2020oi near the lowest end of estimates for a system of initial mass $M_{\rm ZAMS} \approx 6.5-13.0 \;M_{\odot}$, which occurs only in the simulated binary progenitor systems. The mean and median of initial progenitor masses within this subset of models are both $9.5\;M_{\odot}$. Adopting this value and calculating the standard deviation across all viable models, we obtain a most-likely progenitor mass of $M_{\rm ZAMS} = 9.5\,\pm\,1.0\;M_{\odot}$. This value is lower than the initial mass predicted by \citet{2020Rho}, who reports a value of $13 \; M_{\odot}$. As is also noted in \citet{2020Rho} (see their Table~2), the most likely initial progenitor mass predicted for SN~1994I, whose bolometric properties are similar to those of SN~2020oi, is $13-15\;M_{\odot}$ \citep{Iwamoto1994,2006Sauer_1994I}. Adopting our higher Arnett estimate of $M_{\rm ej}=1.00\;M_{\odot}$ results in a higher progenitor mass of $M_{\rm ZAMS} = 10\; M_{\odot}$. This strongly suggests a low-mass binary progenitor origin for SN~2020oi.

Because we have derived a likelihood surface for the properties of our SN cluster from \textit{HST} pre-explosion photometry in \S\ref{sec:IsochroneFits}, we can combine our results with the derived properties of the explosion to extract a most likely age for the SN~2020oi progenitor.

From our likelihood surface, we first marginalize over the cluster metallicity and mass to obtain a probability density \textcolor{black}{function} for the \textcolor{black}{age of the} cluster. We then obtain a histogram of likely progenitor ages from \texttt{BPASS} by considering the ages of only the stellar models that result in a stripped-envelope explosion within the $M_{\rm ej}$ range predicted by the Khatami and Kasen fit to our bolometric light curve. As we note above, these models are all low-mass binary systems. We generate a kernel density estimate associated with this histogram, and then multiply our probability densities and normalize the result to obtain a combined probability density function for the age of the explosion. The resulting distribution is shown in Figure \ref{fig:2020oi_ProgenitorAge}. The most likely age for the SN~2020oi progenitor is found by calculating the peak of the probability density function, and the uncertainty is reported by taking its standard deviation. 

From these estimates, we calculate a final progenitor age of $t_{\rm Age} = 27\,\pm\,7\;\rm{Myr}$.  Although none of the previous SN~2020oi studies constrained the age of the progenitor, this estimate is in general agreement with simulations of stripped-envelope SNe from binary systems (a $3\;M_{\odot}$ helium core pre-explosion is expected to be $\sim$19~Myr old, compared to our $2.1\;M_{\odot}$ density distribution\textcolor{black}{;} see \citealt{2016Rimoldi_BinarySims}). Combined with the explosion parameters from previous sections and the derived progenitor mass of $M_{\rm ZAMS} = 9.5\,\pm\,1.0\;M_{\odot}$, our analysis strongly disfavors a single massive Wolf-Rayet as progenitor for the explosion \citep{2008Crockett_2007gr,2011Dessart_WRprogenitors}.

\section{Discussion and Conclusion}\label{sec:Discussion}
We have presented photometric and spectroscopic observations of the type-Ic SN~2020oi, which resides in the grand-design spiral galaxy M100. Our observations were obtained using Keck, SOAR, and other ground-based telescopes and span $\sim400$~days of the event, allowing us to characterize the explosion in detail. Additional pre-explosion \textit{HST} photometry and MUSE IFU spectroscopy has permitted a detailed investigation of the underlying stellar population at the location of the SN. Table \ref{tbl:finalproperties} lists the properties of both the SN and its host environment derived in previous sections.

 \begin{deluxetable}{lr}\label{tbl:finalproperties}
 \tablecaption{Derived Properties of SN~2020oi}
 \tablecolumns{2}
 \tablewidth{\textwidth}
 \tablehead{
 \colhead{} & \colhead{} 
 }
 \startdata
SFR at SN Site [$M_{\odot}\,yr^{-1}\,\rm{kpc}^{-2}$] & $6.0\,\pm\,1.2\,\times\,10^{-3}$  \\
Metallicity at SN Site ($\left<Z_{*}\right>$) [$Z_{\odot}$]& $\sim0.75$  \\
Total Reddening ($E(B-V)$) [mag] & $0.133\,\pm\,0.03$  \\
Cluster Age ($t_{\rm{Clust}}$) [Myr] & $40^{+30}_{-20}$ \\
Cluster Mass ($M_{\rm Clust}$) [$M_{\odot}$] & $7.24^{+2.33}_{-4.33}\,\times\, 10^{5}$ \\
Cluster Metallicity ($Z_{\rm Clust}$) [$Z_{\odot}$] & $0.03\,\pm\,0.02$\\
Date of Explosion ($t_{\rm exp}$) [MJD] & $58854.0\,\pm\,0.3$  \\
Bolometric Decline Rate ($\Delta m_{15, \rm bol}$) & $1.63\,\pm\,0.14$  \\
Kinetic Energy ($E_{k}$) $[10^{51}\,\rm{erg}]$ & $0.79\,\pm\,0.09$ \\
Ejecta Mass ($M_{\rm ej}$) [$M_{\odot}$]& $0.81\,\pm\,0.03$  \\
Mass of Synthesized $^{56}$Ni ($M_{\rm Ni56}$) [$M_{\odot}$] & $0.08\,\pm\,0.02$  \\
Progenitor ZAMS Mass ($M_{\rm{ZAMS}}$) [$M_{\odot}$] & $9.5\,\pm\,1.0$  \\
Progenitor Pre-Explosion Mass ($M_{f}$) [$M_{\odot}$] & $\sim2.1$  \\
Progenitor Mass-Loss Rate ($\dot{M}$) [$M_{\odot}\,\rm{yr}^{-1}$] & $\sim1.5\,\times\,10^{-4}$ \\
Progenitor Age ($t_{\rm{Age}}$) [Myr] & $27 \pm 7$ 
 \enddata
 \end{deluxetable}
Below, we summarize the primary conclusions associated with our analysis: 
\begin{enumerate}[nolistsep]
    \item Using the bolometric light curve code \texttt{Superbol} in tandem with a Gaussian Process routine to interpolate our photometric observations, we find SN~2020oi to be dimmer than the majority of SNe~Ic and with a photometric evolution similar to that of the type-Ic SN~1994I. We calculate a luminosity decline rate of $\Delta m_{\rm 15,bol} \approx 1.6$, higher than all stripped-envelope SNe analyzed in \textcolor{black}{both} \citet{2016Lyman_bolo} \textcolor{black}{and \citet{2018Taddia}}.
    
    \item We \textcolor{black}{separately} model the bolometric luminosity of the event in the photospheric phase using the modified one-component Arnett model described in \citet{2008_Valenti2003jd} \textcolor{black}{and} following the \citet{2019KhatamiKasen_Ni56} treatment for stripped-envelope SNe. \textcolor{black}{We further use the \texttt{MOSFiT} code \citep{2018AGuillochon_MOSFiT} to model the photoemtry of the event in each observed band. Adopting the results from \citet{2019KhatamiKasen_Ni56},} we find a mass of synthesized nickel of $M_{\rm Ni56} = 0.08\,\pm\,0.02\;M_{\odot}$ and a total ejecta mass of $M_{\rm ej} = 0.81\,\pm\,0.03\;M_{\odot}$. These values \textcolor{black}{fall at the lowest end of the} range reported by \citet{2018Taddia} for SNe~Ic, a result consistent with the faint bolometric light curve and the rapid decline of the explosion. \textcolor{black}{We derive an explosion time of MJD 58854.0 $\pm$ 0.3 using a fireball rise model applied to the first 10 days of photometry.}
    \item Detailed 1D spectral modeling using the radiative transfer code \texttt{TARDIS} reveals a composition near peak in strong agreement with the CO21 model developed to explain the spectral sequence of SN~1994I. We find evidence of \ion{Ca}{2}, \ion{Mg}{2}, \ion{Fe}{2}, \ion{Si}{2}, and \ion{O}{2} features and a best-fit composition that is remains roughly consistent across the epochs simulated, indicating at least partial ejecta mixing.
    
    \item The earliest spectrum obtained ($\delta t = 3.3$d) features an enhanced blue continuum that cannot be explained by the SN~1994I CO21 composition model. Further, we find evidence of \ion{Fe}{2} $\lambda$4500 but not \ion{O}{1} $\lambda$7773, indicating that this material is associated with the outermost layers of the ejecta but contains higher-mass elements typically observed at later epochs. We have obtained reasonable fits to this spectrum by considering an additional high-velocity ($<-23000\;\rm{km\,s^{-1}}$) gas component ($0.1\;M_{\odot}$) to the emission, with a distinct composition to the primary ejecta that includes carbon and potentially helium. 
    
    \item The optical and UV photometry near $\delta t \approx 2.5$ days reveals emission in excess of the expanding-fireball model. This excess is present in data obtained with Las Cumbres Observatory and with \textit{Swift}. We have considered several physical scenarios to explain this emission, including shock-cooling, binary interaction, CSM interaction, and an asymmetric distribution of nickel synthesized from the explosion. We slightly favor the interpretation of ejecta interaction with CSM material\textcolor{black}{, potentially from wave-driven mass-loss or mass transfer onto the companion at the time of the explosion}. Nevertheless, until a more complete picture of the diversity of possible signatures from each of these phenomena is known, we cannot rule out \textcolor{black}{alternative interaction mechanisms}. The flux excess could also potentially be explained by properties intrinsic to the type-Ic explosion; early observations of a statistical sample of events are needed to investigate this possibility.
    
    \item We have identified a marginally-extended source, likely a stellar cluster, coincident with the explosion in \textit{HST} pre-explosion imaging. By combining stellar evolution models from \texttt{BPASS} with modeling of the cluster photometry in \texttt{Prospector}, we derive an age for the SN~2020oi progenitor of $27\,\pm\,7\;\rm{Myr}$. This age is consistent with values predicted from previous starburst evolution models \citep{1995Knapen, 2006Allard}, and with the conceptual picture of the progenitor forming from dynamical interaction of the innermost spiral arms with cold gas in M100's nuclear ring. This is the sole SN of seven discovered in M100 whose location has allowed us to validate the mechanism underlying star formation in the nuclear ring.
    
    \item Our age constraints, coupled with an initial mass of $M_{\rm ZAMS} \approx 9.5\;M_{\odot}$ predicted from \texttt{BPASS} models and a pre-explosion mass of $M_{f} \approx 2.1\;M_{\odot}$ estimated from spectral modeling in \texttt{TARDIS}, present a consistent picture of a low-mass binary progenitor system for SN~2020oi.  \textcolor{black}{An explanation for the optical/UV excess and early spectrum of the explosion must be consistent with a binary progenitor system. The possibility of an explosion during an episode of mass-transfer will be examined in greater detail in a subsequent paper.}
\end{enumerate}

The results of this study highlight the value of early-time observations in constraining the nature of SN progenitors. From its initial discovery, SN~2020oi was closely monitored by the Young Supernovae Experiment \citep[YSE;][]{2021Jones_YSE}, which surveys 1,512~deg$^2$ of sky in $griz$ bands using the Pan-STARRS telescopes to a median 5-$\sigma$ depth of 21.5~mag. Although  surveys such as the Vera Rubin Observatory's Legacy Survey for Space and Time \citep[LSST;][]{2019Ivezic} will vastly expand our understanding of the diversity of stripped-envelope SNe, high-cadence photometry and spectroscopy from additional surveys such as YSE will be critical for distinguishing between progenitor models and expanding our sample of observed short-duration phenomena \citep[as in the case of the type-Ia SN~2018oh, observed by \textit{TESS};][]{Dimitriadis19:18oh_k2}. SN~2020oi is only the fourth spectroscopically-standard SN~Ic with excess flux detected pre-maximum, and this dearth of sufficient analogues for comparison challenges our ability to conclusively characterize this emission. Rapid follow-up of events identified in large surveys will allow us to construct a statistical sample of early-time phenomena and more accurately distinguish between their signatures.

\section*{Acknowledgements}\label{Acknowledgements}
We acknowledge  V.~Alarcon, K.~Clever, A.~Dhara,  J.~Lopez, S.~Medallon, J.~Nunez, C.~Smith, and E.~Strasburger for help with the Nickel observations presented in this paper. We further acknowledge J.~A.~Vilchez, A.~Campillay, Y.~K.~Riveros and N.~Ulloa for their assistance with the Swope observations.

A.G. is supported by the National Science Foundation Graduate Research Fellowship Program under Grant No.~DGE–1746047. A.G. also acknowledges funding from the Center for Astrophysical Surveys Fellowship at UIUC/NCSA and the Illinois Distinguished Fellowship.

W.J-G is supported by the National Science Foundation Graduate Research Fellowship Program under Grant No.~DGE-1842165 and the IDEAS Fellowship Program at Northwestern University. W.J-G acknowledges support through NASA grants in support of {\it Hubble Space Telescope} program GO-16075.

Parts of this research were supported by the Australian Research Council Centre of Excellence for All Sky Astrophysics in 3 Dimensions (ASTRO 3D), through project number CE170100013.

M.R.D. acknowledges support from the NSERC through grant RGPIN-2019-06186, the Canada Research Chairs Program, the Canadian Institute for Advanced Research (CIFAR), and the Dunlap Institute at the University of Toronto.

The UCSC team is supported in part by NASA grant 80NSSC20K0953; NSF grant AST-1815935; the Gordon \& Betty Moore Foundation; the Heising-Simons Foundation; and by a fellowship from the David and Lucile Packard Foundation to R.J.F.

The transient group at Northwestern is partially supported by the Heising-Simons Foundation under grant \#2018-0911 (PI: Margutti). R.M.~acknowledges support by the National Science Foundation under Award No. AST-1909796 and AST-1944985. Support for this work was provided by the National Aeronautics and Space Administration through Chandra Award Number DD0-21114X (PI Stroh) issued by the Chandra X-ray Center, which is operated by the Smithsonian Astrophysical Observatory for and on behalf of the National Aeronautics Space Administration under contract NAS8-03060.

D.O.J is supported by NASA through the NASA Hubble Fellowship grant HF2-51462.001 awarded by the Space Telescope Science Institute, which is operated by the Association of Universities for Research in Astronomy, Inc., for NASA, under contract NAS5-26555. D. A. Coulter acknowledges support from the National Science Foundation Graduate Research Fellowship under Grant DGE1339067. This work was supported by a VILLUM FONDEN Young Investigator Grant to C.G. (project number 25501) and a VILLUM FONDEN Investigator grant to J.H. (project number 16599). K.S.M. acknowledges funding from the European Research Council under the European Union’s Horizon 2020 research and innovation programme (Grant agreement No. 101002652), and from Horizon 2020, EU Grant agreement No. 873089. M.R.S. is supported by the National Science Foundation Graduate Research Fellowship Program under Grant No. 1842400. 

The scientific results reported in this article are based in part on observations made by the Chandra X-ray Observatory. This research has made use of software provided by the Chandra X-ray Center (CXC) in the application packages CIAO. Partial support for this work was provided by the National Aeronautics and Space Administration through Chandra Award Number DD0-21114X issued by the Chandra X-ray Center, which is operated by the Smithsonian Astrophysical Observatory for and on behalf of the National Aeronautics Space Administration under contract NAS8-03060.

The Pan-STARRS1 Surveys (PS1) and the PS1 public science archive have been made possible through contributions by the Institute for Astronomy, the University of Hawaii, the Pan-STARRS Project Office, the Max-Planck Society and its participating institutes, the Max Planck Institute for Astronomy, Heidelberg and the Max Planck Institute for Extraterrestrial Physics, Garching, The Johns Hopkins University, Durham University, the University of Edinburgh, the Queen's University Belfast, the Harvard-Smithsonian Center for Astrophysics, the Las Cumbres Observatory Global Telescope Network Incorporated, the National Central University of Taiwan, the Space Telescope Science Institute, the National Aeronautics and Space Administration under Grant No. NNX08AR22G issued through the Planetary Science Division of the NASA Science Mission Directorate, the National Science Foundation Grant No.\ AST-1238877, the University of Maryland, Eotvos Lorand University (ELTE), the Los Alamos National Laboratory, and the Gordon and Betty Moore Foundation.

This work makes use of observations from the FLOYDS spectrograph on the LCOGT 2m telescope at Siding Spring. This work also makes use of observations from the Las Cumbres Observatory global telescope network (NOIRLab Prop.\ IDs 2019B-0250, 2020A-0334; PI: R.\ Foley). 

A major upgrade of the Kast spectrograph on the Shane 3m telescope at Lick Observatory was made possible through generous gifts from the Heising-Simons Foundation as well as William and Marina Kast.  Research at Lick Observatory is partially supported by a generous gift from Google. 

Some of the data presented herein were obtained at the W.\ M.\ Keck Observatory, which is operated as a scientific partnership among the California Institute of Technology, the University of California and the National Aeronautics and Space Administration. The Observatory was made possible by the generous financial support of the W.\ M.\ Keck Foundation.  

The authors wish to recognize and acknowledge the very significant cultural role and reverence that the summit of Maunakea has always had within the indigenous Hawaiian community.  We are most fortunate to have the opportunity to conduct observations from this mountain. We further acknowledge that Lick Observatory sits on the unceded ancestral homelands of the Chochenyo and Tamyen Ohlone peoples, including the Alson and Socostac tribes, who were the original inhabitants of the area that includes Mt. Hamilton.

Based in part on observations obtained at the Southern Astrophysical Research (SOAR) telescope (NOIRLab Prop.\ IDs 2019B-0251, 2020A-0333; PI: R.\ Foley), which is a joint project of the Minist\'erio da Ci\^encia, Tecnologia e Inova\c{c}\~oes do Brasil (MCTI/LNA), the US National Science Foundation's NOIRLab, the University of North Carolina at Chapel Hill (UNC), and Michigan State University (MSU).  

We acknowledge the use of public data from the Neil Gehrels {\it Swift} Observatory data archive.

Based on observations obtained with the Samuel Oschin 48-inch Telescope at the Palomar Observatory as part of the Zwicky Transient Facility project. ZTF is supported by the National Science Foundation under Grant No. AST-1440341 and a collaboration including Caltech, IPAC, the Weizmann Institute for Science, the Oskar Klein Center at Stockholm University, the University of Maryland, the University of Washington, Deutsches Elektronen-Synchrotron and Humboldt University, Los Alamos National Laboratories, the TANGO Consortium of Taiwan, the University of Wisconsin at Milwaukee, and Lawrence Berkeley National Laboratories. Operations are conducted by COO, IPAC, and UW. The ZTF forced-photometry service was funded under the Heising-Simons Foundation grant \#12540303 (PI: Graham).

\textcolor{black}{\software{BPASS \citep[v2.2;][]{2017Elridge}, CIAO \citep[v4.13;][]{2006Fruscione_CIAO}, DoPhot \citep{Schechter93}, drizzlepac \citep{2012Gonzaga}, dynesty \citep{2020Speagle_dynesty}, emcee \citep{2013Foreman_MCMC}, HOTPANTS \citep{hotpants},  numpy \citep{walt2011numpy}, pandas \citep{reback2020pandas}, photpipe \citep{Rest05:photpipe,2018Kilpatrick_potentialProgenitor}, photutils\citep{photutils}, Prospector \citep{2017Leja},  PSFR \citep{2020Fusco_MUSEResolution},  Scipy \citep{jones2001scipy}, Superbol \citep{2018Nicholl_Superbol}, SWarp \citep{swarp}, STARLIGHT \citep{2005CidFernandes_STARLIGHT}, TARDIS \citep{Kerzendorf2014, Kerzendorf2018}}}

\bibliography{references}

\appendix

\hspace*{-10mm}
\begin{minipage}{0.5\textwidth}
\centering \captionof{table}{Optical and UV Photometry of SN~2020oi}
\vspace*{-1mm}
\label{tbl:optical_phot}
\begin{tabular}{crcrcl}
\toprule
MJD      & Phase   & Band    & Mag.  & Uncertainty & Telescope \\ \hline \hline
58856.46 & +2.46    & $UVW2$ & $>$13.95  & --     & $Swift$   \\
58856.79 & +2.79    & $UVW2$ & $>$13.92  & --     & $Swift$   \\
58857.45 & +3.45    & $UVW2$ & $>$14.17  & --     & $Swift$   \\
58858.85 & +4.85    & $UVW2$ & $>$13.95  & --     & $Swift$   \\
58859.49 & +5.49    & $UVW2$ & $>$13.93  & --     & $Swift$   \\
58860.38 & +6.38    & $UVW2$ & $>$13.96  & --     & $Swift$   \\
58861.25 & +7.25    & $UVW2$ & $>$13.83  & --     & $Swift$   \\
\hline
\end{tabular}
\end{minipage} \hfill
\begin{minipage}{0.5\textwidth}
\hspace*{-5mm}
\vspace*{-10mm}
\begin{tabular}{crcrcl}

\hline
MJD      & Phase   & Band    & Mag.  & Uncertainty & Telescope \\ \hline
58868.27 & +14.27   & $u$    & 15.29                & 0.08   & LCO       \\
58869.27 & +15.27   & $u$    & 15.54                & 0.07   & LCO       \\
58870.27 & +16.27   & $u$    & 15.93                & 0.10    & LCO       \\
58871.26 & +17.26   & $u$    & 16.16                & 0.17   & LCO       \\
58872.29 & +18.29   & $u$    & 16.49                & 0.07   & LCO       \\
58872.35 & +18.35   & $u$    & 16.11                & 0.03   & Swope     \\
58873.09 & +19.09   & $u$    & 16.73                & 0.06   & LCO       \\
\hline
\end{tabular}
\end{minipage} \hfill

\hspace*{-20mm}
\begin{minipage}{0.5\textwidth}
\begin{tabular}{crcrcl}
\hline
MJD      & Phase   & Band    & Mag.  & Uncertainty & Telescope \\ \hline
58943.46 & +89.46   & $u$    & $>$14.67  & --     & LCO       \\
58952.44 & +98.44   & $u$    & $>$16.07  & --     & LCO       \\
58966.35 & +112.35  & $u$    & $>$16.32  & --     & LCO       \\
58974.11 & +120.11  & $u$    & $>$15.64  & --     & LCO       \\
58980.76 & +126.76  & $u$    & $>$16.87  & --     & LCO       \\
58988.31 & +134.31  & $u$    & $>$16.04  & --     & LCO       \\
58996.50  & +142.50   & $u$    & $>$16.69  & --     & LCO       \\
\hline
\end{tabular}
\end{minipage} \hfill

\vspace*{-235.6mm}
\begin{minipage}{0.5\textwidth}
\hspace*{80mm}
\begin{tabular}{crcrcl}
\hline
MJD      & Phase   & Band    & Mag.  & Uncertainty & Telescope \\ \hline
58859.49 & +5.49    & $B$    & $>$14.44  & --     & $Swift$   \\
58860.38 & +6.38    & $B$    & $>$14.25  & --     & $Swift$   \\
58861.25 & +7.25    & $B$    & $>$13.99  & --     & $Swift$   \\
58864.60  & +10.60    & $B$    & 14.30                 & 0.37   & Synthetic \\
58868.60  & +14.60    & $B$    & 14.63                & 0.38   & Synthetic \\
58869.35 & +15.35   & $B$    & 14.58                & 0.03   & Swope     \\
58869.35 & +15.35   & $B$    & 14.73                & 0.03   & Swope     \\
\hline
\end{tabular}
\end{minipage} \hfill

\hspace*{-20mm}
\begin{minipage}{0.5\textwidth}
\begin{tabular}{crcrcl}
\hline
MJD      & Phase   & Band    & Mag.  & Uncertainty & Telescope \\ \hline
58910.26 & +56.26   & $B$    & 17.03                & 0.03   & Swope     \\
58911.21 & +57.21   & $B$    & 17.12                & 0.03   & Swope     \\
58912.28 & +58.28   & $B$    & 17.13                & 0.03   & Swope     \\
58921.26 & +67.26   & $B$    & 16.80                 & 0.10    & Swope     \\
58923.29 & +69.29   & $B$    & 16.71                & 0.04   & Swope     \\
58954.16 & +100.16  & $B$    & $>$18.31  & --     & Nickel    \\
58963.19 & +109.19  & $B$    & $>$17.82  & --     & Nickel    \\
\hline
\end{tabular}
\end{minipage} \hfill
\begin{minipage}{0.5\textwidth}
\hspace*{-5mm}
\begin{tabular}{crcrcl}

\hline
MJD      & Phase   & Band    & Mag.  & Uncertainty & Telescope \\ \hline
58859.30  & +5.30     & $g$    & 14.52                & 0.03   & LCO       \\
58859.52 & +5.52    & $g$    & 14.39                & 0.03   & ZTF       \\
58859.69 & +5.69    & $g$    & 14.39                & 0.03   & LCO       \\
58860.05 & +6.05    & $g$    & 14.65                & 0.07   & LCO       \\
58860.05 & +6.05    & $g$    & $>$16.55  & --     & LCO       \\
58860.05 & +6.05    & $g$    & 14.23                & 0.03   & LCO       \\
58860.07 & +6.07    & $g$    & 14.15                & 0.03   & LCO       \\
\hline
\end{tabular}
\end{minipage} \hfill

\hspace*{-20mm}
\begin{minipage}{0.5\textwidth}
\begin{tabular}{crcrcl}

\hline
MJD      & Phase   & Band    & Mag.  & Uncertainty & Telescope \\ \hline
58872.29 & +18.29   & $g$    & 14.68                & 0.03   & LCO       \\
58872.33 & +18.33   & $g$    & 14.60                 & 0.03   & Thacher   \\
58872.35 & +18.35   & $g$    & 14.73                & 0.03   & Swope     \\
58872.35 & +18.35   & $g$    & 14.61                & 0.03   & Swope     \\
58872.40  & +18.40    & $g$    & 14.85                & 0.38   & Synthetic \\
58873.09 & +19.09   & $g$    & 14.76                & 0.03   & LCO       \\
58873.09 & +19.09   & $g$    & 14.75                & 0.03   & LCO       \\
\hline
\end{tabular}
\end{minipage} \hfill
\begin{minipage}{0.5\textwidth}
\hspace*{-5mm}
\begin{tabular}{crcrcl}

\hline
MJD      & Phase   & Band    & Mag.  & Uncertainty & Telescope \\ \hline
58883.38 & +29.38   & $g$    & 16.00                 & 0.03   & Swope     \\
58883.99 & +29.99   & $g$    & 15.91                & 0.03   & LCO       \\
58884.00  & +30.00    & $g$    & 16.01                & 0.04   & LCO       \\
58884.37 & +30.37   & $g$    & 16.00                 & 0.03   & LCO       \\
58884.37 & +30.37   & $g$    & 15.84                & 0.30    & LCO       \\
58884.37 & +30.37   & $g$    & 16.01                & 0.03   & Swope     \\
58884.37 & +30.37   & $g$    & $>$17.25  & --     & LCO       \\
\hline
\end{tabular}
\end{minipage} \hfill

\hspace*{-20mm}
\begin{minipage}{0.5\textwidth}
\begin{tabular}{crcrcl}

\hline
MJD      & Phase   & Band    & Mag.  & Uncertainty & Telescope \\ \hline
58908.34 & +54.34   & $g$    & 16.55                & 0.03   & ZTF       \\
58909.36 & +55.36   & $g$    & 16.66                & 0.03   & Swope     \\
58910.25 & +56.25   & $g$    & 16.67                & 0.03   & Swope     \\
58911.20  & +57.20    & $g$    & 16.67                & 0.03   & Swope     \\
58911.40  & +57.40    & $g$    & 16.73                & 0.03   & ZTF       \\
58912.27 & +58.27   & $g$    & 16.68                & 0.03   & Swope     \\
58913.33 & +59.33   & $g$    & 16.62                & 0.03   & ZTF       \\
\hline
\end{tabular}
\end{minipage} \hfill

\vspace*{-240.6mm}
\begin{minipage}{0.5\textwidth}
\hspace*{80mm}
\begin{tabular}{crcrcl}
\hline
MJD      & Phase   & Band    & Mag.  & Uncertainty & Telescope \\ \hline
59007.71 & +153.71  & $g$    & $>$18.14  & --     & LCO       \\
59014.20  & +160.20   & $g$    & 20.23                & 0.28   & ZTF       \\
59014.80  & +160.80   & $g$    & $>$18.73  & --     & LCO       \\
59020.20  & +166.20   & $g$    & 19.92                & 0.22   & ZTF       \\
59023.22 & +169.22  & $g$    & 20.15                & 0.31   & ZTF       \\
59026.22 & +172.22  & $g$    & 19.02                & 0.18   & ZTF       \\
59026.35 & +172.35  & $g$    & $>$16.83  & --     & LCO       \\
\hline
\end{tabular}
\end{minipage} \hfill

\hspace*{-20mm}
\begin{minipage}{0.5\textwidth}
\begin{tabular}{crcrcl}

\hline
MJD      & Phase   & Band    & Mag.  & Uncertainty & Telescope \\ \hline
58871.36 & +17.36   & $V$    & 14.12                & 0.03   & Swope     \\
58872.36 & +18.36   & $V$    & 14.30                 & 0.03   & Swope     \\
58872.40  & +18.40    & $V$    & 14.31                & 0.37   & Synthetic \\
58873.09 & +19.09   & $V$    & $>$17.53  & --     & LCO       \\
58875.00  & +21.00    & $V$    & 14.63                & 0.38   & Synthetic \\
58875.38 & +21.38   & $V$    & 14.69                & 0.03   & Swope     \\
58876.37 & +22.37   & $V$    & 14.92                & 0.03   & Swope     \\
\hline
\end{tabular}
\end{minipage} \hfill
\begin{minipage}{0.5\textwidth}
\hspace*{-5mm}
\begin{tabular}{crcrcl}

\hline
MJD      & Phase   & Band    & Mag.  & Uncertainty & Telescope \\ \hline
58912.28 & +58.28   & $V$    & 16.58                & 0.03   & Swope     \\
58921.26 & +67.26   & $V$    & 16.50                 & 0.05   & Swope     \\
58923.29 & +69.29   & $V$    & 16.34                & 0.03   & Swope     \\
58963.20  & +109.20   & $V$    & $>$17.59  & --     & Nickel    \\
58995.60  & +141.60   & $V$    & 16.32                & 0.85   & Synthetic \\
58558.22 & -295.78 & $r$    & 18.58                & 0.18   & ZTF       \\
58575.37 & -278.63 & $r$    & 18.63                & 0.28   & ZTF       \\
\hline
\end{tabular}
\end{minipage} \hfill

\hspace*{-20mm}
\begin{minipage}{0.5\textwidth}
\begin{tabular}{crcrcl}

\hline
MJD      & Phase   & Band    & Mag.  & Uncertainty & Telescope \\ \hline
58859.05 & +5.05    & $r$    & 14.80                 & 0.03   & LCO       \\
58859.30  & +5.30     & $r$    & 14.58                & 0.03   & LCO       \\
58859.48 & +5.48    & $r$    & 14.39                & 0.03   & ZTF       \\
58859.69 & +5.69    & $r$    & 14.46                & 0.03   & LCO       \\
58860.05 & +6.05    & $r$    & 14.33                & 0.03   & LCO       \\
58860.08 & +6.08    & $r$    & 14.31                & 0.03   & LCO       \\
58860.08 & +6.08    & $r$    & 14.32                & 0.03   & LCO       \\
\hline
\end{tabular}
\end{minipage} \hfill
\begin{minipage}{0.5\textwidth}
\hspace*{-5mm}
\begin{tabular}{crcrcl}

\hline
MJD      & Phase   & Band    & Mag.  & Uncertainty & Telescope \\ \hline
58871.34 & +17.34   & $r$    & 13.88                & 0.03   & Thacher   \\
58871.35 & +17.35   & $r$    & 13.79                & 0.03   & Swope     \\
58871.50  & +17.50    & $r$    & 13.86                & 0.03   & ZTF       \\
58872.29 & +18.29   & $r$    & 13.96                & 0.03   & LCO       \\
58872.33 & +18.33   & $r$    & 13.99                & 0.03   & Thacher   \\
58872.35 & +18.35   & $r$    & 14.06                & 0.03   & Swope     \\
58872.40  & +18.40    & $r$    & 14.02                & 0.03   & Synthetic \\
\hline
\end{tabular}
\end{minipage} \hfill

\hspace*{-20mm}
\begin{minipage}{0.5\textwidth}
\begin{tabular}{crcrcl}

\hline
MJD      & Phase   & Band    & Mag.  & Uncertainty & Telescope \\ \hline
58882.32 & +28.32   & $r$    & 15.31                & 0.03   & Swope     \\ 
58883.01 & +29.01   & $r$    & 15.24                & 0.03   & LCO       \\
58883.12 & +29.12   & $r$    & 15.28                & 0.03   & LCO       \\
58883.12 & +29.12   & $r$    & 15.34                & 0.03   & LCO       \\
58883.37 & +29.37   & $r$    & 15.40                 & 0.03   & Swope     \\
58884.00  & +30.00    & $r$    & 15.42                & 0.03   & LCO       \\
58884.00  & +30.00    & $r$    & 15.39                & 0.03   & LCO       \\
\hline
\end{tabular}
\end{minipage} \hfill
\begin{minipage}{0.5\textwidth}
\hspace*{-5mm}
\begin{tabular}{crcrcl}

\hline
MJD      & Phase   & Band    & Mag.  & Uncertainty & Telescope \\ \hline
58904.35 & +50.35   & $r$    & $>$18.74  & --     & Nickel    \\
58904.47 & +50.47   & $r$    & 16.21                & 0.03   & ZTF       \\
58905.28 & +51.28   & $r$    & 16.23                & 0.03   & Swope     \\
58905.28 & +51.28   & $r$    & 16.34                & 0.03   & Swope     \\
58905.57 & +51.57   & $r$    & 16.28                & 0.07   & Thacher   \\
58906.27 & +52.27   & $r$    & 16.24                & 0.03   & Swope     \\
58906.45 & +52.45   & $r$    & 16.25                & 0.03   & ZTF       \\
\hline
\end{tabular}
\end{minipage} \hfill

\hspace*{-20mm}
\begin{minipage}{0.5\textwidth}
\begin{tabular}{crcrcl}

\hline
MJD      & Phase   & Band    & Mag.  & Uncertainty & Telescope \\ \hline
58974.11 & +120.11  & $r$    & $>$16.83  & --     & LCO       \\
58974.25 & +120.25  & $r$    & 19.40                 & 0.30    & ZTF       \\
58976.42 & +122.42  & $r$    & 17.96                & 0.16   & PS1       \\
58978.29 & +124.29  & $r$    & 19.55                & 0.24   & ZTF       \\
58980.25 & +126.25  & $r$    & 19.19                & 0.12   & ZTF       \\
58980.76 & +126.76  & $r$    & $>$16.86  & --     & LCO       \\
58988.32 & +134.32  & $r$    & $>$16.91  & --     & LCO       \\
\hline
\end{tabular}
\end{minipage} \hfill
\begin{minipage}{0.5\textwidth}
\hspace*{-5mm}
\begin{tabular}{crcrcl}

\hline
MJD      & Phase   & Band    & Mag.  & Uncertainty & Telescope \\ \hline
58856.41 & +2.41    & $i$    & 16.12                & 0.04   & LCO       \\
58856.70  & +2.70     & $i$    & $>$14.68  & --     & LCO       \\
58856.70  & +2.70     & $i$    & 15.11                & 0.05   & LCO       \\
58856.70  & +2.70     & $i$    & 15.75                & 0.07   & LCO       \\
58856.74 & +2.74    & $i$    & 16.08                & 0.06   & LCO       \\
58856.74 & +2.74    & $i$    & 16.06                & 0.03   & LCO       \\
58857.30  & +3.30     & $i$    & 15.93                & 0.04   & LCO       \\
\hline
\end{tabular}
\end{minipage} \hfill

\hspace*{-20mm}
\begin{minipage}{0.5\textwidth}
\begin{tabular}{crcrcl}

\hline
MJD      & Phase   & Band    & Mag.  & Uncertainty & Telescope \\ \hline
58867.40  & +13.40    & $i$    & 13.79                & 0.03   & LCO       \\
58868.28 & +14.28   & $i$    & 14.00                 & 0.03   & LCO       \\
58868.60  & +14.60    & $i$    & 13.84                & 0.35   & Synthetic \\
58869.27 & +15.27   & $i$    & 13.98                & 0.03   & LCO       \\
58869.35 & +15.35   & $i$    & 13.85                & 0.03   & Swope     \\
58869.36 & +15.36   & $i$    & 13.92                & 0.03   & LCO       \\
58869.36 & +15.36   & $i$    & 13.94                & 0.03   & LCO       \\
\hline
\end{tabular}
\end{minipage} \hfill
\begin{minipage}{0.5\textwidth}
\hspace*{-5mm}
\begin{tabular}{crcrcl}

\hline
MJD      & Phase   & Band    & Mag.  & Uncertainty & Telescope \\ \hline
58880.12 & +26.12   & $i$    & 15.08                & 0.03   & LCO       \\
58880.12 & +26.12   & $i$    & 15.05                & 0.03   & LCO       \\
58880.29 & +26.29   & $i$    & 15.06                & 0.03   & Thacher   \\
58880.30  & +26.30    & $i$    & 14.99                & 0.38   & Synthetic \\
58880.40  & +26.40    & $i$    & 15.13                & 0.39   & Synthetic \\
58880.44 & +26.44   & $i$    & $>$17.34  & --     & Nickel    \\
58881.28 & +27.28   & $i$    & 15.16                & 0.03   & Thacher   \\
\hline
\end{tabular}
\end{minipage} \hfill

\hspace*{-20mm}
\begin{minipage}{0.5\textwidth}
\begin{tabular}{crcrcl}
\hline
MJD      & Phase   & Band    & Mag.  & Uncertainty & Telescope \\ \hline
58905.58 & +51.58   & $i$    & 16.25                & 0.16   & Thacher   \\
58906.10  & +52.10    & $i$    & $>$17.05  & --     & LCO       \\
58906.10  & +52.10    & $i$    & 15.61                & 0.04   & LCO       \\
58906.28 & +52.28   & $i$    & 16.49                & 0.03   & Swope     \\
58907.36 & +53.36   & $i$    & 16.39                & 0.03   & Swope     \\
58908.30  & +54.30    & $i$    & 16.54                & 0.03   & Swope     \\
58908.30  & +54.30    & $i$    & 16.55                & 0.03   & Swope     \\
\hline
\end{tabular}
\end{minipage} \hfill
\begin{minipage}{0.5\textwidth}
\hspace*{-5mm}
\vspace*{91.8mm}
\begin{tabular}{crcrcl}

\hline
MJD      & Phase   & Band    & Mag.  & Uncertainty & Telescope \\ \hline
59267.12 & +413.12  & $i$    & $>$16.71  & --     & LCO       \\
58857.30  & +3.30     & $z$    & 15.56                & 0.40    & Synthetic \\
58862.53 & +8.53    & $z$    & 14.08                & 0.04   & Thacher   \\
58864.34 & +10.34   & $z$    & 13.84                & 0.03   & Thacher   \\
58864.60  & +10.60    & $z$    & 13.82                & 0.35   & Synthetic \\
58866.33 & +12.33   & $z$    & 13.79                & 0.04   & Thacher   \\
58867.36 & +13.36   & $z$    & 13.78                & 0.04   & Thacher   \\
\hline
\end{tabular}
\end{minipage} \hfill 

\end{document}